\documentclass[12pt,a4wide]{article}
\usepackage{hyperref}
\usepackage{a4wide} 
\usepackage{graphicx}
\graphicspath{ {./images/} }
\usepackage{url}
\usepackage{cite}
\usepackage{textcomp} 
\usepackage{mathrsfs}
\usepackage{mathtools}
\usepackage{amstext,amsmath,amssymb}
\newif\ifshowcitations\showcitationsfalse%
\newif\ifshowlinks\showlinksfalse%

 \usepackage[table]{xcolor}
\usepackage{mathtools}
\makeatletter
\newcommand{\vast}{\bBigg@{4}}
\newcommand{\Vast}{\bBigg@{5}}
\newcommand{\vastl}{\mathopen\vast}

\newcommand{\vastr}{\mathclose\vast}

\makeatother

\usepackage{color}

\usepackage{enumitem}
\usepackage[utf8]{inputenc} 
\ifshowlinks%
  \usepackage[
         colorlinks=true,
         urlcolor=blue,       
         ]{hyperref}
  \newcommand*{\inspireurl}[1]{\\\href{#1}{INSPIRE-HEP entry}}
\else
  \makeatletter
  \newcommand*{\inspireurl}[1]{\@bsphack\@esphack}
  \makeatother
\fi
\ifshowcitations%
  \newcommand*{\citations}[1]{\\* #1}
\else
  \makeatletter
  \newcommand*{\citations}[1]{\@bsphack\@esphack}
  \makeatother
\fi

\def\d{\mathrm{d}}
\def\G{\mathcal{G}}
\def\xih{{\hat \xi}}
\def\O{\mathcal{O}}
\def\C{\mathcal{C}}
\def\V{\mathcal{V}}
\def\L{\mathcal{L}}
\def\be{\begin{equation}}
\def\ee{\end{equation}}
\def\R{\mathcal{R}}
 
\begin{document}
\pagestyle{plain}
\begin{center}
\medskip
{\bf\LARGE{Moduli Stabilization in String Theory}}
\vspace{10mm}

\end{center} 

\begin{center}
Liam McAllister$^{a}$ and Fernando Quevedo$^{b}$
\end{center}

\begin{center}

\vspace{0.15 cm}
{\fontsize{11}{30}
\noindent\textsl{$^{a}$Department of Physics, Cornell University, Ithaca, NY 14853, USA}\\
\noindent\textsl{$^{b}$DAMTP, University of Cambridge, Wilberforce Road, Cambridge, CB3 0WA, UK}}\\
\end{center}

\vspace{1cm}
\noindent We give an 
overview of moduli stabilization in compactifications of string theory.
We summarize current methods for construction and analysis of vacua with stabilized moduli, and we describe applications to cosmology
and particle physics.
This is a contribution to the Handbook on Quantum Gravity.
 
\vspace{8.25cm}
\noindent\today
\newpage

\tableofcontents
\newpage
\section{Introduction}

The fundamental physical laws that govern our Universe must describe gravity and quantum mechanics.  
To discover the laws of quantum gravity, we cannot entirely rely on terrestrial experiments, or even on cosmological observations: the energies of observable processes are far too low to give a complete picture, in contrast to the way that collider experiments eventually revealed the Standard Model of particle physics.  We may hope for some guidance from experiment, but theorists will have to provide a framework.

String theory is such a framework: it is a theory of quantum gravity through which 
we can take a constructive approach to exploring possible laws of quantum gravity in our Universe.
The first obstacle is that the world we observe at low energies is four-dimensional, while the best-understood solutions of string theory are ten-dimensional.  
Kaluza-Klein theory \cite{Kaluza:1921tu,Klein:1926tv}, now more than a century old, provides a way to bridge this gap.  If the extra dimensions correspond to a six-dimensional compact space that is smaller than the reach of any  experimental probe 
then only three spatial dimensions will be seen.

However, the size and shape of the extra dimensions are dynamical: they are parameterized by the expectation values of scalar fields known as \emph{moduli}.
Unless the moduli have large masses, they mediate long-range forces that are not observed in our world.
Thus, a central problem of  Kaluza-Klein theories is to provide a dynamical explanation for the requisite size of the extra dimensions, and to ensure that the moduli masses are consistent with experiment.
Addressing these challenges is the main obstacle in connecting string theory to observations, and it is the subject of this review.\footnote{For previous reviews  on fluxes and the string landscape see \cite{Grana:2005jc,Douglas:2006es,Hebecker:2021egx}. For string cosmology see \cite{Baumann:2014nda}, the recent review \cite{Cicoli:2023opf}, and references therein.
The geometry of string compactifications is treated in \cite{Tomasiello:2022dwe}.
For a more mathematical perspective on flux compactifications, see \cite{MikeEMP}.}

\section{The vacuum problem}
 
To understand quantum gravity in our four-dimensional, non-supersymmetric Universe, we will study compactifications of superstring theory on six-dimensional compact spaces, and seek solutions in which supersymmetry is broken.
In this section we will carefully explain the reasoning that directs us to the class of solutions that are the subject of this chapter: namely, flux compactifications on orientifolds of Calabi-Yau threefolds.

To begin, we take a product ansatz for a ten-dimensional spacetime,
\begin{equation}\label{eq:productansatz}
 ds^2 = g_{\mu\nu}(x)\d x^{\mu}\d x^{\nu}+g_{mn}(y)\d y^m \d y^n\,,   
\end{equation} for $\mu,\nu = 0,\ldots,3$ and $m,n=4,\ldots,9$.  We suppose that $g_{mn}$ is a Riemannian metric on some compact space\footnote{This space may be a proper six-manifold, or it may be some more singular space on which string theory remains well-defined, but in both cases we will use the term `manifold'.} $X_6$. Defining the Ricci tensors $R_{\mu\nu}$ and $R_{mn}$ constructed from $g_{\mu\nu}$ and $g_{mn}$, respectively, the ansatz \eqref{eq:productansatz} solves the ten-dimensional vacuum Einstein equations if and only if  $R_{\mu\nu} = R_{mn} = 0$.  Thus, vacuum solutions of string theory are furnished by Ricci-flat six-manifolds.

Consider a six-manifold $X_6$ with K\"ahler metric $g_{mn}$, and define the Riemannian holonomy group $\text{Hol}(g)$.  
If $\text{Hol}(g) = SU(3) \subset SO(6)$, then
$g$ is Ricci-flat, and we call $X_6$ a Calabi-Yau threefold.
It follows that $X_6$ having holonomy $SU(3)$ is a sufficient condition for a vacuum solution of string theory in compactification on $X_6$.  
One striking property of such solutions is that the compactification preserves some of the supersymmetry found in the ten-dimensional theory.

We are aware of no argument that $\text{Hol}(g) = SU(3)$ is a necessary condition for Ricci-flatness, but every example of a compact Ricci-flat six-manifold constructed to date has holonomy contained in $SU(3)$ (see e.g.~\cite{Acharya:2019mcu}).
Even so, one should bear in mind the possibility that there may exist compact Ricci-flat six-manifolds with holonomy $SO(6)$ that furnish non-supersymmetric vacua of string theory.  
By necessity we will restrict our discussion to the known examples
of vacuum solutions, i.e.~to Calabi-Yau threefolds.  We stress, however, that the presence of supersymmetry in this context is \emph{not} a consequence of an attempt to explain hierarchies in particle physics: it is instead a powerful aid to theoretical control. Specifically, the superstring is our starting point because superstring theories are vastly better-understood than non-supersymmetric string theories, and Calabi-Yau compactifications --- which preserve some supersymmetry at the compactification scale, though this may be broken to nothing at lower energies --- are our geometric focus because they are the only compact Ricci-flat six-manifolds known at the time of this writing.\\

Vacuum solutions of string theory of the form \eqref{eq:productansatz}, with $g_{\mu\nu}=\eta_{\mu\nu}$ and with $X_6$ a Calabi-Yau threefold, are a starting point for discussing cosmology in quantum gravity, but these solutions are far from realistic.  The primary shortcomings are two unwanted features: unbroken supersymmetry, and light scalar fields known as moduli.  

Compactifications of type II string theory on Calabi-Yau threefolds preserve $\mathcal{N}=2$ supersymmetry in four dimensions, and have exact moduli spaces parameterized by the scalar components of vector multiplets and hypermultiplets.  These moduli spaces correspond to geometric deformations of the Ricci-flat metric.  Our world exhibits no supersymmetry at all in the infrared, and in particular the chiral representations seen in the Standard Model are incompatible with $\mathcal{N}=2$ supersymmetry.  Moreover, massless scalar fields with gravitational-strength couplings --- which is precisely how the geometric moduli manifest in four dimensions --- mediate long-range forces\footnote{These limits apply only to spinless fields that are parity-even: pseudoscalars instead mediate spin-dependent forces that are much less constrained: see \S\ref{sec:axions}.} that are highly constrained, and are also subject to stringent limits from 
Big Bang nucleosynthesis.  In sum, type II compactifications on Calabi-Yau threefolds, without any sources of stress-energy, are plagued by massless moduli and extended supersymmetry, and so cannot furnish realistic models of our Universe.

More promising are compactifications that are not vacuum solutions, but instead solve the ten-dimensional equations of motion in the presence of sources of stress-energy in the internal space.
Quantization of the closed string famously reveals a massless spin-2 excitation, the graviton, but there are other important massless fields, including $p$-form gauge potentials with associated $(p+1)$-form field strengths known as fluxes.  These potentials and field strengths generalize the one-form potential and the electric and magnetic fields of  Maxwell theory, respectively.
For example, in type IIB string theory one finds two-form potentials $B_2$ and $C_2$ with associated three-form field strengths $H_3 = \d B_2$ and $F_3 = \d C_2$.  Dirac quantization ensures that in an appropriate normalization, these three-form fluxes are elements of $H^3(X_6,\mathbb{Z})$.  Quantized fluxes are a critical source of stress-energy, and give the subject of flux compactifications its name.  
In the type II and type I theories, consistency conditions require that flux appears in association with other, localized sources: D-branes and orientifold planes.

In the broadest sense, a flux compactification could mean any compactification of string theory containing $p$-form fluxes.
However, in this chapter we will use a restricted definition that nevertheless encompasses much of the work aiming to describe our Universe in string theory.   By a flux compactification, we will mean a compactification of a superstring theory on a Calabi-Yau threefold,\footnote{We briefly discuss compactifications of M-theory on manifolds of $G_2$ holonomy in \S\ref{ss:mth}.} or on an orientifold thereof, that includes $p$-form fluxes and any necessary localized sources.
Configurations of this sort are non-vacuum solutions.

The principal motivation for studying flux compactifications is that the stress-energy of fluxes and localized sources ameliorates the problems of massless moduli and unbroken supersymmetry that are inevitable in Calabi-Yau vacuum configurations.  
A general expectation, which we will explain in detail below, is that flux compactifications that contain sufficiently generic fluxes lead to four-dimensional theories that preserve $\mathcal{N}=1$ or $\mathcal{N}=0$ supersymmetry, and in which all moduli are massive.
We will refer to the set of such isolated vacua as the \emph{landscape of flux vacua}.\footnote{`The Landscape' is often used to refer to the more general set of all consistent quantum gravity theories, in any number of spacetime dimensions, with or without moduli, but in this chapter we focus on isolated four-dimensional vacua.}

In practice it is rarely possible to find vacua that preserve $\mathcal{N}=0$ supersymmetry and have all moduli massive already at the classical level.  Instead one can proceed by finding configurations that preserve $\mathcal{N}=1$ supersymmetry in some approximation --- for example, at leading order in the string loop and $\alpha'$ expansions --- and hence have moduli spaces parameterized by the scalar components of chiral multiplets.  General arguments predict that the moduli acquire mass from quantum effects, and that the exact vacuum configurations are isolated points within the leading-order moduli spaces.  Finding these vacua is feasible in some cases, as we will show.    

To recap, in aiming to describe the Universe within string theory, one seeks isolated and non-supersymmetric solutions of the equations of motion of string theory.  
A primary strategy of recent decades is to begin with Calabi-Yau vacuum configurations of one of the superstring theories and introduce sources of stress-energy --- quantized fluxes, as well as localized sources such as D-branes --- that lead to controllably small corrections to the vacuum solution.  Among these corrections are supersymmetry-breaking mass splittings, as well as masses for moduli fields.

The study of isolated, non-supersymmetric, non-vacuum solutions is necessarily founded on approximations.  No exact solutions are known, and none appears to us to be on the horizon.
In this respect, flux compactification falls into the mainstream of theoretical physics, but differs from some areas of string theory, and more general mathematical physics, in which exact methods are prevalent.  At the same time, the study of flux vacua owes a great deal to approximation schemes that are founded on systematic expansions around exact results, e.g.~using mirror symmetry or nonrenormalization, as we will see.

\section{Type IIB flux compactifications}

The primary class of flux compactifications surveyed in this article are compactifications of type IIB string theory on orientifolds of Calabi-Yau threefolds.  To explain this class of solutions we will first introduce the type IIB supergravity action and the two fundamental expansions, in $g_s$ and $\alpha'$ (\S\ref{sec:pert}).  In \S\ref{ss:isd} we introduce the central ansatz of \emph{imaginary self-dual} (ISD) three-form fluxes, and in \S\ref{sec:sup} and \S\ref{ss:k} we examine the superpotential and K\"ahler potential that describe ISD configurations.\\

The massless bosonic spectrum of type IIB string theory in ten dimensions consists of the metric $g_{MN}$, two-form $B_2$ and dilaton $\phi$ in the Neveu-Schwarz-Neveu-Schwarz sector, and the $p$-form potentials $C_0, C_2, C_4$ in the Ramond-Ramond sector. 
We define the three-form fluxes 
\begin{equation}
    F_3 := \d C_2\,, \qquad H_3 := \d B_2\,,\qquad G_3 := F_3 - \tau H_3\,, 
\end{equation}
the five-form 
\begin{equation}
    \tilde{F}_5 := \d C_4 + \frac{1}{2} B_2 \wedge F_3 -\frac{1}{2} C_2 \wedge H_3\,,
\end{equation}
and the complex axiodilaton  
\begin{equation}
    \tau:=C_0+ie^{-\phi}\,. 
\end{equation}

The full ten-dimensional effective action $S$ for the bosonic fields can be written as
$S=S_{10} +S_{\rm{loc}}$, with $S_{10}$ a bulk action and $S_{\rm{loc}}$ encoding the contributions of localized objects such as D-branes.  
The bulk action $S_{10}^{(0)}$ at leading order in the $g_s$ and $\alpha'$ expansions is 
\begin{equation}\label{eq:s10bulk}
 S_{10}^{(0)} =\frac{1}{2\kappa_{10}^2}\int\sqrt{-g}\left( \mathcal{R}-\frac{\left |\nabla \tau \right |^2}{2\,(\rm{Im}\,\tau)^2} -\frac{\left |G_3\right |^2}{12\,\rm{Im}\,\tau}-\frac{\left |F_5 \right |^2}{4{\cdot} 5!}\right)+\frac{1}{8 i \kappa_{10}^2}\int \frac{C_4\wedge G_3\wedge \overline G_3}{\rm{Im}\,\tau}\,,
\end{equation}
where $\kappa_{10}$ is the ten-dimensional gravitational coupling, and the Einstein-frame metric $g_{MN}$ is related to the string-frame metric $\hat g_{MN}$ by $g_{MN}=\sqrt{\rm{Im}\tau}\,\hat g_{MN}$.
The self-duality condition $\tilde{F}_5 = \star_{10} \tilde{F}_5$ must be imposed as a constraint in addition to the equations of motion that follow from \eqref{eq:s10bulk}.
The $\tilde{F}_5$ Bianchi identity reads
\begin{equation}\label{eq:f5bi}
    \mathrm{d}\tilde{F}_5 = H_3 \wedge F_3 + \rho_{\text{loc}}^{\text{D3}}\,.
\end{equation}
D-branes carry Ramond-Ramond charge, and in particular 
$C_4$ couples to the worldvolume of a D3-brane, which is why 
the source term is denoted by $\rho_{\text{loc}}^{\text{D3}}$: it captures the D3-brane charge $Q_{\text{loc}}^{\text{D3}}:= \int \rho_{\text{loc}}^{\text{D3}}$ from localized sources, including, but not limited to, actual D3-branes.\footnote{Our conventions are such that D3-branes and anti-D3-branes carry positive and negative D3-brane charge $Q_{\text{loc}}^{\text{D3}}$, respectively.}

\subsection{Scalings and perturbative expansions}\label{sec:pert}

The action \eqref{eq:s10bulk} receives corrections in two different expansions: the $\alpha'$ and string loop expansions. 
These two perturbative expansions can be traced to 
the 
two scaling symmetries of the tree-level bulk action \eqref{eq:s10bulk} \cite{Burgess:2020qsc,Cicoli:2021rub}.
We consider the transformations 
\begin{enumerate}[label=(\roman*)]
\item $\qquad\tau \rightarrow a^2 \tau\,, \qquad G_3\rightarrow a G_3\,,\qquad S_{10}\rightarrow S_{10}$\,,
\item $\qquad g_{MN}\rightarrow \lambda^\nu g_{MN}\,,\qquad \tau\rightarrow \lambda^{-2\nu}\tau\,, \qquad \tilde{F}_5\rightarrow \lambda^{2\nu}\tilde{F}_5\,, \qquad S_{10}\rightarrow \lambda^{4\nu}S_{10}$\,.
\end{enumerate}
The transformation (i) is a symmetry contained in
the general $SL(2,\mathbb{R})$ invariance of \eqref{eq:s10bulk}, which acts as
\begin{equation}
  \tau\rightarrow \frac{a\tau+b}{c\tau+d}\,, \qquad G_3\rightarrow \frac{G_3}{c\tau+d}\,.
\end{equation}
On the other hand, although (ii) leaves invariant the tree-level equations of motion, it is not a symmetry of the action: $S_{10}$ is not invariant under (ii), but scales.

The transformations (i) and (ii) can be combined, writing $a=\lambda^\omega$, as:
\be
 g_{MN}\rightarrow \lambda^\nu  g_{MN}\,,\qquad \tau\rightarrow \lambda^{2(\omega-\nu)}\tau\,, \qquad G_3\rightarrow \lambda^{\omega}G_3\,, \qquad F_5\rightarrow \lambda^{2\nu}F_5\,, \qquad S_{10}\rightarrow \lambda^{4\nu}S_{10}\,.
\ee
Both scaling  symmetries are broken by quantum and $\alpha'$ corrections, and as we will now explain, the corresponding perturbative expansions can be organized
by the amount that the scaling symmetries are broken. 

The string loop expansion is in powers of $g_s=\langle e^{\phi}\rangle$, and the $\alpha'$ expansion is essentially a derivative expansion. 
We write the general action, including all perturbative corrections, schematically as: 
\be
S_{10}=\sum_{m,n=0}^\infty (\alpha')^m g_s^n S_{10}^{(m,n)}
\ee
with a similar expansion for $S_{\rm loc}$. Here $S_{10}^{(m,n)} $ can be written as \cite{Burgess:2020qsc}:
\be
S_{10}^{(m,n)}\equiv S_{10}^{(p,r,n)} 
\propto \int  \sqrt{-g} \left( \frac{1}{\hbox{Im}\, \tau} \right)^{(2n-p+r+1)/2} \left({g}^{\circ \circ} \R^\circ_{~\circ \circ\circ} \right)^{p}  \left[ {g}^{\circ \circ}  {g}^{\circ \circ}  {g}^{\circ \circ} G_{\circ\circ\circ} G_{\circ\circ\circ} \right]^r +\ldots\,,  
\label{Smn}
\ee
where $\circ$ determines the appropriate index structure, and we have split $m=p+r-1$ since it is convenient to separate the $m$ dependence in terms of powers of the curvature $\R$ and powers of $G_3$. We have not written explicitly the terms depending on $F_5$ or on higher derivatives of the dilaton, which we include in the ellipses. The $p$ and $r$ powers in $\hbox{Im}\,\tau$ appear because we are writing the action in Einstein frame.

Notice that both $p=1$, $r=0$ and $p=0$, $r=1$ give tree-level results for $m=0$, $n=0$. Also, taking $m=3$ with all combinations of $p,r=0,1,2,3$ satisfying $p+r=4$ reproduces the known $\alpha'^3$ corrections (starting with $\R^4$, etc.) that are known to be the first non-vanishing $\alpha'$ corrections (see for instance \cite{Policastro:2006vt,Policastro:2008hg,Liu:2022bfg}). Terms with smaller values of $m$ cancel due to supersymmetry. The corrected actions $S_{10}^{(m,n)}$ transform under the scalings as:
\be\label{mandn}
S_{10}^{(m,n)} \rightarrow \lambda^{4\nu-2n(w-\nu) +m (w-2\nu)} \, S_{10}^{(m,n)}\,.
\ee

We can then use the scaling transformations as a bookkeeping tool to identify the different corrections $S_{10}^{(m,n)}$. This will be useful in section \S\ref{sss:kpert} in order to uncover the structure of the four-dimensional effective actions in string compactifications.

\subsection{Calabi-Yau compactifications}\label{sec:cyc}

In compactifying type IIB string theory on a Calabi-Yau threefold $X$, one needs to specify the three-form fluxes $H_3$ and $F_3$ \cite{Dasgupta:1999ss, Giddings:2001yu}.  Dirac quantization requires that (in units where $(2\pi)^2\alpha' \equiv \ell_s^2 = 1$)
\begin{equation}
    F_3, H_3 \in H^3(X,\mathbb{Z})\,,
\end{equation} and so the data of a choice of fluxes is a set of integers.

Gauss's law presents an immediate obstacle to constructing flux compactifications.  Many sources of stress energy, including D-branes as well as certain flux configurations, carry positive charge under Ramond-Ramond $p$-form gauge symmetries.
In a compact space, the total charge must be zero, so in the absence of sources with negative charge, a consistent compactification must have no charged sources whatsoever.
Specifically, integrating \eqref{eq:f5bi}, we find the constraint
\begin{equation}\label{eq:gauss}
  0 =   Q_{\text{loc}}^{\text{D3}} + \int H_3 \wedge F_3 = Q_{\text{loc}}^{\text{D3}} + Q_{\text{flux}}^{\text{D3}}\,.
\end{equation}
Thus, a solution in which $Q_{\text{flux}}^{\text{D3}}$ is positive 
--- and we will see soon that these are precisely the solutions of interest for moduli stabilization --- is then possible only in the presence of localized sources of negative D3-brane charge.

A way forward is provided by orientifolds.  
Orientifold planes are non-dynamical objects that have negative tension, and carry negative charge with respect to Ramond-Ramond potentials.  As an example, consider type IIB string theory compactified on a Calabi-Yau threefold $X_6$.
At certain loci in the complex structure moduli space of $X_6$, the manifold may admit a holomorphic involution 
\begin{equation}
\sigma: X_6 \to X_6\,,   
\end{equation}
whose fixed loci are points and/or divisors in $X_6$.  
If the action of $\sigma$ is paired with orientation reversal $\Omega_{\text{ws}}$ on the string worldsheet and multiplication by $(-1)^{F_L}$, with $F_L$ the left-moving fermion number, 
\begin{equation}
\mathcal{O} := \sigma\, \Omega_{\text{ws}} (-1)^{F_L}\,,
\end{equation}
then the configuration that results from projecting onto states with $\mathcal{O}=+1$ is called an O3/O7 orientifold.
The fixed loci, which are known as O3-planes and O7-planes,
carry negative\footnote{In certain cases, O7-planes that break supersymmetry can carry positive charge: see e.g.~\cite{Carta:2020ohw}.} charge with respect to $C_4$, i.e.~they contribute negatively to $Q_{\text{loc}}^{\text{D3}}$.

\begin{table}
    \centering
    \setlength\extrarowheight{5pt}
    \begin{tabular}{ccc}
        Field & Symbol & Range \\
        \hline
         $\phantom{\bigl.}$K\"ahler moduli & $T_a$ & $a=1,\ldots,h^{1,1}_+$\\
          $\phantom{\bigl.}$ complex structure moduli & $z_i$  & $i=1,\ldots,h^{2,1}_-$\\
          $\phantom{\bigl.}$two-forms & $G_{\alpha}$ & $\alpha=1,\ldots,h^{1,1}_-$\\
          $\phantom{\bigl.}$vector multiplets & $V_r$ & $r=1,\ldots,h^{2,1}_+$\\
          $\phantom{\bigl.}$axiodilaton & $\tau$ & $-$
          \\[.6ex]
          \hline
    \end{tabular}
    \caption{$\mathcal{N}=1$ multiplets in an O3/O7 orientifold.}
    \label{tab:multiplets}
\end{table}

The orientifold projection removes half of the supercharges preserved by $X_6$, leading to a theory with $\mathcal{N}=1$ supersymmetry in four dimensions.  
From each hypermultiplet arising in type IIB compactification on $X_6$, a chiral multiplet survives the projection, while from each $\mathcal{N}=2$ vector multiplet either an $\mathcal{N}=1$ vector multiplet or a chiral multiplet survives.  
The action of $\mathcal{O}$ on Dolbeault cohomology classes defines even and odd eigenspaces $H^{p,q}_{\pm}$ with corresponding dimensions $h^{p,q}_{\pm}$.  
The $h^{1,1}$ hypermultiplets yield $h^{1,1}_+$ K\"ahler moduli $T_a$ and $h^{1,1}_-$ two-forms $G_\alpha$, while the  
$h^{2,1}$ hypermultiplets yield $h^{2,1}_-$ complex structure moduli $z_i$ and $h^{2,1}_+$ vector multiplets $V_r$: see Table \ref{tab:multiplets}. 
For simplicity of presentation we will mostly discuss orientifolds with $h^{1,1}_-=h^{2,1}_+=0$,
in which
the massless closed-string scalar fields 
are the geometric moduli $T_a$ and $z_i$, and the axiodilaton $\tau$.\footnote{See e.g.~\cite{Grimm:2004uq,Grimm:2007xm,McAllister:2008hb,Gao:2013pra,Hebecker:2018yxs,Carta:2020ohw,Moritz:2023jdb}
for discussions of the geometry and physics of orientifolds with $h^{1,1}_- \neq 0$.}
 
Compactification of type IIB string theory on an O3/O7 orientifold\footnote{We will discuss other string theories in \S\ref{eq:beyondiib}.  Within type IIB string theory, an advantage of O3/O7 orientifolds compared to O5/O9 orientifolds is the possibility of conformally Calabi-Yau flux compactifications, as we will explain in \S\ref{ss:isd}.} of a Calabi-Yau threefold is a promising starting point for finding flux compactifications that involve nontrivial sources and that preserve at most $\mathcal{N}=1$ supersymmetry. From there, one can further aim to find non-supersymmetric solutions without moduli.

\subsubsection*{Four-dimensional $\mathcal{N}=1$ supersymmetric action}
We can now see how the four-dimensional action transforms under the scaling transformations and how it can be expanded in powers of $\alpha'$ and $g_s$. 
For this we need to use that the volume of $X_6$
scales as
\be \label{eq:vscales}
\V=\int_{X_6}\d^6x \sqrt{g^{(6)}}\rightarrow \lambda^{3\nu}\V\,,
\ee
where the volume is measured in units of the string length $\ell_s=2\pi\sqrt{\alpha'}$. Then the four-dimensional Einstein-frame metric $g^E_{\mu\nu}$, K\"ahler moduli $\tau_a$, complex structure  moduli $z_i$ and Lagrangian $\L$ scale as
\begin{equation} \label{eq:gscales}
g^E_{\mu\nu}=\V  g_{\mu\nu}\rightarrow \lambda^{4\nu}g^E_{\mu\nu}, \qquad \tau_a\rightarrow \lambda^{2\nu}\tau_a, \qquad z_i\rightarrow z_i, \qquad  \L\rightarrow \lambda^{4\nu} \L\,.
\end{equation}

In this article we are interested in $\mathcal{N}=1$ supersymmetric compactifications. The general couplings of four-dimensional supergravity to matter fields were computed in \cite{Cremmer:1982en,Kugo:1982mr}. Up to two derivatives and neglecting the contributions of vector multiplets, the action for an arbitrary number of chiral superfields $\Phi_M$ coupled to supergravity is specified by a superpotential $W(\Phi_M)$ and a K\"ahler potential $K(\Phi_M, \Bar{\Phi}_{\bar N})$, with $W$ a holomorphic function and $K$ a real analytic function of the chiral superfields $\Phi_M$.

We write the four-dimensional Lagrangian in the superconformal formalism \cite{Cremmer:1982en,Kugo:1982mr} in terms of the superpotential $W$ and K\"ahler potential $K$ with conformal compensator $\C$ as
\be
\frac{\L}{\sqrt{-g^E}}=\int  \d^4\theta\,  \overline\C \C e^{-K/3}+\left( \int  \d^2\theta\,  \C^3 W+ \hbox{h.c.}\right)\,.
\ee
Using the scaling of the ten-dimensional actions given in (\ref{mandn}) we infer that the perturbative corrections to $K$ can be written as \cite{Burgess:2020qsc}:
\be\label{perturbativeK}
\left(e^{-K/3}\right)^{(m,n)}\rightarrow \lambda^{\frac{2}{3}(\omega+2\nu)-2(\omega-\nu)n+(\omega-2\nu)m}\left(e^{-K/3}\right)^{(m,n)}\,,
\ee
Here we have used that the weight of $\theta$ is $\nu$ in order for the fermionic kinetic terms to transform appropriately, and thus the compensator $\C$ carries weight $-(\omega+2\nu)/3$. 
 
The superpotential $W$ (which scales as $W\rightarrow \lambda^\omega W$) is holomorphic and therefore receives no perturbative corrections whatsoever,\footnote{The original argument uses the shift symmetry for the dilaton field \cite{Witten:1985bz, Burgess:1985zz,Dine:1986vd}, but for flux compactifications the dilaton appears in the superpotential and so a more refined proof of the non-renormalization theorem, 
given in \cite{Burgess:2005jx}, is needed.} whereas the K\"ahler potential $K$ is not holomorphic and is subject to both perturbative 
corrections --- as can be seen from \eqref{perturbativeK} --- as well as non-perturbative 
corrections. 
Thus we can write
\begin{equation}\label{eq:wkform}
    W = W_{\text{tree}} + W_{\text{np}}\,, \qquad K = K_{\text{tree}} + K_{\text{pert}} + K_{\text{np}}\,.
\end{equation}
After introducing the central ansatz for type IIB flux compactifications in \S\ref{ss:isd}, we will examine each of 
$W_{\text{tree}}$ (\S\ref{sss:wtree}),
$W_{\text{np}}$ (\S\ref{sss:wnp}),
$K_{\text{tree}}$ (\S\ref{sss:ktree}),
and
$K_{\text{pert}}$
(\S\ref{sss:kpert}).  We will not treat $K_{\text{np}}$, which is poorly understood.

Knowing $K$ and $W$ we can compute all the relevant couplings for the bosonic and fermionic components of the chiral superfields $\Phi_M$.
In particular, the F-term scalar potential, which is the key quantity to compute in order to find the vacuum states, depends on $W$ and $K$ as follows:
\begin{equation}\label{eq:vkw}
    V_F=e^K\left[K^{M\bar N}D_M W \bar{D}_{\bar N}\overline{W}-3|W|^2\right]\,,
\end{equation}
where $K^{M \bar N}$ is the inverse of the K\"ahler metric $K_{M\bar N}=\partial_M\partial_{\bar N}K$, and $D_M W=\partial_M W+K_M W$ is the K\"ahler covariant derivative, with $K_M=\partial_M K$. Here $\partial_M$ refers to the derivative with respect to the scalar component $\phi_M$ of the superfield $\Phi_M$, and we have adopted units in which $M_{\text{pl}}=1$.

\subsection{ISD flux compactifications}\label{ss:isd}

In a flux compactification,\footnote{In this review we concentrate on fluxes of antisymmetric tensor fields. More general fluxes, including 
geometric fluxes (see e.g.~\cite{Kachru:2002sk}) and 
non-geometric fluxes \cite{Wecht:2007wu,Andriot:2012wx}
are possible: see the review
\cite{Plauschinn:2018wbo}.}
the metric of the internal space is not Ricci-flat, because the stress-energy of fluxes drives deviations from the Calabi-Yau vacuum configuration.  However, in an important class of type IIB flux compactifications, the metric is conformal to a Calabi-Yau metric, differing only by a warp factor.  To see this, we consider the warped ansatz 
\begin{equation}\label{eq:warpedproductansatz}
 ds^2 = e^{2A(y)}\,g_{\mu\nu}(x)\d x^{\mu}\d x^{\nu}+e^{-2A(y)}g_{mn}(y)\d y^m \d y^n\,,   
\end{equation} with $g_{mn}$ a Riemannian metric on a compact space $X_6$ that admits a Calabi-Yau metric $g_{mn}^{\text{CY}}$.
In a Calabi-Yau vacuum solution, $A(y)$ is trivial and $g_{mn}=g_{mn}^{\text{CY}}$ on $X_6$, whereas in the presence of general sources, $g_{mn}$ is unrelated to $g_{mn}^{\text{CY}}$.  Defining the  Hodge star $\star$ constructed from the metric $g_{mn}$, one finds that $\star^2=-1$, so $\star$ has eigenvalues $\pm i$.  Writing the three-form fluxes $F_3$ and $H_3$ in the complex combination
\begin{equation}
 G_3 := F_3 - \tau H_3\,,   
\end{equation}
with $\tau$ the axiodilaton, we can decompose $G_3$ into $+i$ and $-i$ eigenspaces of $\star$,
\begin{equation}
 G_\pm := G_3 \mp i \star G_3\,,   
\end{equation}
which are termed \emph{imaginary self-dual (ISD)} and \emph{imaginary anti-self-dual (IASD)}, respectively.

Consider type IIB string theory compactified on an O3/O7 orientifold of a Calabi-Yau threefold, and containing only ISD fluxes, D3-branes, D7-branes, O3-planes, and O7-planes, without IASD fluxes and without antibranes.  Such a configuration is called an ISD compactification. 

Several key properties of ISD compactifications were recognized by Giddings, Kachru, and Polchinski \cite{Giddings:2001yu}.
First, the Einstein equations for \eqref{eq:warpedproductansatz} are solved by $g_{mn}=g_{mn}^{\text{CY}}$ with a generally nontrivial\footnote{
The dynamics of the warp factor is analyzed in \cite{Giddings:2005ff,Douglas:2009zn}.} warp factor $A(y)$.  That is, the metric $e^{-2A(y)}g_{mn}(y)$ on the internal space is conformally Calabi-Yau.  Second, the classical solution at leading order in the $\alpha'$ expansion enjoys a dilatation symmetry: the size of $X_6$ is a modulus (see \S\ref{sss:ktree}).  Third, generic ISD fluxes give masses to the complex structure moduli of $X_6$, and to the axiodilaton.  This is easy to see from the ten-dimensional action,
\begin{equation}
S_{10} \supset \int_{X_6} G_3 \wedge \star \overline{G_3}\,,
\end{equation}
in which the Hodge star, which depends on the metric $g_{mn}$, couples to the fluxes.\footnote{One might wonder how to choose quantized fluxes that are sufficiently generic to stabilize all, rather than just some, of the complex structure moduli.  We will address this below.}

Dimensional reduction of an ISD compactification leads to an
$\mathcal{N}=1$ supersymmetric effective action in four dimensions.  On general grounds the resulting superpotential $W$ and K\"ahler potential $K$ depend on the moduli as
\begin{equation}\label{eq:wandk}
W=W(z_i,T_a,\tau)\,, \qquad K = K(z_i,\bar{z}_i, T_a,\overline{T}_a,\tau,\bar{\tau})\,,
\end{equation} 
However, as we will now see, there is a great deal of structure in $W$ and $K$ that can be exploited in searching for vacua. 

\subsection{Superpotential}\label{sec:sup}

\subsubsection{Flux superpotential}\label{sss:wtree}
The classical Gukov-Vafa-Witten flux superpotential is \cite{Gukov:1999ya}
\begin{equation}\label{eq:gvw}
W_{\text{tree}} \equiv W_{\text{flux}} := \sqrt{\tfrac{2}{\pi}} \int_{X_6} G_3 \wedge \Omega\,,
\end{equation}
where $\Omega$ is the holomorphic $(3,0)$ form on $X_6$, and we have adopted a convenient normalization.  

To better understand this expression,
we take $\alpha_A, \beta^A$ to be a symplectic basis of $H^3(X_6,\mathbb{Z})$, with $A=0,\ldots, h^{2,1}$ and with $\int_{X_6}\alpha_A \wedge \beta^B = \delta_{A}^{~B}$, and
we introduce the periods
\begin{equation}\label{eq:perioddef}
\vec{\Pi}:=\begin{pmatrix}
\int\Omega\wedge \beta_A \\
\phantom{\Bigl.}\int\Omega\wedge \alpha^A
\end{pmatrix}=\begin{pmatrix}
\mathcal{F}_A\\
z^A
\end{pmatrix}\, .
\end{equation}
The periods $z^A$ furnish local projective coordinates on complex structure moduli space: in a patch where $z^0 \neq 0$, we can fix the normalization of $\Omega$ to set $z^0=1$, and then use $\{z_i\}, i=1,\ldots,h^{2,1}$ as independent coordinates.

We write the integrals of the three-form fluxes as
\begin{equation}\label{eq:fhdef}
    \vec{f} := \biggl(\int F_3 \wedge \alpha_A, \int F_3 \wedge \alpha^A\biggr)\,, \qquad  \vec{h} := \biggl(\int H_3 \wedge \alpha_A, \int H_3 \wedge \alpha^A\biggr)\,.
\end{equation}
By Dirac quantization, $\vec{f}, \vec{h} \in \mathbb{Z}^{2h^{2,1}+2}$.
Introducing the symplectic matrix
\begin{equation} \label{eq:sigdef}
\Sigma:=
\begin{pmatrix} 0 & {\mathbb{I}}\\
-{\mathbb{I}} & 0
\end{pmatrix}\,,
\end{equation}
we can express \eqref{eq:gvw} as the pairing
\begin{equation}\label{eq:fdotpi}
   W_{\text{flux}}(\tau,z_i) = \sqrt{\tfrac{2}{\pi}}\,\vec{\Pi}^{\mathsf{T}}{\cdot}\Sigma{\cdot}\bigl(\vec{f}-\tau\vec{h}\bigr)\,.
\end{equation}
Thus, the flux superpotential is determined by the periods, the integer flux quanta $\vec{f}$ and $\vec{h}$, and the axiodilaton $\tau$.

\subsubsection{Non-perturbative superpotential}\label{sss:wnp}

The K\"ahler moduli $T_a$ do not appear in the classical flux superpotential $W_{\text{flux}}$, and indeed do not appear in $W$ at any perturbative order in $\alpha'$ or $g_s$.  We explained this fact in general terms in \S\ref{sec:pert}, but we will now be more specific about the dependence on $T_a$.

Suppose that $\{D_a\}, a=1,\ldots, h^{1,1}_+$ is a basis of $H_4(X_6,\mathbb{Z})$ consisting of effective divisors (holomorphic four-cycles), with complexified volumes $T_a$ given by
\begin{equation}\label{eq:tdef}
    T_a := \frac{1}{2}\int_{D_a}  J \wedge J + i \int_{D_a} C_4 \equiv \tau_a + i \theta_a\,,
\end{equation}
where $J$ is the K\"ahler form, and we have introduced 
the axion $\theta_a$ and the volume modulus (or saxion) $\tau_a$.

The action \eqref{eq:s10bulk} is invariant under continuous shifts
\begin{equation}\label{eq:pqc4}
\theta_a \to \theta_a + c\,, \qquad c\in \mathbb{R}\,.
\end{equation}
This shift symmetry descends from the gauge redundancy of the $C_4$ kinetic term proportional to $|\tilde{F}_5|^2$, and is known as a Peccei-Quinn (PQ) symmetry.\footnote{The Peccei-Quinn symmetry is discussed from a slightly different perspective in \S\ref{sec:axionstring}.} 
Because the string worldsheet carries no Ramond-Ramond charge, the PQ symmetry \eqref{eq:pqc4} of $C_4$ is not broken at any order in string perturbation theory, or at any order in $\alpha'$.

Any correction to the effective action at some perturbative order in the $\alpha'$ expansion would have to scale as a negative power of (some of) the $\tau_a$, in order to vanish in the limit of infinite volume.  
But on general grounds the superpotential must be holomorphic, and so it can only depend on cycle volumes via the complex combination $T_a = \tau_a + i\theta_a$: the $T_a$ are the `good K\"ahler coordinates' on K\"ahler moduli space. 
But no term polynomial in $1/T_a$ is invariant under 
the PQ symmetry \eqref{eq:pqc4}, which remains unbroken to all perturbative orders, and so the superpotential must have no perturbative dependence on the K\"ahler moduli $T_a$.
 
The symmetry \eqref{eq:pqc4} can be broken by non-perturbative effects from Euclidean D3-branes, causing $W$ to depend on exponentials of the $T_a$, as we now explain.
We suppose that $D = c^a D_a$, with $c^a \in \mathbb{Z}$, is 
a four-cycle in $X_6$, but is not necessarily an effective divisor.
Then the semiclassical action of a Euclidean D3-brane wrapping $D$ is 
\begin{equation}\label{eq:sed3}
    \frac{1}{2\pi}\,S_{\text{ED3}} = \text{Vol}(D) + i \int_{D} C_4\,.
\end{equation}  
If $D$ is an effective divisor, it is calibrated by the K\"ahler form $J$, and so obeys 
\begin{equation}
\text{Vol}(D)=\frac{1}{2}\int_{D}  J \wedge J\,,   
\end{equation} so that 
\begin{equation}\label{eq:sed3eff}
    \frac{1}{2\pi}\,S_{\text{ED3}} =c^a T_a\,.
\end{equation}  
Euclidean D3-branes wrapping such a divisor contribute to the superpotential, rather than to the K\"ahler potential or higher F-terms, if and only if the associated Dirac operator on $D$ has exactly two zero modes.  These zero modes can be counted in terms of the dimensions of the orientifold-graded sheaf cohomology groups $H^{i}_{\pm}(D,\mathcal{O}_D)$, with $i = 0, 1, 2$.
If $D$ is smooth, effective, and obeys the \emph{rigidity} condition\footnote{Beware that in some settings $D$ is called rigid if it fulfills the weaker condition $\text{dim}\, H^{2}(D,\mathcal{O}_D) = 0$.}  
\begin{equation}\label{eq:fullrigid}
    \text{dim}\, H^{\bullet}_+(D,\mathcal{O}_D) = (1,0,0)\,, \qquad \text{dim}\, H^{\bullet}_-(D,\mathcal{O}_D) = 0\,,
\end{equation} with $\bullet$ standing for $i=0,1,2$, 
then\footnote{The rigidity condition \eqref{eq:fullrigid} alone is neither necessary nor sufficient for a non-perturbative superpotential contribution on an effective divisor $D$.  If $D$ is smooth then \eqref{eq:fullrigid} is indeed sufficient \cite{Witten:1996bn}.
For certain singular $D$, \eqref{eq:fullrigid} is not necessary, but a corrected sufficient condition that accounts for the singular loci is applicable \cite{Gendler:2022qof}.  For certain other $D$ that do not fulfill \eqref{eq:fullrigid} by fault of having $\text{dim}\, H^{2}(D,\mathcal{O}_D)>0$, rigidification by flux can lead to a superpotential term: Euclidean D3-branes wrapping such a $D$ and bearing magnetic flux can contribute to $W$
(see e.g.~\cite{Grimm:2011dj,Bianchi:2011qh,Bianchi:2012pn,Bianchi:2012kt}).  However, in the examples given below we will be able to consistently neglect contributions from singular divisors or divisors rigidified by flux.}
Euclidean D3-branes wrapping  $D$ generate
the non-perturbative superpotential term
\begin{equation}\label{eq:wrigidnotpure}
W_{\text{ED3}|D,\text{rigid}} = \mathcal{A}(z_i,z_{D3},z_{D7},\tau) e^{-2\pi c^a T_a}\,.
\end{equation}
The prefactor $\mathcal{A}(z_i,z_{D3},z_{D7},\tau)$ is a one-loop Pfaffian that can depend on the complex structure moduli, the axiodilaton, and the positions $z_{D3}$ and $z_{D7}$ of any spacetime-filling D3-branes (see e.g.~\cite{Berg:2004ek,Baumann:2006cd,Kim:2022uni,Kim:2023cbh})
and D7-branes \cite{Cvetic:2012ts}.  Rigidity of $D$ implies that $\mathcal{A}(z_i,z_{D3},z_{D7},\tau)$ is not identically vanishing, though it is generally a section of a nontrivial bundle on moduli space, with zeros on certain subloci --- for example, if a spacetime-filling D3-brane coincides with $D$ \cite{Ganor:1996pe}.

If a smooth divisor $D$ fulfilling \eqref{eq:fullrigid}
obeys the further condition that its uplift to F-theory\footnote{For reviews on F-theory see \cite{Denef:2008wq,Weigand:2018rez}, and for 
relations between Euclidean D3-branes and Euclidean M5-branes see e.g.~\cite{Blumenhagen:2010ja}.
Euclidean M5-branes are considered from the perspective of the heterotic dual in \cite{Cvetic:2011gp}.  A comprehensive review of D-brane instantons, including many phenomena we have omitted here, appears in \cite{Blumenhagen:2009qh}.} has trivial intermediate Jacobian, we call $D$ pure rigid, and its Pfaffian has no dependence on $z_i$, $z_{D7}$, or $\tau$: the Pfaffian of a pure rigid divisor is a section of the trivial bundle over the moduli space of the associated fourfold \cite{Witten:1996hc}.

Thus, a pure rigid divisor in a compactification that contains no spacetime-filling D3-branes generates a simple exponential,
\begin{equation}\label{eq:wrigidpure}
W_{\text{ED3}|D,\text{pure rigid}} = \mathcal{A}\, e^{-2\pi c^a T_a}\,,
\end{equation}
with $\mathcal{A}$ a constant.

The gaugino condensate superpotential is rather analogous to $W_{\text{ED3}}$.  Consider a stack of D7-branes that wrap an effective divisor $D = c^a D_a$ and generate an $\mathcal{N}=1$ supersymmetric Yang-Mills theory with gauge group $G$ and with $h^{0,2}(D)+h^{0,1}(D)$
chiral multiplets charged in the adjoint of $G$.
If $h^{0,2}(D)=h^{0,1}(D)=0$ then the theory is pure super-Yang-Mills, and at low energies it generates a gaugino condensate superpotential,
\begin{equation}\label{eq:ll}
W_{\lambda\lambda|D} = \mathcal{A}(z_i,z_{D3},\tau) e^{-2\pi c^a T_a/c(G)}\,,
\end{equation} where  
$c(G)$ is the dual Coxeter number of $G$.
The Pfaffian now has the interpretation of resulting from a threshold correction to the gauge coupling.  In particular, the dependence on $z_{D3}$ occurs because strings stretched between the D7-brane stack and a D3-brane produce chiral multiplets charged in the fundamental representation of $G$, and whose masses affect the low-energy condensate.

The full superpotential thus takes the form
\begin{equation}\label{eq:wtot}
W = W_{\text{flux}}(\tau,z_i) + W_{\text{np}}(\tau,z_i,T_a)\,,
\end{equation} with
\begin{equation}\label{eq:wnpgen}
W_{\text{np}} = W_{\text{ED3}} + W_{{\lambda\lambda}} +W_{\text{ED(-1)}}\,.
\end{equation}
Here $W_{\text{ED3}}$ denotes the sum of all Euclidean D3-brane terms, 
$W_{{\lambda\lambda}}$ is the sum of all gaugino condensate terms,
and $W_{\text{ED(-1)}}$ is generated by Euclidean D(-1)-branes, i.e.~D-instantons, and obeys
\begin{equation}\label{eq:wedminus}
W_{\text{ED(-1)}} = \mathcal{O}(e^{-\pi \tau})\,.
\end{equation}

Because $W_{\text{ED3}}$ and $W_{{\lambda\lambda}}$ are exponential in four-cycle volumes, and $W_{\text{ED(-1)}}$ is exponential in $\tau$, we conclude that to all perturbative orders in the $\alpha'$ and $g_s$ expansions, the superpotential is given by
\begin{equation}
W \approx W_{\text{flux}} \,,
\end{equation}
up to corrections that are exponentially small when four-cycle volumes $T_a$ are large and the string coupling $\text{Im}(\tau)$ is weak.

\subsection{K\"ahler potential}\label{ss:k}
\subsubsection{Tree level}\label{sss:ktree}
 
At the sphere level, the moduli space of closed string fields in an O3/O7 orientifold of a Calabi-Yau threefold $X_6$ factorizes into the complex structure, K\"ahler, and axiodilaton moduli spaces:
\begin{equation}
{\mathcal M}={\mathcal M}_{\text{cs}}(X_6)\times {\mathcal M}_{\text{K}}(X_6) \times {\mathcal M}_{\tau}\,.
\end{equation}
Therefore the
metric for the moduli is block diagonal, and the K\"ahler potential splits into dilaton, complex structure and K\"ahler moduli terms \cite{Grimm:2004uq}:
\begin{equation}\label{eq:ktree}
K_{\rm {tree}}=-\ln\Bigl(-{i}(\tau-\bar\tau)\Bigr)-\ln\left(-{i}\int_{X_6}\Omega(z_i)\wedge\overline{\Omega}({\bar z}_i )\right)-2\ln\Bigl(\V(T_a,{\overline{T}}_a)\Bigr)\,.
\end{equation}
The holomorphic $(3,0)$-form $\Omega$ encodes the dependence on complex structure moduli, as can be seen by expanding in a  basis of three-forms.  
At large complex structure (LCS), 
we have
\begin{equation}
\int_{X_6}\Omega \wedge\overline{\Omega} = \vec{\Pi}^{\dagger}{\cdot}\Sigma{\cdot}\vec{\Pi}\,,
\end{equation}
in terms of the periods $\vec{\Pi}$ defined in \eqref{eq:perioddef} and  the symplectic matrix $\Sigma$ defined in \eqref{eq:sigdef}.

To understand the dependence of $K_{\rm {tree}}$ on the K\"ahler moduli,
we recall from \eqref{eq:tdef} that the complexified four-cycle volumes are $T_a=\tau_a + i\theta_a$, with axion fields $\theta_a$  appearing from the compactification of the Ramond-Ramond four-form $C_4$,  $\theta_a=\int_{D_a}C_4$.
As we explained in \S\ref{sss:wnp}, the axions $\theta_a$ 
enjoy PQ symmetries \eqref{eq:pqc4} inherited from the gauge symmetry of $C_4$.  In combination with holomorphy of $W$, the PQ symmetries \eqref{eq:pqc4} forbid the $T_a$ from appearing in $W$ at any perturbative order.
But the K\"ahler potential 
is not holomorphic,
and so $K_{\rm {tree}}$ and $K_{\rm {pert}}$ do depend on the $\tau_a$, but not on the $\theta_a$.
 
We introduce a basis $\{\omega_a\}$ of $H^2(X,\mathbb{Z})$ and
write the K\"ahler form as 
\begin{equation}
    J=t^a \omega_a\,,
\end{equation}
where the K\"ahler parameters $t^a$ measure the volumes of two-cycles.
Then the volume $\V$ of $X_6$ can be written
\begin{equation}\label{eq:calvolis}
    \V=\frac{1}{6}\int_{X_6} J\wedge J \wedge J=\frac{1}{6}\kappa_{abc}t^a t^b t^c\,,
\end{equation}
where 
\begin{equation}
   \kappa_{abc}=\int_X \omega_a\wedge\omega_b\wedge \omega_c\,
\end{equation}
are the triple intersection numbers.
The four-cycle volumes $\tau_a$ are then given by
\begin{equation}\label{eq:tauis}
    \tau_a=\frac{\partial \V}{\partial t^a}=\frac{1}{2}\kappa_{abc}t^b t^c\,.
\end{equation}
Note that $\V$ is written explicitly as a function of the $t^a$, but the K\"ahler coordinates are $T_a = \tau_a + i \theta_a$. The relation between $t^a$ and $\tau_a$ is not in general analytically invertible, and so the dependence of $K$ on $\tau_a$ is usually only implicit, via $K(\tau_a)=K(t^b(\tau_a))$.

Even though the dependence of $K_{\text{tree}}$   on the  $\tau_a$ is complicated and model-dependent, its dependence on 
$\V$ is simple.  Under the scaling \eqref{eq:vscales}, \eqref{eq:gscales} with $\nu=1/2$,
\begin{equation}
    \tau_a\rightarrow \lambda \tau_a\,,
\end{equation} we have
\begin{equation}
    \V\rightarrow \lambda^{3/2}\V\,, \qquad  K_{\text{tree}}\rightarrow K_{\text{tree}}-3\ln\lambda\,.
\end{equation}
By taking derivatives with respect to $\lambda$ and evaluating at $\lambda=1$ we  derive the identity
\begin{equation}\label{eq:noscaleprop}
    K^{a\bar b}_{\text{tree}}K_a^{\text{tree}} K_{\bar b}^{\text{tree}}=3\,.
\end{equation}
This powerful result is known as the {\emph{no-scale} property} \cite{Cremmer:1983bf}. 
 
A key consequence of no-scale structure is that
the negative term $-3|W|^2$ in the scalar potential \eqref{eq:vkw} is precisely cancelled by the
term $K^{a\bar b}_{\text{tree}}K_a^{\text{tree}} K_{\bar b}^{\text{tree}} |W|^2$. 
Thus, if the superpotential is independent of the
K\"ahler moduli --- which is true at tree level, cf.~$W_{\text{tree}}$ in \eqref{eq:gvw} ---  the scalar potential reduces to  
\begin{equation}\label{eq:vtvns}
    V_{\text{tree}}=V_{\text{no-scale}}=e^K K^{M\bar N}D_M W {\bar{D}}_{\bar N}\overline{W}\,,
\end{equation}
where now $M,N=1,\ldots, h^{2,1}+1$, corresponding to the complex structure moduli $z_i$ and the dilaton $\tau$, and everything on the right-hand side of \eqref{eq:vtvns} is understood to be evaluated using $W_{\text{tree}}$ and $K_{\text{tree}}$.

The importance of this result cannot be overemphasized. Some of its implications are:
\begin{itemize}
    \item The tree-level scalar potential $V_{\text{no-scale}}$ is positive-definite,   
    and so the minima in the $z_i$ and $\tau$ directions have vanishing F-terms for these fields: $D_MW=0$.
    \item Supersymmetry is broken in the K\"ahler moduli directions, since $D_{T_a}W$ is arbitrary in the minimum of $V_{\text{no-scale}}$.
    \item The value of $V_{\text{no-scale}}$ is zero at the minimum, which means that the cosmological constant vanishes,  even though supersymmetry is broken (by the F-terms $D_{T_a}W$).
     \item The presence of any source of positive energy will drive a runaway to large volume: recall that $e^K \propto \V^{-2}$ times higher-order terms in the $1/\V$ expansion.
    \item The K\"ahler moduli $T_a$ are not stabilized by $V_{\text{no-scale}}$.  Thus, stabilization of the $T_a$ is possible only once
    quantum corrections to the scalar potential --- through one or more of $W_{\text{np}}$, $K_{\text{pert}}$, and $K_{\text{np}}$ --- impact the vacuum structure.
\end{itemize}

\subsubsection{Perturbative corrections}\label{sss:kpert}

The K\"ahler potential beyond tree-level is the least understood of the 
quantities that are important in moduli stabilization. 
The tree-level expression $K_{\text{tree}}$ in \eqref{eq:ktree} is the dominant contribution to the kinetic terms but, due to the no-scale property \eqref{eq:noscaleprop}, 
the scalar potential computed with $K=K_{\text{tree}}$ and $W=W_{\text{tree}}=W_{\text{flux}}$ vanishes.
Thus, the leading contributions to $V$ will come from higher-order perturbative (or non-perturbative) corrections to $K$, or from the non-perturbative superpotential.

We have seen in \S\ref{sec:pert} that the K\"ahler potential receives corrections order by order in perturbation theory in both the string loop and $\alpha'$ expansions.  The 
scaling properties 
\eqref{perturbativeK} serve to organize these corrections:
the tree-level expression $K_{\rm {tree}}$ given in \eqref{eq:ktree} satisfies the $(m,n)=(0,0)$ scaling, and
for general values of $m,n$ we use  
\eqref{perturbativeK} to write the expression for the K\"ahler potential to all orders in perturbation theory as:
\be \label{eq:ek3}
e^{-K/3}= \left(\hbox{Im}\tau \right)^{1/3}\V^{2/3}\sum_{m,n} A^{(m,n)}\left(\frac{1}{\hbox{Im}\tau} \right)^n \left[\frac{\left(\hbox{Im}\tau\right)^{1/2}}{\V^{1/3}}\right]^m\, .
\ee
The coefficients $A^{(m,n)}$ depend on scale-invariant combinations of the fields, such as the complex structure moduli $z_i$ or ratios of K\"ahler moduli.

In terms of the tree-level superpotential $W_0:= \langle W_{\text{flux}}\rangle$, cf.~\eqref{eq:gvw}, we write the scalar potential at order $m,n$ as:
\be
V^{(m,n)}=B^{(m,n)}\left(\frac{1}{\hbox{Im}\tau} \right)^n \left[\frac{\left(\hbox{Im}\tau\right)^{1/2}}{\V^{1/3}}\right]^m \frac{\left |W_0\right |^2}{\V^2\,\hbox{Im}\tau}\, ,
\ee
where $B^{(m,n)} $ are scale-invariant combinations of the fields.

The expression \eqref{eq:vkw} for the scalar potential in terms of $K$ and $W$ is the most general possibility for an $\mathcal{N}=1$ supersymmetric Lagrangian in which each term has at most two spacetime derivatives of the fields.
However, in general there are higher-derivative corrections to the Lagrangian, for example involving higher powers of the curvature.  By supersymmetry, such higher-derivative corrections can also affect the structure of the scalar potential, and need to be considered.

In string theory, these corrections to the scalar potential include higher powers of the fluxes  $G_3$, corresponding to higher powers of $W_0$, which in turn reflect higher-order superspace covariant derivatives, i.e.~F-terms. The scaling analysis easily captures these higher powers of $W_0$ by writing:
\be
B^{(m,n)}=B^{(m,n,r)}\left(\frac{\left |W_0\right |^2}{\V^{2/3}\,\hbox{Im}\tau}\right)^{r-1},\qquad m=p+r-1
\ee 
where $ B^{(m,n,r)}$ is a scale-invariant function of the fields that is independent of $W_0$. 
The quantity  $\epsilon :=\left |W_0\right |^2/\left(\V^{2/3}\hbox{Im}\tau \right)$ is a scale-invariant function of the fields proportional to   $\left(gF/M^2 \right)^2\simeq \left(m_{3/2}/M\right)^2$, where $F$ is the auxiliary field that breaks supersymmetry,  $M$ is the cutoff scale, here identified with the Kaluza-Klein scale, $m_{3/2}$ is the gravitino mass, and $g$ is the coupling  between heavy Kaluza-Klein states and light modes. Then $\epsilon \simeq (m_{3/2}/M)^2$  is a natural small parameter,  as  required for the validity of the EFT at scales below the cutoff $M$. 

The scalar potential at each given order $(m,n,r)$ in the $\alpha'$, string loop, and F-term expansions, 
with
small parameters $1/\V$, $1/\hbox{Im}\tau$, and $\epsilon$, respectively, 
can then be written as: 
\be \label{eq:vmnr}
V^{(m,n,r)}=\frac{B^{(m,n,r)}}{\V^{4/3}}\left(\frac{1}{\hbox{Im}\tau} \right)^n \left[\frac{\left(\hbox{Im}\tau\right)^{1/2}}{\V^{1/3}}\right]^m \left(\frac{\left |W_0\right |^2}{\V^{2/3} \hbox{Im}\tau}\right)^r\,.
\ee 

The leading-order terms can be classified as follows.

\begin{itemize}
\item $m=n=0$, $r=1$. This is the standard tree-level potential,
\be
V^{(0,0,1)}=\frac{B^{(0,0,1)} \left |W_0\right |^2}{\hbox{Im}\tau\, \V^2}\,, \qquad \hbox{no-scale}\, \implies B^{(0,0,1)}=0\,.
\ee

\item 
$m=1$, $n=0$, $r=1$. This $\alpha'^1$ term may in principle appear, and the corresponding scalar potential,
\be
V^{(1,0,1)}=\frac{B^{(1,0,1)} \left | W_0\right |^2}{\sqrt{\hbox{Im}\tau}\,\V^{7/3}}\,,
\ee
could be dominant in a $1/\V$ expansion, because the tree-level potential $V^{(0,0,1)}$ vanishes.
However, general dimensional analysis arguments show that $ B^{(1,0,1)}=0$ 
\cite{Cicoli:2021rub}. Even though corrections that lead to   $\V^{-7/3}$ behaviour for the scalar potential have not yet been identified, it is an open question whether they may occur at higher string loops, i.e.~for $(m,n,r)=(1,n,1)$ with $n>0$.

\item
$m=2$, $n=2$, $r=1$. This is an $\alpha'^2$ correction at loop order in the string expansion, either as one-loop open strings (the open string loop counting parameter is $n-1$) or as tree-level exchange of Kaluza-Klein states  among D-branes \cite{Berg:2005ja,Berg:2007wt, Cicoli:2007xp,Cicoli:2008va, Berg:2014ama} (see also \cite{vonGersdorff:2005bf}).
However, since the terms in $e^{-K/3}$ proportional to $A^{(2,n)}$ 
 are independent of $\V$ for all $n$, cf.~\eqref{eq:ek3}, the 
 associated
contribution to the scalar potential vanishes, corresponding to what has been called \emph{extended no-scale structure} \cite{Cicoli:2007xp, Cicoli:2008va}. It follows that  $B^{(2,2,1)}=0$. 
\be
V^{(2,2,1)}=\frac{B^{(2,2,1)} \left | W_0 \right |^2}{(\hbox{Im}\tau)^2 \, \V^{8/3}}\,,\qquad \hbox{extended no-scale}\, \implies B^{(2,2,1)}=0\,.
\ee
The same argument  applies to any other allowed values for $n$ and $r$ for fixed $m=2$, i.e.~$B^{(2,n,r)}=0$.  

\item $m=3$, $n=0$, $r=1$. These are string tree-level, $\alpha'^3$ corrections that have been explicitly calculated \cite{Liu:2022bfg} and come from the ten-dimensional $\mathcal{R}^4$, $\mathcal{R}^3 |G_3|^2$, and $\mathcal{R}^2 (\nabla G_3)^2$ terms:
\be\label{bbhk}
V^{(3,0,1)}=\frac{B^{(3,0,1)} \sqrt{\hbox{Im}\tau} \left |W_0 \right |^2}{\V^{3}}\,.
\ee
Explicit calculations determine the coefficient $B^{(3,0,1)}$ to be  proportional to the Euler number of the original Calabi-Yau manifold.
This correction is the best understood: it is an $\mathcal{N}=2$ supersymmetric correction that is present even without orientifolding \cite{GRISARU1986409,GROSS19861,Antoniadis:1997eg,Becker:2002nn}. This result is expected to receive $\mathcal{N}=1$ corrections.

\item 
$m=4$, $n=2$, $r=1$. These are $\alpha'^4$, open string one-loop and $\O(F^2)$ corrections,
\be
V^{(4,2,1)}=\frac{B^{(4,2,1)} \left | W_0\right |^2}{{\hbox{Im}\tau} \, \V^{10/3}}\,.
\ee
These corrections have also been computed in particular cases \cite{Berg:2005ja}, and are subdominant compared to the $\alpha'^3$ correction  $V^{(3,0,1)}$ when $\V$ is large enough to trust the $\alpha'$ expansion.

\item 
$m=3$, $n=0$, $r=2$. These are string tree-level, $\alpha'^3$ corrections at order $\O(F^4)$ in auxiliary fields,
\be
V^{(3,0,2)}=\frac{B^{(3,0,2)} \left | W_0\right |^4}{\sqrt{\hbox{Im}\tau} \, \V^{11/3}}\,.
\ee
\end{itemize}

We could continue with terms that depend on higher powers of the string coupling and inverse volume, but typically these will be  subdominant at large volume and weak coupling.\footnote{A variety of techniques have been used to extract information about corrections to the K\"ahler potential, including explicit string worldsheet calculations and M-theory and F-theory analyses. Both $\alpha'^2$ and $\alpha'^3$ corrections have been obtained, consistent with the scaling behavior outlined above. Some such contributions shift the associated coefficient $B^{(m,n,r)}$, while others can
be absorbed by field redefinitions. See for instance \cite{Halverson:2013qca,Grimm:2013gma,Grimm:2013bha,Junghans:2014zla,Minasian:2015bxa,Haack:2018ufg}.}  Moreover, we have considered only a power-law expansion of $e^{-K/3}$, and in general the series expansion may include logarithmic terms.  Such corrections may result  from loops of light fields rather than of heavy Kaluza-Klein modes. Corrections of this type were recently considered in \cite{Conlon:2010ji, Grimm:2017pid, Weissenbacher:2019mef, Antoniadis:2019rkh,Klaewer:2020lfg,Burgess:2022nbx}.
 
The expansion  \eqref{eq:vmnr} we have given for the scalar potential is in terms of powers of $1/\V$, $1/\hbox{Im}\tau$, and $\epsilon$. 
However, once D-branes are incorporated, there is a new expansion parameter: 
the gauge coupling in the D-brane theory, which is inversely proportional to the volume of the cycle wrapped by  the D-brane.
Thus each of the K\"ahler moduli may provide an extra expansion parameter, just as 
in \S\ref{sss:wnp} the superpotential received non-perturbative corrections in an expansion in cycle volumes.

\section{Vacua in type IIB compactifications}

We have now introduced enough of the structure of type IIB flux compactifications on O3/O7 orientifolds to be able to detail the construction of isolated vacua, i.e.~solutions with stabilized moduli.

\subsection{KKLT}\label{ss:kklt}

We will begin with the KKLT scenario, proposed twenty years ago by Kachru, Kallosh, Linde, and Trivedi \cite{Kachru:2003aw}.

At leading order in the $\alpha'$ and $g_s$ expansions, the effective theory of an O3/O7 orientifold flux compactification is described by
\begin{equation}
W \approx W_{\text{flux}}(z_i,\tau)\,, \qquad K \approx K_{\text{tree}}\,,
\end{equation}
where $W_{\text{flux}}$  and $K_{\text{tree}}$
are given in given in \eqref{eq:gvw} and \eqref{eq:ktree}, respectively.
In such theories there typically exist points $z_{i}^{\star}, \tau^{\star}$ in the joint complex structure and axiodilaton moduli space at which the F-terms of these fields vanish,
\begin{equation}
D_{z_i}W\bigr|_{z_{i}^{\star}, \tau^{\star}} = D_{\tau} W\bigr|_{z_{i}^{\star}, \tau^{\star}} = 0\,,
\end{equation}
and the complex structure moduli and axiodilaton are stabilized by supersymmetric mass terms.
We define the expectation value of the flux superpotential,
\begin{equation}
    W_0 := \langle W_{\text{flux}} \rangle \equiv W_{\text{flux}}(z_{i}^{\star}, \tau^{\star})\,,
\end{equation}
where the brackets denote evaluation on the configuration $z_{i}^{\star}, \tau^{\star}$ at which the complex structure moduli and axiodilaton are stabilized.

The F-terms of the K\"ahler moduli $T_a$, again at leading order in $\alpha'$ and $g_s$, read
\begin{equation}\label{eq:kfterm}
D_{T_a}W\bigr|_{z_{i}^{\star}, \tau^{\star}} \approx W_{0}\,\partial_{T_a} K_{\text{tree}} \,,
\end{equation}
and so $W_0$ is the order parameter measuring supersymmetry breaking.
Except in the highly non-generic situation where
$W_0=0$,
the F-terms \eqref{eq:kfterm} do not vanish anywhere inside the moduli space, and in the classical theory at leading order in $\alpha'$, supersymmetry is broken by the K\"ahler moduli F-terms. 

The first key idea of the KKLT scenario is that if $W_0$ is exponentially small, then the correspondingly small breaking of supersymmetry by fluxes can be compensated by a non-perturbative superpotential term, leading to supersymmetric stabilization of all moduli in an AdS$_4$ vacuum, in a parameter regime where computational control is possible.

The second key idea is that if the complex structure moduli are stabilized near a point in moduli space where a conifold singularity occurs, the compactification can contain a strongly-warped region, and the presence of an anti-D3-brane in this region can lead to controllable breaking of supersymmetry.

We will first discuss supersymmetric stabilization, deferring a treatment of anti-D3-brane supersymmetry breaking to \S\ref{sec:antibrane}.  For notational simplicity, we will illustrate the stabilization idea in a hypothetical example with $h^{1,1}_+=1$.  Explicit models in actual Calabi-Yau orientifolds, whose only important difference from the toy model is that they have multiple K\"ahler moduli, are presented in \S\ref{sec:explicitkklt}, following \cite{Demirtas:2021nlu}.

Suppose that $X_6$ is a Calabi-Yau threefold with $h^{1,1}=1$, and that $\mathcal{O} = \sigma \Omega_{\text{ws}} (-1)^{F_L}$ is an O3/O7 orientifold involution on $X_6$ such that $X_6/\mathcal{O}$ has $h^{1,1}_+=1$.  We denote by $T$ the K\"ahler modulus of $X_6$, and by $D$ the corresponding effective divisor.
The K\"ahler potential reads
\begin{equation}
    K \approx K_{\text{tree}} = - 3\,\text{log}(T+\overline{T})\,,
\end{equation} and if $D$ supports a pure $\mathcal{N}=1$ super-Yang-Mills theory on D7-branes, with gauge group $G$, the superpotential is
\begin{equation}\label{eq:kkltw1}
    W = W_0 + \mathcal{A}(z_i,\tau)e^{-2\pi T/c(G)} + \ldots \,.
\end{equation}   
If $|W_0| \ll 1$, the F-term of the K\"ahler modulus, as computed with $K=K_{\text{tree}}$,
\begin{equation}\label{eq:dtw0}
D^{{\text{tree}}}_{T}W = W \partial_{T} K_{\text{tree}} - \frac{2\pi}{c(G)}\mathcal{A}(z_i,\tau)e^{-2\pi T/c(G)}\,,
\end{equation} then has a zero at a point $T^\star$ inside the moduli space,
\begin{equation}\label{eq:tstar}
\text{Re}\,T^\star = \frac{c(G)}{2\pi}\,\text{log}\bigl(|W_0|^{-1}\bigr)+\ldots \,,
\end{equation}
where the omitted corrections, easily obtained from \eqref{eq:dtw0}, are subleading for $W_0 \ll 1$.
At $T=T^\star$, the supersymmetry-breaking effects of the flux superpotential $W_{\text{flux}}$ and the gaugino condensate superpotential  $W_{{\lambda\lambda}}$ cancel each other.

For sufficiently small $|W_0|$, the approximation $K \approx K_{\text{tree}}$ is consistent at $T=T^\star$, and there exists a field configuration very near\footnote{We discuss the small shifts away from
$z_{i}^{\star}, \tau^{\star}, T^\star$ in the example of \S\ref{sec:explicitkklt}.} $z_{i}^{\star}, \tau^{\star}, T^\star$ at which the F-terms of all moduli vanish exactly.  The resulting solution is an $\mathcal{N}=1$ supersymmetric AdS$_4$ vacuum in which all moduli --- the K\"ahler modulus $T$, the complex structure moduli $z_a$, and the axiodilaton $\tau$ --- are stabilized.  

If any spacetime-filling D3-branes are present, the open string moduli $z_{D3}$ that measure their positions inside $X_6$ appear in the Pfaffian $\mathcal{A}$, because 3-7 strings are charged under $G$ and so affect the gaugino condensate:
\begin{equation}
    W = W_0 + \mathcal{A}(z_i,\tau, z_{D3})e^{-2\pi T/c(G)} + \ldots \,.
\end{equation}
At the level of equation counting, this dependence is sufficient to stabilize the D3-brane positions $z_{D3}$: the D3-brane moduli space would have been all of $X_6$ in the absence of the gaugino condensate superpotential, but the actual moduli space is a set of isolated points.  This has been verified explicitly in toroidal orientifold compactifications and in certain local models \cite{DeWolfe:2007hd}.  It also accords with the general expectation that generic $\mathcal{N}=1$ supergravity theories do not have exact moduli spaces.

We made two crucial assumptions in the above discussion: we supposed that there exists a consistent choice of quantized flux such that $W_0$ is exponentially small, and we supposed that the divisor $D$ supports a gaugino condensate on D7-branes.
A natural question is whether there exist actual Calabi-Yau orientifolds, and explicit choices of quantized flux, that fulfill these conditions and give rise to a supersymmetric AdS$_4$ vacuum.  This question has been resolved in the affirmative by direct construction \cite{Demirtas:2021ote,Demirtas:2021nlu, Demirtas:2019sip},
as we will now explain. 

\subsubsection{Mechanism for small flux superpotential}\label{sec:pfv}

The quantized fluxes $F_3$ and $H_3$ correspond to classes in $H^3(X_6,\mathbb{Z})$, which can be understood as a lattice of dimension $2h^{2,1}+2$.  A specification of fluxes then corresponds to the choice of two lattice vectors $\vec{f}, \vec{h} \in \mathbb{Z}^{2h^{2,1}+2}$.  A choice of fluxes carries the D3-brane charge   
\begin{equation}
    Q^{\text{D3}}_{\text{flux}} = \int_{X_6} H_3 \wedge F_3  =\tfrac{1}{2}{\vec{f}}\,\,^{\mathsf{T}}{\cdot} \Sigma{\cdot} \vec{h}\,. 
\end{equation}
Thus, not all $\vec{f}, \vec{h}$ are consistent: Gauss's law \eqref{eq:gauss} limits fluxes to those with $Q^{\text{D3}}_{\text{flux}}$ no larger than the negative D3-brane charge of the orientifold planes.  Flux vacua thus correspond to choices of lattice vectors in a bounded region.  For large $h^{2,1}$, the number of allowed choices is vast.

The KKLT scenario relies on the existence of fluxes $\vec{f}, \vec{h}$ such that the expectation value $|W_0|$ of the flux superpotential
is exponentially small.  
Important early work on the statistics of flux vacua found the expected distribution of $|W_0|$ in the approximation that the fluxes are continuous \cite{Denef:2004ze}.  The outcome was that the smallest expected value of $|W_0|$ could be exponentially small in $h^{2,1}$, but that at the same time such flux configurations were exponentially rare.

For another perspective, \eqref{eq:fdotpi} shows that the problem of finding small $|W_0|$ is akin to finding a short vector in a high-dimensional lattice.  The shortest vector problem is a well-known, cryptographically hard problem \cite{NPhard}.  This is consistent with a picture in which finding exponentially small $|W_0|$ may be possible, but would require $h^{2,1} \gg 1$, and moreover would be exponentially costly \cite{Denef:2006ad,Halverson:2018cio}.\footnote{For a counterargument, see \cite{Bao:2017thx}.}

This state of affairs appears discouraging,
but
there is important additional structure that can be exploited \cite{Demirtas:2019sip}.
Length is measured in \eqref{eq:fdotpi} by pairing with the period vector $\vec{\Pi}$.  So suppose we can decompose 
\begin{equation}\label{eq:dec}
    \vec{\Pi} = \vec{\Pi}_{\text{poly}} + \vec{\Pi}_{\text{exp}}\,,
\end{equation}
where $\vec{\Pi}_{\text{poly}}$ is polynomial  and 
$\vec{\Pi}_{\text{exp}}$ is exponentially small, in a sense we will make precise.
Then the idea is to find configurations $\vec{f},\vec{h}$ such that
\begin{equation}\label{eq:pfvgen}   
   (\vec{f}-\tau\vec{h})^{\mathsf{T}} {\cdot}  \Sigma {\cdot}\vec{\Pi}_{\text{poly}}\equiv 0\,,
\end{equation} 
i.e.~flux choices that are automatically orthogonal to the `big' part of $\vec{\Pi}$.
In such a case one will have
\begin{equation}
    |W_0| =  \sqrt{\tfrac{2}{\pi}}\,\Bigl|(\vec{f}-\tau\vec{h})^{\mathsf{T}} {\cdot}\Sigma {\cdot} \vec{\Pi}_{\text{exp}}\Bigr| \ll 1\,.
\end{equation}
To realize this idea, one needs a decomposition of the form of \eqref{eq:dec}, and a means of choosing fluxes that obey the orthogonality condition \eqref{eq:pfvgen}.  

Near the large complex structure (LCS) point in moduli space, the periods automatically obey  \eqref{eq:dec}.  To see this, we recall that the prepotential $\mathcal{F}$ for the complex structure moduli $z_i$ takes the form
\begin{equation}\label{eq:prep}
 \mathcal{F} =  
 \mathcal{F}_{\text{poly}}(z_i) + \mathcal{F}_{\text{exp}}(z_i) = \mathcal{F}_{\text{poly}}(z_i) -\frac{1}{(2\pi i)^3} \sum_{\beta = c^i q_i} \text{GV}(\beta)\,\text{Li}_3\bigl(e^{2\pi i c^i z_i}\bigr)\,.
\end{equation} 
Here $\widetilde{X}_6$ is the threefold that is mirror to $X_6$; the sum runs over curve classes $\beta \in H_2(\widetilde{X}_6,\mathbb{Z})$,
expressed in terms of coefficients $c^i$ and a basis $q_i$; $\text{GV}(\beta)$ is the genus-0 Gopakumar-Vafa invariant of $\beta$; $\text{Li}_3$ is the trilogarithm; and 
$\mathcal{F}_{\text{poly}}(z_i)$ is a polynomial of degree three in the $z_i$.
The sum of exponentials corresponds to the contributions of worldsheet instantons in compactification of type IIA string theory on $\widetilde{X}_6$.  

As usual in mirror symmetry, the practical approach is to compute $\mathcal{F}$ exactly in the duality frame where it is purely classical geometry: here, that means evaluating the periods of $\Omega$ on $\widetilde{X}_6$, which reduces to the problem of evaluating (rather elaborate) contour integrals, and can be automated.  Thus, all the constants appearing in \eqref{eq:prep} are computable: the coefficients in $\mathcal{F}_{\text{poly}}$ are determined by the intersection numbers and Chern classes of $\widetilde{X}_6$, and the Gopakumar-Vafa invariants can be obtained from the mirror map \cite{Demirtas:2023als}.

On relating the periods $\vec{\Pi}$ to the prepotential, one easily shows that the decomposition \eqref{eq:prep} implies a decomposition \eqref{eq:dec}, with 
$\vec{\Pi}_{\text{poly}}$ resulting from approximating $\mathcal{F} \approx \mathcal{F}_{\text{poly}}$, and $\vec{\Pi}_{\text{exp}}$ the correction resulting from the instanton series $\mathcal{F}_{\text{exp}}$.

The orthogonality condition \eqref{eq:pfvgen} corresponds to a Diophantine equation in the flux integers.  To see this, we change variables from $\vec{f},\vec{h} \in \mathbb{Z}^{2h^{2,1}+2}$ to $\mathbf{M},\mathbf{K} \in \mathbb{Z}^{h^{2,1}}$ 
(see \cite{Demirtas:2021nlu} for details)
and define
\begin{equation}
p^i := (\tilde{\kappa}_{ijk} M^k)^{-1} K_j\,,
\end{equation} where $\tilde{\kappa}_{ijk}$ are the triple intersection numbers of the mirror $\widetilde{X}_6$.
Then one finds that \eqref{eq:pfvgen} holds for choices of flux obeying
\begin{equation}\label{eq:pfpk}
    \mathbf{p}\cdot \mathbf{K} = 0\,, \qquad \text{with} \qquad \mathbf{p} \in \mathcal{K}_{\widetilde{X}_6}\,,
\end{equation}
where $\mathcal{K}_{\widetilde{X}_6}$ is the K\"ahler cone of $\widetilde{X}_6$.
As always one must also impose the tadpole condition:
we have $-\frac{1}{2} \mathbf{M}\cdot\mathbf{K} \equiv Q^{\text{D3}}_{\text{flux}}$, 
and so $\mathbf{M}$, $\mathbf{K}$ are constrained to obey
\begin{equation}\label{eq:mktad}
\frac{1}{2} \mathbf{M}\cdot\mathbf{K} = Q^{\text{D3}}_{\text{loc}}  \,.
\end{equation}
  
A choice of fluxes obeying \eqref{eq:pfpk} and \eqref{eq:mktad}, and thus obeying \eqref{eq:pfvgen}, is called a \emph{perturbatively flat vacuum}, or PFV.
In a PFV, one linear combination of the complex structure moduli and axiodilaton remains unfixed at the perturbative level (i.e., for $\mathcal{F} \approx \mathcal{F}_{\text{poly}}$), while the $h^{2,1}$ orthogonal directions are supersymmetrically stabilized by fluxes. We can choose to parameterize the perturbatively flat direction by $\tau$.  The effective superpotential for $\tau$, which is dictated by the quantized fluxes and the Gopakumar-Vafa invariants, is a sum of exponentials:
\begin{equation}\label{eq:weffgv}
     W_{\text{eff}}(\tau) = -\frac{1}{2^{3/2}\pi^{5/2}}\biggl( 
     \sum_\beta \mathbf{M}{\cdot}\beta\,\text{GV}(\beta)\,\text{Li}_2\bigl(e^{2\pi i \tau \mathbf{p}\cdot \beta} \bigr) \biggr)\,.
\end{equation}
In some examples one finds a racetrack structure, where two terms compete in a region of weak coupling where all other terms are negligible \cite{Demirtas:2019sip}.

For example, there exists an O3/O7 orientifold with $h^{2,1}_-=5$ and $h^{1,1}_+=113$, and a choice of quantized fluxes therein, for which, in the normalizations of \cite{Demirtas:2021nlu},
\begin{equation} 
   \sqrt{8\pi^5}\, W_{\text{eff}}(\tau) = - 2\,e^{2\pi i \tau \cdot \frac{7}{29}} + 252\,e^{2\pi i \tau \cdot \frac{7}{28}} + \mathcal{O}\bigl(e^{2\pi i \tau \cdot \frac{43}{126}}\bigr)\,.
\end{equation} 

The F-term for $\tau$ vanishes at $\tau= \tau^{\star}$ with $\text{Im}(\tau^\star) \approx 0.011$.
This example therefore has an exponentially small flux superpotential 
\cite{Demirtas:2021ote,Demirtas:2021nlu},
\begin{equation}\label{eq:example5113}
|W_0| = |W_{\text{eff}}(\tau_\star)| \approx \biggl(\frac{2}{252}\biggr)^{29} \approx 10^{-61}\,.
\end{equation}

To recap: by finding fluxes that obey the Diophantine equation \eqref{eq:pfvgen}, one ensures that 
in the joint complex structure and axiodilaton moduli space, a single complex field remains
massless at the perturbative level, while all other directions acquire large supersymmetric masses.\footnote{For a variation on the PFV setup that allows further structure, see \cite{Cicoli:2022vny}, while for a mechanism using asymptotic Hodge theory, see \cite{Bastian:2021hpc}.}
For the remaining PFV direction $\tau$, the effective superpotential is a sum of exponentials, and so when a minimum exists at some $\tau^{\star}$, the vev of the superpotential $|W_0| = |W_{\text{eff}}(\tau_\star)|$ is \emph{naturally} exponentially small.
To engineer such a minimum, one can search through the discrete choices of topology --- of $X_6$ and correspondingly its mirror $\widetilde{X}_6$ --- and fluxes $\vec{f},\vec{h}$ to find a case where $W_{\text{eff}}$ is a racetrack, with two terms having nearly equal exponents and hierarchical prefactors.  Examples of this sort have been found on a large scale: many billions of PFVs have been identified in Calabi-Yau threefolds \cite{coni}.

Small values of $|W_0|$ in PFVs are far more prevalent than continuous flux models predict, and occur at much smaller values of $h^{2,1}$, because
the smallness of $|W_0|$ 
does not rely on searching in a high-dimensional lattice.\footnote{The reason for the disparity compared to continuous flux predictions is that quantized fluxes are a set of measure zero in the space of continuous fluxes, and statements about distributions in the continuous ensemble only hold up to sets of measure zero \cite{Demirtas:2021nlu}.
A statistical comparison appears in \cite{Broeckel:2021uty}.}
The example of \eqref{eq:example5113} has $h^{2,1}=5$, and it is the values of the Gopakumar-Vafa invariants and flux integers that dictate the exponential structure.
 
In summary, exponentially small values of $|W_0|$ can be found systematically by constructing PFVs.
This relies on computing the topology of mirror pairs $(X_6,\widetilde{X}_6)$ of Calabi-Yau threefolds; computing the periods; finding O3/O7 orientifold projections; and solving the Diophantine equation \eqref{eq:pfvgen} in the flux integers.  Searching in a high-dimensional lattice is not required, and so the PFV approach represents a dramatic reduction in complexity.

\subsubsection{Non-perturbative superpotential}\label{sec:wnpkklt}

We now turn to computing the non-perturbative superpotential $W_{\text{np}}$ for the K\"ahler moduli, as defined in \eqref{eq:wnpgen}.
In \S\ref{sss:wnp} we explained the conditions under which an effective divisor $D$ supports a Euclidean D3-brane or gaugino condensate superpotential term.  
To stabilize all $h^{1,1}_+$ K\"ahler moduli $T_a$ in a compactification,  one needs to ensure that there are contributions to $W_{\text{np}}$ that depend on each of the $T_a$.\footnote{See 
\cite{Halverson:2019vmd} for an analysis of the prospects for K\"ahler moduli stabilization in ensembles of geometries.}
Moreover, there are infinitely many candidate divisors, and to conclusively demonstrate the existence of a vacuum for the K\"ahler moduli one must show that the terms that one computes and includes are in fact dominant over all omitted terms.

To this end, suppose that $X_6$ is a Calabi-Yau orientifold, $\{D_a\}, a=1,\ldots, h^{1,1}$
is a basis of $H_4(X_6,\mathbb{Z})$,
and $\{D_A\}, A=1,\ldots, N \ge h^{1,1}$
is a set of irreducible divisors that generate the semigroup of effective divisors, i.e.~such that every effective divisor $D$ on $X_6$ can be written as a nonnegative integer linear combination of the $D_A$.  We write $D_A = c^a_{~A} D_a$ with $c^a_{~A} \in \mathbb{Z}$.
We can specify a point in the K\"ahler moduli space of $X_6$ by the complexified volumes $T_a$ of the $D_a$.  

The orientifold action determines which divisors support D7-brane gauge groups.  In favorable classes of compactifications, each O7-plane coincides with exactly four D7-branes, and gives rise to the gauge algebra $\mathfrak{so}(8)$, whose dual Coxeter number is 
 $c\bigl(\mathfrak{so}(8)\bigr)=6$.
When the associated divisor is rigid in the sense of \eqref{eq:fullrigid}, the low-energy theory is pure glue super-Yang-Mills and generates a gaugino condensate superpotential.
Suppose that on $X_6$, a set $D_1,\ldots, D_{k}\,(k<h^{1,1})$ of divisors are rigid and support  $\mathfrak{so}(8)$ gauge algebras, while the remainder of the semigroup generators
$D_{k+1},\ldots, D_{N}$ are either not wrapped by D7-branes, or else support non-rigid D7-brane stacks (e.g., with $\text{dim}\, H_+^{\bullet}(D,\mathcal{O}_D)=(1,0,1)$).

There very often exists a point  $T_a^{\star}$ at which
\begin{equation} \label{eq:kkltpt}
   \frac{1}{6}\text{Re}(T_1) \approx   \ldots \approx 
   \frac{1}{6}\text{Re}(T_k) \approx \text{Re}(T_{k+1}) \approx \ldots \approx \text{Re}(T_{h^{1,1}}) \ll \text{Re}(T_{h^{1,1}+1})\,,   \ldots, \text{Re}(T_{N})\,.
\end{equation} 
We will call a point obeying \eqref{eq:kkltpt} a \emph{KKLT point}.

At a KKLT point $T_a = T_a^{\star}$, Euclidean D3-branes wrapping any of the $D_A$ with $A \in \{k+1,\ldots, h^{1,1}\}$ are hierarchically more important than Euclidean D3-branes wrapping 
the $D_A$ with $A > h^{1,1}$, or wrapping nontrivial linear combinations of the $D_A$.  The task of evaluating the leading Euclidean D3-brane superpotential terms then reduces to the finite task of counting the fermion zero modes of the $D_A$ for $A \in \{k+1,\ldots h^{1,1}\}$,
i.e.~of computing
$\text{dim}\,H^{i}_{\pm}(D_A,\mathcal{O}_{D_A})$ for $i = 0, 1, 2$.

If all of $D_{k+1}, \ldots, D_{h^{1,1}}$ are rigid and not wrapped by D7-branes, the total superpotential takes the form
\begin{equation}\label{eq:kkltptW}
W = W_{\text{flux}} + \sum_{a=1}^{k} \mathcal{A}_a(z_i,\tau, z_{D3}) e^{-2\pi T_a/6} +
 \sum_{a=k+1}^{h^{1,1}} \mathcal{A}_a(z_i,\tau, z_{D3}) e^{-2\pi T_a} + \ldots
\end{equation}
where the omitted terms are hierarchically smaller than the terms shown.\footnote{The terms in the Euclidean D(-1)-brane superpotential 
\eqref{eq:wedminus} are no larger than $e^{-\pi\tau}$, so if $g_s \ll 1$ and if $W_{\text{flux}}$ arises from a PFV in which 
the dominant terms in \eqref{eq:weffgv} have $\mathbf{p}{\cdot}\beta<1/2$, we can safely neglect $W_{\text{ED(-1)}}$.
Moreover, it was shown in \cite{Kim:2022jvv} that $W_{\text{ED(-1)}}$ vanishes identically in the case that all D7-brane gauge algebras are $\mathfrak{so}(8)$. See \cite{Demirtas:2021nlu}, \cite{Kim:2022jvv} for discussions of this point.}

We call \eqref{eq:kkltptW} a \emph{KKLT superpotential}.
In the special case that all the divisors $D_1,\ldots,D_{h^{1,1}}$ are not just rigid, but also pure rigid
--- meaning that their uplifts to divisors of the F-theory fourfold have trivial intermediate Jacobians --- and that the compactification contains no spacetime-filling D3-branes, the Pfaffian prefactors of the pure rigid superpotential terms are simply constants $\mathcal{A}_a$, and so we have
\begin{equation}\label{eq:kkltptWrig}
W = W_{\text{flux}} + \sum_{a=1}^{k} \mathcal{A}_a\, e^{-2\pi T_a/6} +
 \sum_{a=k+1}^{h^{1,1}} \mathcal{A}_a\,e^{-2\pi T_a} + \ldots
\end{equation}
The conditions leading to \eqref{eq:kkltptWrig} may appear highly restrictive, but compactifications furnishing pure rigid KKLT superpotentials have been found at large $h^{1,1}$ \cite{Demirtas:2021nlu,coni}.

In a compactification with a KKLT point $T_a^{\star}$ \eqref{eq:kkltpt} and a KKLT superpotential \eqref{eq:kkltptW},
the dominant terms in $W_{\text{np}}$
are exactly those shown in \eqref{eq:kkltptW}.  In such a case, the effective theory for the K\"ahler moduli $T_a$ is well-approximated by \eqref{eq:kkltptW} and \eqref{eq:ktree}, and the task of stabilizing the K\"ahler moduli  is purely computational: one needs to check whether a supersymmetric vacuum exists in a regime of control, at some point $T_a^{\text{vac}}$ very near $T_a^{\star}$.
 
\subsubsection{Requirements for KKLT vacua}\label{sec:explicitkkltgen}

The KKLT scenario for de Sitter vacua requires a Calabi-Yau orientifold flux compactification that fulfills the following criteria:
\begin{enumerate}
    \item[a.] The fluxes stabilize the complex structure moduli and axiodilaton at a point $(z^{\star},\tau^{\star})$ in moduli space where $|W_0|$ is exponentially small. 
    \item[b.] The total superpotential, including non-perturbative terms, takes the form\footnote{A form more general than
    \eqref{eq:kkltptW}, involving multiple distinct dual Coxeter numbers, would suffice for the KKLT mechanism, but because examples of \eqref{eq:kkltptW} exist and are simpler to describe, we will restrict to them here.} \eqref{eq:kkltptW}.
    \item[c.] The point $(z^{\star},\tau^{\star})$ is near a conifold singularity in the moduli space, and the fluxes threading the conifold give rise to a Klebanov-Strassler \cite{Klebanov:2000hb} throat region.
    \item[d.] The warp factor and geometry of the throat are such that the inclusion of one or more anti-D3-branes metastably breaks supersymmetry \cite{Kachru:2002gs} and leads to a de Sitter vacuum.
    \item[e.] The resulting vacuum occurs in a parameter regime where all the approximations being made are valid: $g_s \ll 1$; the K\"ahler moduli are stabilized at a point $T^{\star}$ where the $\alpha'$ expansion is under control; and the backreaction effects of D-branes, fluxes, and orientifolds  on the leading-order solution are controllably small.
\end{enumerate}
Conditions (a), (b), and (e) together are sufficient for an $\mathcal{N}=1$ supersymmetric AdS$_4$ vacuum, while (a), (b), (c) and (e) suffice for an $\mathcal{N}=1$ supersymmetric AdS$_4$ vacuum with a conifold region.
The further condition (d) is necessary for a de Sitter vacuum.

\subsubsection{Explicit constructions}\label{sec:explicitkklt}

To illustrate
the ideas explained above,
we will now describe
explicit examples of flux compactifications on orientifolds of Calabi-Yau threefolds, with all moduli stabilized, that 
are incarnations of the KKLT scenario.  We will begin with the case of 
$\mathcal{N}=1$ supersymmetric AdS$_4$ vacua  
without warped throats \cite{Demirtas:2021nlu}, i.e.~vacua fulfilling (a), (b), and (e).

Suppose that $X_6$ is a Calabi-Yau threefold hypersurface in a toric variety obtained from the Kreuzer-Skarke list.  
Computational advances summarized in \S\ref{sec:computation} make it possible to compute intersection numbers \cite{Demirtas:2022hqf}, O3/O7 orientifold actions \cite{Moritz:2023jdb}, the topology of divisors and their uplifts to F-theory \cite{Kim:2021lwo,Jefferson:2022ssj}, and the Gopakumar-Vafa invariants of curves \cite{Demirtas:2023als} in $X_6$ and $\widetilde{X}_6$.  Combined with a choice of quantized fluxes \eqref{eq:fhdef}, these data suffice to determine the flux superpotential \eqref{eq:fdotpi} and the non-perturbative superpotential terms  from Euclidean D3-branes \eqref{eq:fullrigid} and gaugino condensates \eqref{eq:ll}, so that the full superpotential \eqref{eq:wtot} is known.

In \S\ref{sec:pfv} we described a mechanism for achieving (a): one chooses quantized
fluxes obeying \eqref{eq:pfpk} and \eqref{eq:mktad}, leading to a PFV whose effective superpotential $W_{\text{eff}}(\tau)$ is a sum of exponentials \eqref{eq:weffgv},
and selects a case in which the leading terms in \eqref{eq:weffgv} compete in a two-term racetrack \cite{Demirtas:2019sip}, so that $|W_0| \ll 1$.
One can accomplish this in practice for $X_6$  a Calabi-Yau threefold hypersurface with $h^{2,1} \lesssim 8$.

In \S\ref{sec:wnpkklt} we summarized the zero mode computation that determines the non-perturbative superpotential for the K\"ahler moduli, and so allows one to check condition (b).  Calabi-Yau threefold hypersurfaces with $h^{1,1} \gg 1$ tend to have many --- often $h^{1,1}$ or $h^{1,1}+1$ --- pure rigid divisors, in which case $W$ takes the KKLT form \eqref{eq:kkltptWrig}.
One can then check whether a KKLT point $T_a^{\star}$ \eqref{eq:kkltpt} exists in a region of control.\footnote{Furthermore, one can systematically incorporate the leading corrections to the condition \eqref{eq:tstar}. These corrections, which result from the finite size of $1/|W_0|$, and also from the $\alpha'^3$ and worldsheet instanton contributions to the K\"ahler potential and K\"ahler coordinates inherited from the parent $\mathcal{N}=2$ compactification, can all be computed explicitly \cite{Demirtas:2021nlu}.}

Examples have been found that fulfill all these conditions, in threefolds with $4 \le h^{2,1} \le 7$ and $51 \le h^{1,1} \le 214$.
In the example of \eqref{eq:example5113}, with $h^{2,1}=5$ and $h^{1,1}=113$, the string coupling is polynomially small, $g_s \approx 0.011$, and the vacuum energy is proportional to
\begin{equation}\label{eq:5113}
\biggl(\frac{2}{252}\biggr)^{58} \approx 10^{-122}\,,
\end{equation}
corresponding to an extraordinary degree of scale separation \cite{Demirtas:2021nlu}.  Extensive analysis of the control of corrections due to worldsheet instantons appears in \cite{Demirtas:2021nlu}, and rests on direct computation of Gopakumar-Vafa invariants to high degree.\footnote{String loop corrections to the K\"ahler potential have not been computed explicitly in the examples of 
\cite{Demirtas:2021nlu}, but their effects are parametrically  suppressed 
at weak coupling and large volume, and so will be negligible in the vacua of \cite{Demirtas:2021nlu} unless there are accidental enhancements by many orders of magnitude.  Advances in computing string loop corrections to the K\"ahler potential appear in \cite{Kim:2023eut,Kim:2023sfs}.}

We now turn to examples with warped throat regions.
The condition (c) was studied in \cite{Alvarez-Garcia:2020pxd,Demirtas:2020ffz}, and was shown to be compatible with (a): one can extend the PFV mechanism of \cite{Demirtas:2019sip} to the vicinity of conifolds.  
Stabilization of the K\"ahler moduli, condition (b), was not 
addressed in \cite{Alvarez-Garcia:2020pxd,Demirtas:2020ffz}, though no obstacle to stabilization was identified.

The full set of conditions, (a)-(e), are addressed in \cite{coni}: explicit examples have been found in which (a), (b), and (c) are fulfilled, and (d) and (e) hold to leading order in the $\alpha'$ expansion, i.e.~omitting $\alpha'$ corrections \cite{Hebecker:2022zme,Schreyer:2022len} to the potential of the anti-D3-brane in the warped throat.  
Thus, at leading order in $\alpha'$, these configurations are metastable KKLT de Sitter vacua.
We comment on related issues in
\S\ref{sec:antibrane}.

\subsubsection{Ten-dimensional description}\label{sec:10dkklt}

To understand how a KKLT vacuum can arise from the interplay of localized sources, fluxes, and the quantum effects of Euclidean D3-branes and gaugino condensates, it is instructive to obtain a ten-dimensional description of the solution, as originally proposed in
\cite{Koerber:2007xk},  and to compare this with the EFT that has been determined from general studies in non-perturbative supersymmetric field theory \cite{Amati:1988ft,Veneziano:1982ah,Affleck:1984xz,Burgess:1995aa}. A consistent ten-dimensional configuration should agree with the four-dimensional results.

Gaugino condensation is a four-dimensional effect, but in string compactifications the gaugino bilinear expectation value $\langle \lambda\lambda\rangle$ can be computed in the four-dimensional theory and included as a source term localized on D7-branes.
This requires knowing the couplings of D7-brane gauginos to bulk fields \cite{Camara:2004jj,Dymarsky:2010mf}, such as 
\begin{equation}\label{eq:d7source}
\mathcal{L} \supset c \int_{D} \sqrt{g}\, G_3{\cdot}\Omega\, \bar{\lambda} \bar{\lambda}\,,
\end{equation} where $c$ is a constant and $D$ is the divisor wrapped by the D7-branes.
Setting $\bar{\lambda} \bar{\lambda} \to \langle \bar{\lambda} \bar{\lambda}\rangle \neq 0$, the coupling \eqref{eq:d7source} becomes a source term for $G_3$ that is localized on $D$.
 
One can then seek a solution of the ten-dimensional supergravity equations of motion with these localized sources.
The appropriate framework is compactification on manifolds with dynamic SU(2) structure \cite{Grana:2005sn}: the gaugino condensate sources a localized deviation away from the conformally Calabi-Yau solution of the classical theory \cite{Koerber:2007xk,Heidenreich:2010ad,Baumann:2010sx,Dymarsky:2010mf},
including IASD flux profiles.

Consistency requires that the four-dimensional curvature $\mathcal{R}_4^{\text{10d sol.}}$ obtained from the ten-dimensional equations of motion should match the curvature $\mathcal{R}_4^{\text{EFT}}$ obtained from the four-dimensional Einstein equations equipped with the KKLT scalar potential obtained from \eqref{eq:vkw},\eqref{eq:kkltw1}.
The undertaking of precisely matching the ten-dimensional and four-dimensional results by computing $\mathcal{R}_4^{\text{10d sol.}}$ and comparing to $\mathcal{R}_4^{\text{EFT}}$ was initiated in 
\cite{Moritz:2017xto} and pursued in \cite{Hamada:2018qef,Gautason:2018gln,Carta:2019rhx,Gautason:2019jwq,Hamada:2019ack}.  Essential data for making this comparison comes from the ten-dimensional supersymmetry conditions (i.e., Killing spinor equations) that account for the gaugino condensate:
these conditions were proposed in \cite{Koerber:2007xk}, building on \cite{Grana:2005sn}, and further studied in \cite{Bena:2019mte}.

\begin{figure}[t]
\begin{center}
\includegraphics[width=140mm,height=82mm]{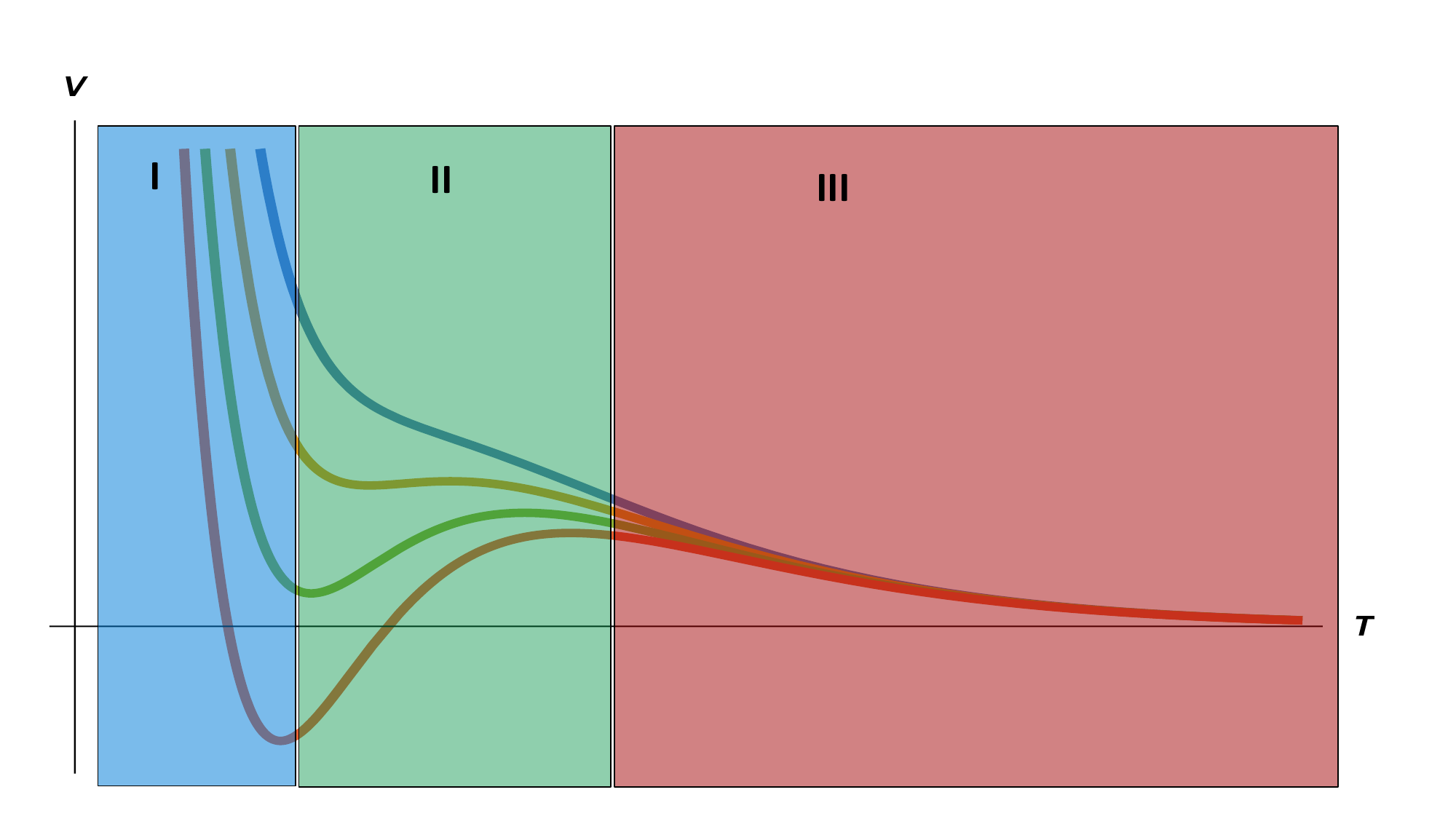}  
\caption{The Dine-Seiberg problem. For large volumes and weak coupling the scalar potential runs away towards the infinite volume or zero coupling limit. Region III is the region in which calculations in the EFT can be arbitrarily well controlled. The natural place to find a minimum, if one exists, is in region I, since it is here that quantum effects could most readily counterbalance the leading-order runaway behaviour. Obtaining a minimum in the physically interesting region II, with finitely large volume and weak coupling, is more challenging and requires extra sources of hierarchies.  Examples of such hierarchies are those achieved by a choice of fluxes in the KKLT scenario, or by competition of perturbative and non-perturbative terms in LVS. } \label{Fig:DS} 
\end{center}
\end{figure}
It was shown in \cite{Kachru:2019dvo} and \cite{Grana:2022nyp} that the supersymmetry conditions of \cite{Koerber:2007xk} do not provide a consistent representation of the gaugino condensate.  The corrected Killing spinor equations necessary for consistency were obtained in \cite{Kachru:2019dvo} at leading order in $\langle \lambda\lambda\rangle$, and then in general in \cite{Grana:2022nyp}.
Equipped with the corrected supersymmetry conditions, \cite{Kachru:2019dvo} found an exact match,
\begin{equation}
\mathcal{R}_4^{\text{10d sol.}}=\mathcal{R}_4^{\text{EFT}}\,,
\end{equation}
which is strong evidence that the ten-dimensional solution is a consistent representation of the four-dimensional effective theory.  The question of localized divergences in this setting remains open 
\cite{Hamada:2021ryq,Grana:2022nyp}.

\subsection{LVS}\label{ss:lvs}

We reviewed in \S\ref{sss:ktree} that the tree-level effective theory in type IIB flux compactifications has no-scale structure, cf.~\eqref{eq:vtvns}, and so the K\"ahler moduli $T_a$ are unstabilized at tree level.  Dependence of the scalar potential on the K\"ahler moduli arises from corrections to the tree-level data,
\begin{equation}
W = W_{\text{tree}} + \delta W\,, \qquad K = K_{\text{tree}} + \delta K\,.
\end{equation}
Recalling the general form \eqref{eq:wkform}, we have $\delta W = W_{\text{np}}$ and $\delta K = K_{\text{pert.}}+K_{\text{np}}$, and as before we use $W_0$ to denote the expectation value of $W_{\text{tree}}$ in the supersymmetric vacuum configuration of the complex structure moduli and axiodilaton.

Including the corrections $\delta K$ and $\delta W$ in the  scalar potential will lift the original flatness of the potential:
\be
V= V_{\delta K} + V_{\delta W}+\ldots  
\ee
The tree-level contribution proportional to $|W_0|^2$ vanishes thanks to the no-scale property \eqref{eq:vtvns}, and the corrections due to $\delta K$ and $\delta W$ are of order:
\be
V_{\delta K}\sim |W_0|^2\,\delta K, \qquad V_{\delta W}\sim |W_0|\,\delta W+(\delta W)^2\,.
\ee

One would like to find regions of moduli space where the portions of $V_{\delta K}$ and $V_{\delta W}$ that are computable actually suffice to ensure a controlled minimum for the moduli potential.\footnote{Computing $K_{\text{np}}$ systematically is currently out of reach, so we will only study regions where $\delta K \approx K_{\text{pert.}}$, and we therefore omit $K_{\text{np}}$ henceforth.}
The possible hierarchies that will be relevant are then\footnote{The cases
\eqref{eq:thisiskklt} and \eqref{eq:thisislvs} could be seen not as two alternative scenarios, but instead as different regimes for exploring the same class of models, in the sense that a scan of values of the flux superpotential can smoothly interpolate from one class  to the other, as was shown in the example of $\mathbb{C}{\mathbb P}^4[{1,1,1,6,9}]$ in \cite{AbdusSalam:2007pm}.}
\begin{eqnarray}
|W_0| \sim\delta W \gg \delta K\,,    \label{eq:thisiskklt} \\ 
|W_0|  \gg \delta K \sim \delta W\,,    \label{eq:thisislvs}\\
|W_0|  \gg \delta K \gg \delta W     \label{eq:thisispertk}\,.  
\end{eqnarray} 
We have already discussed \eqref{eq:thisiskklt}: the KKLT scenario relies on choosing fluxes such that $\delta W\sim |W_0|$. 
In this case $V_{\delta K}\sim  |W_0|^2 \delta K\ll V_{\delta W}\sim |W_0|^2$, and so we can neglect the corrections $\delta K$. 
The leading-order contribution to the scalar potential $V_{\delta W}$ can then give rise to  a minimum, as we explained in \S\ref{ss:kklt}.

However, because $\delta K$ is polynomial in inverse volumes $1/T_a$, whereas $\delta W \equiv W_{\text{np}}$ is exponential in volumes, it is arguably more natural to suppose that $\delta K \gtrsim \delta W$.  That is, although we showed explicitly in \S\ref{sec:explicitkklt} that one can find configurations obeying \eqref{eq:thisiskklt} and stabilizing all moduli, one might expect \eqref{eq:thisislvs} and \eqref{eq:thisispertk} to occur more often in the landscape of flux vacua.

Stabilization achieved exclusively through perturbative corrections to $K$, as in \eqref{eq:thisispertk}, is an interesting possibility explored in \cite{Berg:2005yu}.
For weak coupling and large volumes the first non-vanishing term in the perturbative expansion will dominate, giving rise to a runaway potential and therefore no non-trivial minima at weak coupling. This is the celebrated Dine-Seiberg problem (see Figure \ref{Fig:DS}).

To avoid this outcome in the regime \eqref{eq:thisispertk}, one would need to understand $\delta K$ well enough to engineer a competition between two or more terms that suffices to stabilize all moduli in a region of control.  This difficult task has not yet been achieved.\footnote{An interesting perturbative approach to moduli stabilization was outlined recently in \cite{Burgess:2022nbx}. There is a concrete case in QFT in which a perturbation expansion can be re-summed: as is well-known, the renormalization group provides logarithmic corrections to the EFT. For string moduli these are re-summations in an expansion in $\alpha \log \V$, with $\alpha$ a weak coupling  parameter.  Following the Dine-Seiberg logic, this structure would naturally lead to a minimum in which the expansion parameter is  $\O(1)$, which would imply $\V\sim e^{1/\alpha}\gg 1$. This would be an alternative way to fix the volume modulus consistent with weak coupling, and would naturally lead to exponentially large volumes. It would be very interesting to achieve explicit string realizations of this scenario.}

We now turn to the more fruitful case \eqref{eq:thisislvs}, in which  $\delta K$ and  $\delta W$ are in competition. 
The main point is that the generic Calabi-Yau manifold has many K\"ahler moduli.  One combination of the moduli  gives the volume $\V$ that controls the overall $\alpha'$ expansion. However, as we have seen, each of the K\"ahler  moduli may be an expansion parameter: Euclidean D3-branes 
on a rigid four-cycle $D$ with $\text{Vol}(D)=c^a \tau_a$ give $\delta W \sim e^{-2\pi c^a\tau_a}$. 
If $W_0=\O(1)$ we have $V_{\delta K}\sim \delta K$ and $V_{\delta W}\sim \delta W$.
So if $\delta K\sim 1/\V $ and $\delta W\sim e^{-2\pi c^a\tau_a}$ the scalar potential may develop a minimum where $\delta K \sim  \delta W$, i.e.~at 
\begin{equation}\label{eq:lvsrel}
    \V\sim e^{2\pi c^a\tau_a}\,,
\end{equation} and for  
$\tau_a \gg 1$ the volume $\V$ will be exponentially large in the minimum.  This is the origin of the name Large Volume Scenario (LVS) \cite{Balasubramanian:2005zx, Conlon:2005jm}.

Let us now be more explicit.  Following our discussion of perturbative corrections to the K\"ahler potential in \S\ref{sss:kpert}, we consider the leading-order $\alpha'$ correction at string tree level: this is the $\alpha'^3$ correction of \cite{Becker:2002nn}, cf.~\eqref{bbhk}:
\be\label{eq:kwithbbhl}
K_{\alpha'^3}=-\ln\left(-{i}\int_{X_6}\Omega(z_i)\wedge\overline{\Omega}({\bar z}_i )\right)-\ln\Bigl(-i(\tau-\bar\tau)\Bigr)-2\ln\left(\V+\frac{\xi}{2}\left(\frac{\tau-\overline\tau}{2i}\right)^{3/2}\right),  
\ee
where 
\be
\xi =-\frac{\zeta(3)\chi(X_6)}{2(2\pi)^3}.
\ee
Here $\chi(X_6)$ is the Euler number of the corresponding Calabi-Yau manifold and $\zeta$ is the Riemann $\zeta$ function. $\mathcal{N}=1$ corrections to the K\"ahler potential, yet to be computed, are expected to modify this coefficient.

The simple modification \eqref{eq:kwithbbhl} of the K\"ahler potential changes the structure of the scalar potential, once it is combined with the flux plus non-perturbative superpotential \eqref{eq:wtot},
\be\label{eq:wlvs}
W=W_0+\sum_{a=1}^n \mathcal{A}_a e^{-2\pi c^a T_a}
\ee
where the index $a$ runs over all effective divisors,
with the understanding that $\mathcal{A}_a$ may vanish for some of them if the corresponding cycle is not rigid.  Thus, $n$ may be smaller than, equal to, or larger than $h^{1,1}$, depending on the threefold in question.  
 
The leading-order scalar potential is positive-definite in the dilaton and complex structure moduli directions, and therefore to leading order these fields are stabilized by the fluxes, as in KKLT, at $D_{z_i}W=0$ and $D_\tau W=0$.

The remaining potential as a function of the K\"ahler moduli reads:
\be\label{eq:vsteplvs}
V=e^K\left[ K_{T_a}K^{T_a\overline{T}_b}W\overline{W}_{\overline{T}b}+W_{T_a}K^{T_a\overline{T}_b}\left(\overline{W}_{\overline{T_b}}+K_{\overline{T_b}}\overline{W}\right)+\left(K_{T_a}K^{T_a\overline{T}_b}K_{\overline{T_b}}-3 \right)\left | W \right |^2\right].
\ee
Inserting the expressions \eqref{eq:wlvs} and \eqref{eq:kwithbbhl} in \eqref{eq:vsteplvs}, the scalar potential can be written as a sum of three terms (quadratic, linear and independent of $W_0$, respectively) \cite{AbdusSalam:2020ywo},
\be \label{eq:mastersummed}
V=V_{\alpha'^3}+V_{\text{np1}}+ V_{\text{np2}}\,,
\ee
where:  
\begin{eqnarray}\label{master}
    V_{\alpha'^3}&=& e^K\left | W_0 \right |^2 \times \frac{3\xih\left(\V^2+7\V\xih +\xih^2\right)}{\left(\V-\xih \right)\left(2\V+\xih \right)^2}\,, \label{eq:ftmast}\\
    V_{\text{np1}} &=&  e^K \left | W_0\right | \times \sum_a 2 \left | \mathcal{A}_a\right | e^{-c_a\tau_a}\cos\bigl( b_a\theta_a -\phi_a +\theta_W\bigr) \nonumber  \\[-2pt] 
    &&\times \left[\frac{\left(4\V^2+\xih\V +4\xih^2 \right)}{\left( \V-\xih\right)\left( 2\V+\xih\right)}c_a\tau_a
    +\frac{3\xih \left(\V^2+7\xih \V+\xih^2\right)}{\left(\V-\xih\right)\left(2\V+\xih \right)^2}\right]\,,  \label{eq:stmast} \\[8pt]
    V_{\text{np2}} &=& e^K\sum_{a,b}\left |\mathcal{A}_a\right | \left |\mathcal{A}_b\right | e^{-\left(c_a\tau_a + c_b\tau_b\right)}\cos\bigl(c_a\theta_a-c_b\theta_b -\phi_a +\phi_b \bigr)\nonumber \\
    &&\times \vastl[-\left(4\V+2\xih\right) \left(\kappa_{abc}t^c\right) c_a c_b + \frac{4\V-\xih}{\V-\xih}\left( c_a\tau_a\right)\left(c_b\tau_b\right)\nonumber \\
     && + \frac{4\V^2+\xih \V + 4 \xih^2}{\left (\V-\xih \right)\left(2\V+\xih \right)}\left(c_a\tau_a+c_b \tau_b \right) +\frac{3\xih \left ( \V^2 + 7\xih \V + \xih^2 \right)}{\left (\V-\xih \right)\left(2\V+\xih \right)^2}\vastr]\,. \label{eq:ttmast}
\end{eqnarray}
Here $\xih:= \xi/g_s^{3/2}$, and $\phi_a$ and $\theta_W$ are the phases of $\mathcal{A}_a$ and $W_0$, respectively.

The scalar potential \eqref{eq:mastersummed} is singular at $\V=\xih$ for $\xih>0$ and $\V=-\xih/2$ for $\xih<0$,
but in any case validity of the perturbative expansion requires $\V \gg \xih$, so these singularities will not be relevant.  
In general the domain of validity of the scalar potential is also limited by the fact that the K\"ahler moduli need to lie in the K\"ahler cone.

The first term \eqref{eq:ftmast} vanishes as $\xi \to 0$, as it should in order to recover the tree-level result in this case. 
For $\V\gg \xih$ we see from \eqref{eq:kwithbbhl} that $e^{K}$ scales as $1/\V^2$, so the first term \eqref{eq:ftmast} scales as $1/\V^3$, whereas \eqref{eq:stmast} and \eqref{eq:ttmast} scale as $1/\V^2$ and $1/\V$, respectively, 
but also depend on powers of $e^{-\tau_a}$, making the moduli stabilization calculation very rich. 
An important technical point is that $V_{\text{np2}}$
depends explicitly on the two-cycle moduli $t^a$, and in general it is not possible to determine $t^a(\tau_b)$ analytically. In practical terms it is therefore easier to consider the scalar potential as a function of the $t^a$ fields.

The formula (\ref{master}) is very general and holds for any number of moduli.\footnote{Here we have neglected the orientifold-odd moduli. A more general expression including these moduli can be found in \cite{Cicoli:2021tzt}.} 
Let us consider key illustrative cases: 
\begin{itemize}
\item $h^{1,1}=1$. This is the simplest case considered by KKLT. Here there is only one K\"ahler modulus, $T_1$, and one triple intersection number $\kappa_{111}:=k$. We assume that there is a non-perturbative contribution, i.e.~we take $n=h^{1,1}=1$, and
\be
\V=\frac16 kt^3, \qquad \tau_1=\frac12 kt_1^2, \qquad \V=\frac13 \sqrt{\frac{2}{k}}\tau_1^{3/2, \qquad }W=W_0+\mathcal{A}e^{-cT_1}.
\ee
Since it is possible to express $\V$ as a function of $\tau_1$ we can work with the variable $\tau_1$. Equation (\ref{eq:mastersummed}) reproduces   the KKLT potential for $\xih=0$. 
\be
V=\frac{9a k g_se^{K_{\text{cs}}}}{\tau_1^2}\left |\mathcal{A}\right |e^{-c\tau_1}\left [ \left | W_0 \right | \cos (c\theta_1+\theta_W -\phi_1)+\frac{|\mathcal{A}|}{3}e^{-c\tau_1}\left(c\tau_1+3\right)\right]
\ee
which reproduces the supersymmetric AdS minimum of KKLT.
Non-vanishing values of $\xih$ add corrections to the KKLT potential that have been considered in \cite{Balasubramanian:2004uy}, and that can contribute to uplifting the original AdS minimum. Since there is only one K\"ahler modulus there is no LVS minimum.

\item $h^{1,1}=2$. This is the simplest case allowing an LVS minimum. Depending on the nature of the divisors and the triple intersection numbers, the volume may or may not be expressed explicitly in terms of the variables $\tau_1$ and $\tau_2$. Also in this case we may or may not have non-perturbative dependence of $W$ on both moduli ($n=1,2$). If $n=2$ this allows a generalization of KKLT for two moduli. If $n=1$ the second modulus needs to be stabilized by perturbative effects.

The simplest example has $X_6$ an orientifold of the Calabi-Yau threefold hypersurface in weighted projective space   $\mathbb{C}{\mathbb P}^4[{1,1,1,6,9}]$. In this case,
\be
\tau_1=\frac{t_1^2}{2}, \qquad \tau_2=\frac{\left(t_1+6t_2\right)^2}{2}, \qquad 6\V=3t_1^2 t_2+18t_1t_2^2+36t_2^3=\tau_2^{3/2}-\tau_1^{3/2}\,.
\ee
The scalar potential for $\V\gg \xih$ takes the form:
\be
V\simeq g_se^{K_{\text{cs}}}\left[\frac{\alpha}{\V^3}+\frac{\beta \tau_2 e^{-c_1\tau_1}\cos\left(c_1\theta_1+\theta_W-\phi_1\right)}{\V^2}+\frac{\gamma \sqrt{\tau_1}e^{-2c_1\tau_1}}{\V}\right]
\ee
where
\be
\alpha=\frac{3\xih \left | W_0\right |^2}{4}, \qquad \beta= 2 |\mathcal{A}| \left | W_0 \right | c_1, \qquad \gamma= \frac{4}{3}c_1^2 |\mathcal{A}|^2\,.
\ee
Extremizing with respect to $\theta_1$ sets the cosine term to $-1$, and then extremizing with respect to $\tau_1$ and the volume $\V$ (or $\tau_2$) gives rise to exponentially large volume if $\alpha>0$ (i.e., if the Euler characteristic $\chi(X_6)$ is negative):
\be
\langle \V\rangle \simeq\langle \tau_2^{3/2}\rangle \simeq  e^{c_1\tau_1}, \qquad \langle \tau_1\rangle \simeq \frac{\xi^{2/3}}{g_s}\gg 1\,.
\ee
Here we considered only the case $n=1$, without non-perturbative superpotential dependence on $\tau_2$. This is justified for the hypersurface in $\mathbb{C}{\mathbb P}^4[{1,1,1,6,9}]$ since the divisor with volume $\tau_2$ is not rigid. The presence of fluxes may modify this conclusion and give rise to non-perturbative contributions (see for instance \cite{Louis:2012nb}). In any case, such terms are highly suppressed compared to the $1/\V$ dependence and can be neglected. 

The above vacuum is the prototypical example of LVS. 
In this example, it was essential that the volume took the form
\begin{equation}\label{eq:simplescf}
\V\propto \tau_2^{3/2}-\tau_1^{3/2}\,,
\end{equation}
so that $\V$ decreases as $\tau_1$ increases.
Since the cycle of volume $\tau_1$ behaves like a hole in $X_6$, volumes $\V$ with the structure \eqref{eq:simplescf} are said to be of Swiss cheese form.
The generalization to more than two K\"ahler moduli is
\be \label{eq:scf}
\V\propto F(\tau_{\tilde a})-G(\tau_{\hat a})
\ee
where we have split the moduli $\tau_a$ into two classes, $\tau_{\tilde a}$ and $\tau_{\hat a}$, and $F$ and $G$ are positive homogeneous functions of degree $3/2$ of their arguments, such that the volume $\V$ decreases with increasing $\tau_{\hat a}$.
In the case \eqref{eq:simplescf}, 
the corresponding divisor corresponds to a del Pezzo surface that can be shrunk to zero size without affecting the positivity of the volume. 

When $G$ is given by
\begin{equation}\label{eq:ddp}
    G=\sum_{\hat a} \tau_{\hat a}^{3/2}\,,
\end{equation} then the shrinking divisors are called diagonal del Pezzo surfaces. 
Although the form \eqref{eq:ddp} appears to be quite special, it is common at   $h^{1,1}=2$, appearing in $22$ out of $39$  Calabi-Yau threefolds from the Kreuzer-Skarke list at $h^{1,1}=2$ \cite{AbdusSalam:2020ywo}.\footnote{Only $27$ of the $39$ examples are topologically inequivalent, according to the systematic counting of inequivalent threefolds in the Kreuzer-Skarke list in \cite{Chandra:2023afu,Gendler:2023ujl}.} 
Furthermore, this structure  can easily be generalized to the $h^{1,1}>2$ cases.

In the models with $h^{1,1}=2$ for which the Swiss cheese structure is not present, equation (\ref{eq:mastersummed}) remains applicable. 
A systematic analysis of all the models with $h^{1,1}=2$  in the Kreuzer-Skarke list shows that seven additional models may give rise to LVS minima even though they lack diagonal del Pezzo divisors. 
In these cases it is more convenient to extremize the scalar potential with respect to the two-cycle moduli $t^a$. 
The remaining $10$ models with $h^{1,1}=2$ correspond to K3 or $T^4$ fibrations over $\mathbb{P}_1$. In these cases the volume modulus is linear in one of the $t^a$ moduli and can be written as $\V\sim \tau_1\sqrt{\tau_2}$. These cases do not give rise to LVS vacua .

\item $h^{1,1}=3$. Again formula (\ref{eq:mastersummed}) may be used to identify vacua in which all the moduli are stabilized. There are again extensions of KKLT and LVS vacua. Out of $305$ different Calabi-Yau orientifolds more than $120$ allow for LVS minima, either by having a Swiss cheese structure $\V=\tau_3^{3/2}-\tau_2^{3/2} -\tau_1^{3/2}$, or a slightly modified form $\V=\tau_3^{3/2}-(\alpha \tau_2+\beta\tau_1)^{3/2} -\tau_1^{3/2}$. 

The case of $h^{1,1}=3$ illustrates much richer structure than the previous cases. A substantial fraction of models cannot be minimized in terms of the $\tau_a$ and one needs to minimize (\ref{master}) with respect to the $t^a$. Even though analytic results are difficult to extract, numerical analysis is enough to identify minima, several of them of the LVS type. This illustrates that there are more LVS minima than naively expected.

Some of these models have one rigid divisor, such as a del Pezzo surface,
while the third modulus is a K3 fibration modulus that is not fixed at this order of perturbation theory, and whose fixing relies on higher-order corrections (such as those discussed in \S\ref{sss:kpert}) that scale with higher powers of $1/\V$, for example as $\delta V\sim \O(\V^{-10/3})$. These fibrations allow for interesting anisotropic compactifications
\cite{Cicoli:2011yy} in which two of the dimensions are much larger than the other four. The flatness of the fiber moduli potentials has also been used as one of the leading models for string moduli inflation \cite{Cicoli:2008gp}: see \S\ref{ss:inflationst}.

\item $h^{1,1}\geq 4$. Cases with larger numbers of K\"ahler moduli have been less  systematically explored, mostly due to computational limitations (however, see \S\ref{sec:computation}), but they are all captured by formula (\ref{eq:mastersummed}). 
In cases where there are $n<h^{1,1}$  rigid divisors, 
the existence of an isolated vacuum
relies on understanding higher-order perturbative corrections to the K\"ahler potential.
 
The structures that appeared at small $h^{1,1}$, such as the appearance of
del Pezzo surfaces and fiber moduli, continue at larger $h^{1,1}$:
in fact, it is known that most Calabi-Yau manifolds are fibrations \cite{Rohsiepe:2005qg,Johnson:2014xpa,Candelas:2012uu,Gray:2014fla}.  
LVS minima are currently limited to $h^{1,1}<h^{2,1}$,  which sets a bound on the exploration of these vacua.
 
\end{itemize}

\subsection{Supersymmetry breaking and de Sitter uplift}

The vacua with stabilized moduli that we have described so far are AdS$_4$ solutions: these are $\mathcal{N}=1$ supersymmetric in the case of KKLT, and non-supersymmetric in the case of LVS.
The \emph{uplift} paradigm \cite{Kachru:2003aw} proposes that from an AdS$_4$ solution with $V=V_{\text{AdS}}<0$, one can incorporate a source of supersymmetry breaking such that
\begin{equation}\label{eq:uplift}
V \to V_{\text{tot}} = V_{\text{AdS}} + V_{\text{up}}
\end{equation} has a metastable de Sitter minimum.

Let us illustrate the idea in the simple example of a KKLT model with $h^{1,1}_{+}=1$, as in 
\S\ref{ss:kklt}.  In the supersymmetric AdS$_4$ vacuum, the F-term potential \eqref{eq:vkw} is
\begin{equation}\label{eq:vadsk}
    V_{\text{AdS}} = V_F = -3e^{K_{\text{tree}}}|W|^2+\ldots\,,
\end{equation} with $W$ given by \eqref{eq:kkltw1}, $T$ obeying \eqref{eq:tstar}, and the omitted terms representing higher corrections that can be neglected for $|W_0|\ll 1$.

The first idea for an uplift, due to Kachru, Pearson, and Verlinde (KPV) \cite{Kachru:2002gs}, proposes that $p$ anti-D3-branes at the bottom of a Klebanov-Strassler warped throat region can break supersymmetry, with\footnote{The denominator was given as $(T+\overline{T})^3$ in \cite{Kachru:2003aw} and corrected to $(T+\overline{T})^2$, which properly accounts for the warping, in \cite{Kachru:2003sx}: see the treatment in \cite{Junghans:2014xfa}.}
\begin{equation}\label{eq:antibraneuplift}
V_{\text{up}} = \frac{\mathscr{D}\,T_{\text{D3}}}{(T+\overline{T})^2}>0\,,
\end{equation}  where
\begin{equation}\label{eq:antibraneupliftD}
\mathscr{D} = 2p\,  e^{4A(y_{\text{IR}})} \,.
\end{equation} Here $T_{\text{D3}}$ is the D3-brane tension, $y_{\text{IR}}$ is the location of the tip of the warped throat inside $X_6$, and $e^{A(y)}$ is the warp factor from \eqref{eq:warpedproductansatz}, evaluated in the Klebanov-Strassler solution \cite{Klebanov:2000hb}.  The exponential smallness of the warp factor ensures that the supersymmetry breaking effect is small in string units, i.e.~$\mathscr{D} \ll 1$.

For suitable values of $\mathscr{D}$, the total potential for $T$, 
\begin{equation}\label{eq:vtotkklt}
V_{\text{KKLT}} := V_{\text{AdS}} + V_{\text{up}}\,,    
\end{equation}
with $V_{\text{AdS}}$ given by \eqref{eq:vadsk} and $V_{\text{up}}$ given by \eqref{eq:antibraneuplift}, has a metastable de Sitter minimum at a point $T=T^{\text{dS}}$.  With $T^{\star}$ the minimum \eqref{eq:tstar} obtained from \eqref{eq:vadsk} alone, one finds that 
\begin{equation}
    T^{\text{dS}}-T^{\star} = \mathcal{O}\bigl(1/{T^{\star}}\bigr) = \mathcal{O}\Bigl(1/\,\text{log}\bigl(|W_0|^{-1}\bigr)\Bigr) \ll 1\,,
\end{equation}
so the K\"ahler modulus shifts by a small amount toward larger volume.  

We remark that the expression \eqref{eq:vtotkklt} arose from asserting that the F-term potential $V_{\text{AdS}}$ and the anti-D3-brane energy $V_{\text{up}}$ combine by simple addition.  This is easily argued to be a good approximation in the four-dimensional EFT \cite{Kachru:2003aw}, and the ten-dimensional computations summarized in \S\ref{sec:10dkklt} provide independent evidence from a very different perspective \cite{Kachru:2019dvo}.   

The expressions \eqref{eq:antibraneuplift} and \eqref{eq:antibraneupliftD} for the anti-D3-brane potential were obtained at leading order in $\alpha'$ by KPV \cite{Kachru:2002gs}.  At this level of approximation, 
explicit flux compactifications furnishing candidates for KKLT de Sitter vacua have been obtained in \cite{coni} (cf.~\cite{dstalk}).  These vacua are built on the PFV mechanism reviewed in \S\ref{sec:pfv} and the resulting framework for supersymmetric AdS$_4$ vacua described in \S\ref{sec:explicitkklt}.  However, computations of $\alpha^\prime$ corrections to the KPV potential \cite{Hebecker:2022zme,Schreyer:2022len} indicate that 
in order to realize metastable de Sitter vacua that survive $\alpha^\prime$ corrections, the curvature of the throat region must fulfill a condition that is more stringent than that assumed in KPV and achieved in \cite{coni}.

\subsubsection{Anti-D3-branes} \label{sec:antibrane}

Before we describe other ideas for sources of supersymmetry-breaking uplifts (in \S\ref{ss:otheruplift}), we will summarize the status of anti-D3-brane supersymmetry breaking.  

The analysis of KPV relied on treating the anti-D3-brane as a probe of the Klebanov-Strassler geometry, and subsequent work pursued a backreacted solution \cite{DeWolfe:2008zy,McGuirk:2009xx}.
The consistency of the KPV proposal was questioned in \cite{Bena:2009xk}, on the basis that the linearized solution produced by anti-D3-branes smeared on the $S^3$ has singularities resulting from three-form fluxes.
Dymarsky obtained an $SU(2) \times SU(2)$-invariant linearized solution with appropriate boundary conditions \cite{Dymarsky:2011pm} (see also \cite{Bena:2011hz,Bena:2011wh}), showed that the ADM mass and the Coulomb potential for a D3-brane matched the probe computations \cite{Kachru:2003sx}, and further argued that the singularities in the fluxes were artifacts of linearization.
Singularities and polarization were studied in e.g.~\cite{Blaback:2011pn,Gautason:2013zw,Junghans:2014xfa},
and the effective field theory description of antibrane sources
was used in \cite{Michel:2014lva,Polchinski:2015bea} to argue that instabilities are absent.
It was then shown in \cite{Cohen-Maldonado:2015lyb,Cohen-Maldonado:2016cjh} (see also \cite{Hartnett:2015oda}) that the singularities were a consequence of the ansatz being too restrictive, and not allowing the polarized brane. 
The blackfold formalism was applied in \cite{Armas:2018rsy} to study the anti-D3-brane in a broader region of parameter space, including finite temperature, and the results gave strong evidence for consistency of the metastable KPV state (see also \cite{Blaback:2019ucp,Nguyen:2019syc,Nguyen:2021srl}).

A further issue arises in KKLT models as a result of the fine-tuning needed to achieve a de Sitter uplift \cite{Carta:2019rhx,Gao:2020xqh}.
For the anti-D3-brane energy  \eqref{eq:antibraneuplift} to fall in the range such that
\eqref{eq:vtotkklt} has a metastable minimum with positive vacuum energy, one needs the warp factor in \eqref{eq:antibraneupliftD} to obey
\begin{equation}\label{eq:upliftreq}
e^{4A(y_{\text{IR}})} \sim |W_0|^2\,.
\end{equation}
The warp factor can be expressed in terms of the D3-brane charge $N_{\text{D3}}^{\text{throat}}$ and the curvature radius $R_{\text{throat}}$ at the tip of the throat as \cite{Giddings:2001yu}
\begin{equation}\label{eq:efour}
    e^{4A(y_{\text{IR}})} = \text{exp}\Biggl(-\frac{8\pi}{3} \frac{N_{\text{D3}}^{\text{throat}}}{R_{\text{throat}}^4}\Biggr)\,,
\end{equation}
while at the KKLT point $T_a^{\star}$ the K\"ahler moduli and $W_0$ are related by \eqref{eq:tstar}, i.e.~
\begin{equation}\label{eq:newret}
\text{Re}\,T_a^\star = \frac{c_a}{2\pi}\,\text{log}\bigl(|W_0|^{-1}\bigr)\,,
\end{equation}
with $c_a$ the dual Coxeter number for the gauge group on $D_i$.  Because $c_a \in \{1,6\}$ in all the examples of \S\ref{sec:explicitkklt}, and $6$ is not a large number for the considerations at hand, we will set $c_a \to 1$ to simplify the discussion.
Combining the relations \eqref{eq:upliftreq}, \eqref{eq:efour}, and \eqref{eq:newret}, and requiring $R_{\text{throat}} \gtrsim 1$ so that the supergravity description of the throat is valid, we find
\begin{equation}\label{eq:singbulk}
N_{\text{D3}}^{\text{throat}} \gtrsim \text{Re}\,T_a^\star\,.
\end{equation}
The observation made in \cite{Carta:2019rhx} and further investigated in \cite{Gao:2020xqh} is that for some compactification geometries $X_6$, \eqref{eq:singbulk} implies that the warp factor $e^{A}$ becomes negative in the bulk of $X_6$, rather than just becoming negative in an exponentially small neighborhood of the O7-planes, where supergravity is in any case not the appropriate description.  This issue is called the \emph{singular bulk problem} \cite{Gao:2020xqh}.

Whether the singular bulk problem is present in a given $X_6$ depends on how the overall volume $\mathcal{V}$ and the four-cycle volumes are related.  Roughly speaking, $\mathcal{V}$ controls the size of the throat that can fit into $X_6$ without causing warp factor singularities to appear (see \cite{Carta:2019rhx,Gao:2020xqh,Carta:2021lqg} for a precise statement).
Taking the volumes to be related by
\begin{equation}\label{eq:vnaive}
    \mathcal{V} = \mathcal{P}\,(\text{Re}\,T_a^\star)^{3/2}\,,
\end{equation} with $\mathcal{P}$ a constant,
one can check that if  $\mathcal{P}$ is
of order unity, the singular bulk problem is severe \cite{Carta:2019rhx,Gao:2020xqh}.

Taking $\mathcal{P} \sim 1$ is a reasonable first guess for a highly isotropic Calabi-Yau threefold, but given an explicit compactification one can simply compute $\mathcal{P}$ at the KKLT point $T_a^{\star}$.  In the example given in \eqref{eq:5113} of \S\ref{sec:explicitkklt}, one finds $\mathcal{P} \approx 8000$ \cite{Demirtas:2021nlu}.  The large value of $\mathcal{P}$ is a geometric fact about four-cycle sizes at KKLT points in threefolds with many moduli --- cf.~\eqref{eq:ksa} --- that very significantly ameliorates the singular bulk problem.  This mechanism persists in de Sitter configurations: in the vacua of \cite{coni}, in which the fine-tuning  \eqref{eq:upliftreq} is explicitly engineered and a de Sitter uplift occurs at leading order in $\alpha'$, the singular bulk problem is again alleviated by geometric factors.

\subsubsection{Non-linear supersymmetry and nilpotent superfields} 

In order to study anti-D3-brane supersymmetry breaking, it is sometimes useful to repackage the
supersymmetry-breaking effects 
in terms of a nilpotent chiral superfield $X$
(see for instance \cite{Ferrara:2014kva,Kallosh:2014wsa,Bergshoeff:2015jxa, Kallosh:2015nia}):
\be
X=X_0+\sqrt{2}\psi\theta+F\theta \overline \theta, \qquad X^2=0 \implies X_0=\frac{\psi\psi}{2F}\,.
\ee
The only propagating component of $X$ is the fermion $\psi$,
which is identified with the goldstino of nonlinearly-realized supersymmetry.  The nilpotent constraint substantially limits the possible $X$ dependence of the superpotential and K\"ahler potential to be
\be
W=W_0+\rho X, \qquad K=K_0+K_1 X+\overline{K}_1\overline X+K_2 X\overline X\,,
\ee
with $W_0, \rho, K_i$ functions of the other fields. In the case of a single K\"ahler modulus $T$, $ K_0=-3\log \left(T+\overline T\right)$, so in an expansion in $1/\left(T+\overline T \right)$ we can write\footnote{This can be also argued in terms of the $SL(2,\mathbb R)$ modular symmetry 
$T \rightarrow \left(aT-ib\right)/\left(icT+d\right)$, if $X$ transforms as a modular form of  weight $\lambda$.} 
$K_2=\left( T+\overline T\right)^{-\lambda}$, so that
the scalar potential takes the general form:
\be
V=V_0+V_X, \qquad V_X=K^{-1}_{X\overline X}\left |\frac{\partial W}{\partial X}\right |^2=\frac{\left |\rho \right |^2}{\left(T + \overline T \right)^{3-\lambda}}\,,
\ee
with $V_0$ the KKLT or LVS scalar potential. 
Notice that modular weight $\lambda =1$ corresponds to a K\"ahler potential of the no-scale type, $K=-3\log \left(T+\overline T -X\overline X\right)$, and also for $\lambda =1$ we have $V_X=V_{\text{up}}$. This implies that the nilpotent superfield formalism naturally captures the effects of an anti-D3-brane. 
In sum, the Volkov-Akulov formalism of nonlinearly-realized supersymmetry provides a convenient representation of anti-D3-brane supersymmetry breaking.
 
Explicit orientifold compactifications have been found in which the spectrum of the anti-D3-brane at the tip of a throat consists of only the goldstino field, with the nilpotent superfield formalism describing the goldstino EFT couplings\cite{Kallosh:2015nia, Garcia-Etxebarria:2015lif,Aparicio:2015psl}: see also \cite{Antoniadis:2014oya,Dudas:2015eha,GarciadelMoral:2017vnz,Cribiori:2019hod}.

\subsubsection{Other uplifting mechanisms }\label{ss:otheruplift}
Anti-D3-brane uplifting has been the best-studied mechanism to break supersymmetry in KKLT and to uplift the supersymmetric vacuum to a de Sitter vacuum.
However, many other proposals exist, of which we will highlight a few.\footnote{An alternative to the uplifts discussed here was proposed in \cite{deAlwis:2021zab}: quantum effects from the Standard Model sector may uplift an AdS vacuum with sufficiently small vacuum energy.}

\begin{itemize}
\item{\bf Uplift from $\alpha'$ corrections.} The 
$\alpha'^3$ corrections to the K\"ahler potential given in \eqref{eq:kwithbbhl} contribute
\be \label{eq:vbbhl}
V_{\alpha'^3}= \frac{\xi \left|W_0\right|^2}{\V^3}, \qquad \xi \propto -\chi(X_6)\,.
\ee
For $\xi>0$ (which, in light of mirror symmetry, holds in half of all Calabi-Yau manifolds, setting aside cases with $\chi=0$) this term contributes positively to the KKLT potential and may lift the minimum towards de Sitter. The main issue is control of the approximation: the term \eqref{eq:vbbhl} needs to compete with lower-order terms present in the KKLT potential. For LVS the term  \eqref{eq:vbbhl} is responsible for the original non-supersymmetric AdS minimum. But scanning over different values of the flux superpotential, from the very small values required for KKLT to order $\O(1-100)$ as appears in the original LVS, other minima appear in different regimes, some of them with positive vacuum energy.  However, although the volume is relatively large, it is usually not large enough for the approximations to be trusted \cite{AbdusSalam:2007pm}. For concrete examples of this mechanism see for instance \cite{Balasubramanian:2004uy,Westphal:2006tn,Rummel:2011cd,Louis:2012nb}.

\item {\bf D-term uplift}. The scalar potential for $\mathcal{N}=1$ supersymmetric theories has two contributions: the F-term potential \eqref{eq:vkw} that we have been considering so far, and the D-term potential resulting from integrating out the D-term of a gauge superfield $\hat V^\alpha$:
\be
V=V_F+V_D, \qquad V_D=\frac{1}{2}D^\alpha D_\alpha\,.
\ee
Here
\be\label{dterm}
D_\alpha=i\G_m X^m_\alpha,\qquad \G=K+\log|W|^2,
\ee
and $X^m_\alpha$ define the Killing vectors such that under  
a gauge transformation the matter fields $\psi_m$ transform as $\delta \psi^m = X^m_\alpha \epsilon^\alpha$.
For linearly-realized gauge symmetry, $X^m_\alpha=\left( -i T_\alpha\right)^m_k\psi^k$.

Since the  D-term potentials  are manifestly positive, unlike F-terms, it is natural to inquire if they can be used for de Sitter uplift \cite{Burgess:2005jx}.  In generic orientifold constructions, D-terms that can potentially lead to de Sitter vacua can be identified. The point  is that the O7-plane fixed loci carry D7-brane charge and so require the presence of D7-branes to cancel D7-brane tadpoles. Stacks of  D7-branes wrapping a divisor $D_a$ host a corresponding gauge theory, with gauge coupling $g_a$ determined by $\langle \hbox{Re}\,T_a\rangle =1/g_a^2$. 
The D7-brane Chern-Simons coupling
\be
\int_{\mathcal{M}_8}C_4\wedge\hbox{Tr}\, e^{\frac{i F}{2\pi}}
\ee
gives rise to a $\hat B\wedge F$ term in four dimensions, with $\hat B_{\mu\nu}\propto C_{\mu\nu mn}$, after expanding the exponential and turning on magnetic fluxes $\langle F^{mn}\rangle \neq 0, m,n=4,\ldots, 7$. The $\hat B\wedge F$ term, in turn,  
leads to the cross term of a St\"uckelberg Lagrangian  $\left(A_\mu +\partial_\mu a\right)^2$, with $a$ the axion component of $T_a$ (dual to the antisymmetric tensor  in the sense that  $\partial^\mu a=\epsilon^{\mu\nu\rho\sigma}\partial_\nu \hat B_{\rho\sigma}$),  
which implies that the K\"ahler potential $K$ depends on the combination $T+\overline T+q\hat V$. This K\"ahler potential gives rise to a (field-dependent) Fayet-Iliopoulos (FI)
D-term  proportional to $\partial K/\partial \hat V\big |_{\hat V=0}$. This can be seen as the modulus $T_a$ having a charge $q$ under the corresponding $U(1)$ symmetry.

In general, FI terms $\xi_{FI}$ contribute to a $U(1)$ D-term potential as:
\be\label{eq:dtermgen}
V_D\propto \left(\xi_{FI}-\sum_mq_mK_m \varphi_m\right)^2\,,
\ee
where $q_m$ are the charges of matter fields $\varphi_m$. Since $V_D$ is 
positive-definite, the D-term potential \eqref{eq:dtermgen} may uplift the F-term scalar potential unless the minimum of \eqref{eq:dtermgen} occurs at $V_D=0$.

Unfortunately, as can be seen from (\ref{dterm}), in supergravity the vanishing of the F-term ($F_m\propto D_m W$) implies the vanishing of the D-term. So the D-term uplift mechanism does not work for KKLT \cite{Choi:2005ge, Villadoro:2005yq}, since the AdS vacuum is supersymmetric.   This mechanism may work in LVS since in that case the AdS vacuum is not supersymmetric, and so the F-term is non-zero.
In explicit string realizations the D7-branes support not just gauge fields but also matter fields,\footnote{Chiral matter on D7-branes, for example from a realization of the MSSM, can interfere with moduli stabilization \cite{Blumenhagen:2007sm}.} naturally  leading to the T-brane mechanism that we discuss next (see also \cite{Parameswaran:2007kf,Braun:2015pza} for related approaches).

\item{\bf T-brane uplift}. 
If supersymmetry is broken, the dependence of the scalar potential on the matter fields $\varphi_m$ will include not only the D-term contribution $V_D$ but also the contribution from soft supersymmetry breaking terms, such as scalar masses:
\begin{equation}
    V_{\text{soft}}=\sum_n m_n^2 \left |\varphi_n \right |^2\,.
\end{equation}
Then minimizing the potential
$V_D+ V_{\text{soft}}$ with respect to $\varphi_n$ will lead to a positive contribution to the overall scalar potential, since both terms cannot vanish simultaneously.  The resulting positive energy can cause uplifting, with $V_{\text{up}}\sim |W_0|^2/\V^{8/3}$ \cite{Cicoli:2015ylx}. In this sense the uplifting comes from a combination of D-terms and F-terms.

This mechanism has an interesting string theory realization in terms of magnetic and three-form fluxes as well as T-branes \cite{Cicoli:2015ylx}. T-branes are D-brane configurations or bound states defined  by the fact that the Higgs fields $\Phi$ in the adjoint representation of the brane non-abelian gauge group, written in a matrix form, are not diagonalizable.  Thus $[\Phi,\Phi^\dagger]\neq 0$, and $\Phi$ can be written not in a diagonal but in a triangular form, hence the term T-brane \cite{Donagi:2003hh,Cecotti:2010bp,Donagi:2011jy}. 

The eight-dimensional equations of motion for an orientifolded  stack of $N$ D7-branes are:
\be\label{tbraneeq}
J\wedge \mathcal{F}-\left[\Phi,\Phi^\dagger\right]=0\,,
\ee
where $J$ is the pullback of the K\"ahler form, the two-form flux is  ${\mathcal{F}}=\left(2\pi \alpha' \right) F - \iota^*B$ with $\iota^*B$ the pullback of the NS-NS B-field, and $\Phi$ is the scalar field in the adjoint representation of the corresponding $U(N)$ gauge group. Equation (\ref{tbraneeq}) corresponds to the four-dimensional D-term equation with $\Phi$ the canonically-normalized field. 
If the flux $\mathcal F$ is non-primitive, i.e.~if $J \wedge \mathcal{F} \neq 0$, then $\left[\Phi,\Phi^\dagger\right]\neq 0$, and hence $\Phi$ cannot be a diagonal matrix. On the other hand, a triangular matrix satisfies  
this condition, thus defining the D-brane configuration as a T-brane. 

Furthermore the non-vanishing $(0,3)$ fluxes break supersymmetry, and the soft masses 
$m^2 |\Phi|^2$ can be computed explicitly from the D7-brane DBI action, leading to a stringy realization of the T-brane uplift scenario \cite{Cicoli:2015ylx}.

\item{\bf Complex structure moduli uplift.}
As we have seen, the complex structure moduli part of the scalar potential $V_{\text{cs}}$ is positive semi-definite. It is then naturally minimized at the supersymmetric point  
$D_{z_i}=0$ with $V_{\text{cs}}=0$. However, considering the full scalar potential and knowing that $V_{\text{cs}}$ is also a function of the K\"ahler moduli, and that the rest of the scalar potential is not positive, $V_{\text{cs}}$ can be considered as an uplift term \cite{Saltman:2004sn,Gallego:2017dvd,Krippendorf:2023idy}.

This can be seen clearly in LVS, where $V_{\text{cs}}\sim \O\left(1/\V^2\right)$ whereas $V_{\text{K}}\sim \O\left(1/\V^3\right)$. Therefore, instead of concluding that positivity implies $V_{\text{cs}}=0$, we should say that positivity implies 
$V_{\text{cs}}\sim \O\left(1/\V^3\right)$, and if $V_{\text{cs}}$ is positive it could play the role of an uplifting term.

This proposal requires a large amount of tuning to get the desired size for $V_{\text{cs}}$, but such tuning may be possible given the large number of fluxes. A concrete example using the continuous flux approximation was presented in 
\cite{Gallego:2017dvd}.

\item{\bf Non-perturbative dilaton superpotential uplift.}
In type IIB string theory, matter fields can be located either on D3-branes or D7-branes, which can host either the Standard Model or a hidden sector.  In both KKLT and LVS we have considered hidden sectors on D7-branes, for which the gauge coupling $g_a$ is determined by the volume $\text{Re}\,T_a$ of the divisor $D_a$ wrapped by the D7-branes,
and the non-perturbative superpotential terms $W_{\text{np}}$ are proportional to 
$e^{-cT_a}$. 
However, hidden sectors may also be located on D3-branes, in which case
the gauge coupling is determined by the axiodilaton $\tau$. 
The non-perturbative superpotential is then of the form:
\be \label{eq:d3wnp}
W_{\text{np}}= B(z_i,\tau)e^{-b(\tau-\rho)}
\ee
where $\rho$ is the blow-up of the singularity where D3-branes are located. After fixing $\rho$ at the singularity $\langle \rho \rangle=0$ from its D-term field equations and also using $\langle \tau \rangle$ from the flux superpotential,\footnote{Recall that the dilaton is fixed at leading order by the fluxes, as it appears in the tree-level flux superpotential.  Thus, we can replace $\tau$ by its vev at this order.} the term \eqref{eq:d3wnp} makes a contribution to the total F-term scalar potential of the form \cite{Cicoli:2015ylx}:
\be
V_{\text{up}}=C \frac{e^{-2b\langle \tau\rangle}}{\V}\geq 0
\ee
with $C$ a positive constant.
Adding the contribution of these superpotentials to the LVS potential provides a potential uplift term \cite{Cicoli:2015ylx}. This proposal is model-dependent and still lacks an explicit string example. It would be interesting to search for such a model since it works in a way very similar to the anti-D3-brane uplift, in the sense that the uplifting term can be easily tuned to match the $\O\left(\V^{-3}\right)$ of the LVS potential, with the non-perturbative term $ e^{-2b\langle \tau \rangle}$ playing the role of the warp factor.

\item{\bf Flux superpotential uplift.} 
Finally, structure in the flux superpotential itself can lead to spontaneous breaking of supersymmetry in the complex structure moduli and axiodilaton sector.  Paired with a non-perturbative superpotential depending on --- and serving to stabilize --- the K\"ahler moduli, as in the KKLT scenario, the result can be a metastable de Sitter minimum.  

Specifically, suppose that in compactification of type IIB string theory on an O3/O7 orientifold of a Calabi-Yau threefold $X_6$, three-form fluxes are chosen to produce a PFV, as in \eqref{eq:pfvgen}, and are particularly chosen so that the effective superpotential \eqref{eq:weffgv} for the axiodilaton coordinate along the PFV is a sum of $N$
dilogarithms,
\begin{equation}\label{eq:weffgvnow}
     W_{\text{eff}}(\tau) = -\frac{1}{2^{3/2}\pi^{5/2}}\biggl( 
     \sum_{I=1}^{N} \mathbf{M}{\cdot}\beta_I\,\text{GV}(\beta_I)\,\text{Li}_2\bigl(e^{2\pi i \tau \mathbf{p}\cdot \beta_I} \bigr) \biggr)+\ldots \,,
\end{equation} 
and the omitted terms are subleading.
Here $\mathbf{M}$ is a fixed set of fluxes, $\beta_I$ are effective curve classes of the mirror threefold $\widetilde{X}_6$, and $\mathbf{p}$ is a vector in the K\"ahler cone of $\widetilde{X}_6$.
For most choices of $X_6$, $\mathbf{M}$, and $\mathbf{p}$, the $N$ terms in \eqref{eq:weffgvnow} will not effectively compete to produce a minimum at finite $\tau$.  However, for $N>3$ the structure in \eqref{eq:weffgvnow} can be sufficient to produce a metastable de Sitter minimum \cite{deSitterpaper}.
This mechanism has not yet been realized in an explicit compactification.

\end{itemize}

The preceding sections illustrate some of the significant efforts 
that have been dedicated to achieving
moduli stabilization in anti-de Sitter and de Sitter vacua of type IIB string theory (other string theories will be discussed in \S\ref{eq:beyondiib}).
All of the components of the KKLT and LVS scenarios have undergone continuing testing in extremely detailed realizations. 
Critical scrutiny has been been directed at
anti-D3-brane supersymmetry breaking (\S\ref{sec:antibrane}),
the ten-dimensional description of non-perturbative effects (\S\ref{sec:10dkklt}), and corrections that may
ruin the validity of various approximations, among many other issues.  
There have even been speculations that string theory may have no de Sitter solutions \cite{Danielsson:2018ztv,Obied:2018sgi}. For a sample of recent work emphasizing and addressing  these challenges see \cite{Moritz:2017xto, Cicoli:2018kdo, Carta:2019rhx,Hamada:2018qef,Hamada:2019ack,Kachru:2019dvo,Kachru:2018aqn, Kallosh:2019axr, Hebecker:2022zme,Schreyer:2022len, Grana:2022nyp,Bena:2022ive,
Grana:2022dfw,
Bena:2021wyr,
Bena:2020xrh,
Grana:2020hyu,
Bena:2019mte,
Bena:2011wh,
Bena:2011hz,
Bena:2009xk, Crino:2020qwk,Crino:2022zjk,Junghans:2022kxg,Junghans:2022exo,Gao:2020xqh,Blaback:2014tfa,Junghans:2014wda,Junghans:2014xfa, 
Blumenhagen:2016bfp,
Lust:2022xoq,Lust:2022lfc,
ValeixoBento:2023nbv,Bento:2021nbb, Polchinski:2015bea, Cicoli:2023njy,Junghans:2023lpo}. 
The intensity and focus of the recent literature illustrates that the field is very much alive and active.

\subsection{Moduli stabilization and the Standard Model}

As the examples above illustrate, there are 
many approaches to the problems of stabilizing moduli and obtaining  metastable de Sitter vacua in type IIB string theory. 
However, finding a de Sitter vacuum is not the final goal: to describe quantum gravity in our Universe we need a de Sitter solution that also incorporates the Standard Model.  

The tasks of moduli stabilization and of embedding realistic particle physics in string theory are intertwined. 
The gauge and Yukawa couplings of the Standard Model particles, and the masses of their superparters, are free parameters in the EFT, and it is the dynamics of moduli stabilization that fixes their values.
Thus, moduli stabilization is a prerequisite for extracting meaningful four-dimensional particle physics from string compactifications.
At the same time, the presence of chiral matter can impact moduli stabilization \cite{Blumenhagen:2007sm}.

A full treatment of particle physics in string theory is beyond the scope of this work (see for instance \cite{Marchesano:2022qbx} for a recent review, and \cite{Halverson:2018vbo} for lectures on possible signatures).   However, we can mention several concrete examples of quasi-realistic Calabi-Yau orientifold models with stabilized moduli \cite{Cicoli:2011qg,Cicoli:2012vw,Cicoli:2013cha,Cicoli:2017shd,Cicoli:2021dhg}.  These models are examples of the {\it modular} or bottom-up approach to string model building \cite{Aldazabal:2000sa,Conlon:2008wa,Donagi:2008ca, Beasley:2008dc},
which separates local questions, such as the realization of the Standard Model on a set of D-branes, from global issues such as moduli stabilization and supersymmetry breaking. 
Each question can be addressed separately, and at the end the two can be put together into consistent Calabi-Yau compactifications. 

Even though considerable success has been achieved in this direction, it is fair to say that much more work will be required to establish
well-controlled 
de Sitter vacua 
in string theory or F-theory models with 
realistic particle physics.

\subsubsection{Soft supersymmetry breaking terms}
Supersymmetry breaking in the moduli sector naturally connects to the physics of the Standard Model. In type IIB string theory the Standard Model may arise on D3-branes at a singularity, or on D7-branes wrapping four-cycles (which can also  be generalized to F-theory constructions). Mediation of supersymmetry breaking 
to
the Standard Model sector is model-dependent.  However, because the moduli fields originate from the gravitational sector and so have gravitational-strength couplings, moduli-mediated supersymmetry breaking is a variant of 
gravity-mediated supersymmetry breaking.

The general structure of soft terms from moduli stabilization has been known for many years (see for instance  \cite{Kaplunovsky:1993rd,Brignole:1993dj}). Writing the gaugino masses as $M_{1/2}$, the scalar masses as $m_0$, and the trilinear terms as $A_{\alpha\beta\gamma}$ with indices $\alpha,\beta,\gamma$ running over the Standard Model matter fields, one finds
\begin{eqnarray}
M_{1/2}& = & \frac{1}{f+\bar{f}}\,F^I\partial_I f\,, \nonumber \\
m_\alpha^2& = & V_0 +m_{3/2}^2-F^I {\bar F}^{{\bar J}}\partial_I\partial_{{\bar J}}\log \hat K_\alpha\,,\\
A_{\alpha\beta\gamma} & = & F^IK_I+F^I\partial_I \log Y_{\alpha\beta\gamma}- F^I \partial_I \Bigl(\hat K_\alpha \hat K_\beta \hat K_\gamma\Bigr)\,.\nonumber
\end{eqnarray}
Here $Y_{\alpha\beta\gamma}$ are the Yukawa couplings of the matter fields; $f$ is the moduli-dependent gauge kinetic function; $F^I$ are the moduli F-terms, with the indices $I,J$ running  over the moduli; and $\hat K_i$ is the moduli-dependent K\"ahler potential for the matter fields $\varphi_\alpha$. The full K\"ahler potential $\mathcal K$  is the sum of the moduli K\"ahler potential $K$ and the leading-order 
matter K\"ahler potential $\hat K_\alpha |\varphi_\alpha|^2$:
\be
\mathcal{K}= K(\tau, T_a, z_i) +\hat K_\alpha |\varphi_\alpha|^2+\ldots \,.
\ee

The above quantities have been computed in various D-brane  configurations representing the Standard Model \cite{Choi:2005ge, Conlon:2005ki,Conlon:2006wz,Blumenhagen:2009gk,Aparicio:2014wxa,Aparicio:2015psl}. In the table below we summarize the order of magnitude of the soft terms for the KKLT and LVS models, taking the Standard Model on D3-branes or D7-branes, assuming anti-D3-brane uplift, and using the nilpotent superfield formalism to capture the breaking of supersymmetry \cite{Aparicio:2015psl} (for a recent discussion using dilaton superpotential uplifting see \cite{Cicoli:2023gbd}). Notice that the generic expectation that soft terms should be of order $m_{3/2}$ does not hold for string models in which there are suppressions either by $1/\text{log}(|W_0|^{-1})$ or by powers of $1/\V$.

In particular, scenarios such as split supersymmetry, in which the fermionic  superpartners (gauginos) are hierarchically lighter than the scalar superpartners (squarks and sleptons), can be obtained from string models. 
The Standard Model on D3-branes 
manifests 
a variant of 
sequestered supersymmetry breaking,  
in the sense that the source of supersymmetry breaking couples very weakly to the Standard Model fields and there are  usually first-order cancellations for the leading contributions to the soft terms. This is similar to the no-scale structure of the moduli  potential in that the non-trivial structure comes from higher-order corrections. However, the corrections to the matter K\"ahler potential are poorly understood.

\begingroup

\setlength{\tabcolsep}{10pt} 
\renewcommand{\arraystretch}{1.5}  
\begin{table}[h!]
\begin{center}
\centering
\begin{tabular}{ | c || c | c || c | c |}
\hline
 \rowcolor{lightgray!}  {\bf Soft term} &   \multicolumn{2}{c||}{\bf KKLT} &   \multicolumn{2}{c|}{\bf LVS} \\
 \hline
 \hline
 & D3 & D7 & D3 & D7 \\
 \hline \hline
 $M_{1/2}$ & $\frac{m_{3/2}}{\left | \log |W_0| \right |} $ & $\frac{m_{3/2}}{\left | \log |W_0| \right |} $ & $\frac{m_{3/2}}{g_s^3\V}$ & $\frac{m_{3/2}}{c\tau_a}$ \\  \hline
 $m_\alpha^2$ & $\frac{m_{3/2}^2}{g_s^3\V}$ & $m_{3/2}^2$ & $\frac{m_{3/2}^2}{g_s^3\V}$ & $\frac{m_{3/2}^2}{\left(c\tau_a\right)^2}$ \\  \hline
 $A_{\alpha\beta\gamma}$ & $M_{1/2}$ & $M_{1/2}$ & $M_{1/2}$ & $M_{1/2}$ \\  \hline
\end{tabular}
\end{center} 
\caption {Structure of soft supersymmetry breaking terms (in orders of magnitude) for both KKLT  and LVS.  The D3 and D7 column labels indicate that the  Standard Model is hosted either on D3-branes at a singularity or on D7-branes wrapping a four-cycle with size $\tau_a$.} 
\label{tab:soft}
\end{table}
\endgroup

In general the soft terms depend on the location of the Standard Model (D3-brane or D7-branes), on the de Sitter uplifting mechanism, and on corrections to the matter K\"ahler potential. For instance, in LVS with the Standard Model on D3-branes at singularities, 
the masses in Planck units are are follows: the gravitino mass is
$m_{3/2}\sim \O(1/\V)$; the gaugino masses are further suppressed, $M_{1/2}\sim \O(1/\V^{2})$; and scalar masses can be
$m_0\sim \O(1/\V^{3/2})$ or $m_0\sim \O(1/\V^{2})$. In the first case it is clear that $m_{3/2}\gg m_0 \gg M_{1/2}$, which is a typical case of split supersymmetry. The second case would then correspond to high scale supersymmetry breaking.

The hierarchies can be reduced by several sources of de-sequestering \cite{Berg:2010ha, Conlon:2011jq, Berg:2012aq}. De-sequestering can occur if flavor D7-branes intersect the D3-brane singularity, or if the singularity is orientifolded (and not just orbifolded).\footnote{In this case it has been argued that moduli redefinitions are needed at one loop that would cause de-sequestering of soft terms\cite{Conlon:2010ji}.} Furthermore, superpotential de-sequestering can be caused by couplings of the form \cite{Berg:2010ha}
\begin{equation}\label{eq:wdeseq}
    \Delta W = \mathcal{O}_{\text{vis}}\,e^{-2\pi T}\,,
\end{equation} where $\mathcal{O}_{\text{vis}}$ is a gauge-invariant chiral operator constructed from visible sector chiral superfields, and $T$ is a K\"ahler modulus.  The coupling \eqref{eq:wdeseq} can be viewed as a result of closed string exchange: thus, de-sequestering can occur even if the visible sector D-branes and the four-cycle supporting Euclidean D3-branes do not intersect.
See  \cite{Cicoli:2023gbd} for an up-to-date discussion of sequestering in LVS.

\section{Beyond IIB}\label{eq:beyondiib}

There is no fundamental reason to restrict to compactifications of type IIB string theory as an arena for studying
moduli stabilization, and as we will soon see, there are very interesting results in other string theories and in M-theory.
However, some practical and technical considerations are responsible for the fact that type IIB string is over-represented in the last twenty years of work on moduli stabilization.  Let us briefly note these causes. 

The first reason is the fact that the leading backreaction of the fluxes on the metric is only a scaling of the original Calabi-Yau metric: one can find conformally Calabi-Yau solutions \eqref{eq:warpedproductansatz} in type IIB flux compactifications \cite{Giddings:2001yu}.  This allows one to extend the use of Calabi-Yau geometry from $\mathcal{N}=2$ to $\mathcal{N}=1$ solutions.

A second reason is that arguably the most robust mechanism for breaking supersymmetry at a parametrically low scale \cite{Kachru:2002gs} involves an exponentially warped throat region that caps off smoothly in the infrared: this is in contrast to solutions with singularities, in which the supersymmetry breaking is incalculable.
At the time of writing, the Klebanov-Strassler solution \cite{Klebanov:2000hb} of type IIB supergravity (and close relatives involving orbifolding or deforming it) is the only known solution with these properties.  Changing this state of affairs would be an important advance.

A third reason is that in some of the other corners of the duality web, constructing solutions with $\mathcal{N}=1$ supersymmetry involves 
significantly greater mathematical challenges.
For example, M-theory compactifications on $G_2$ manifolds require working in real differential geometry, while for F-theory compactifications on Calabi-Yau fourfolds the K\"ahler potential is poorly understood.

These observations should be understood as explanations for the state of the literature, or indeed as calls to arms, not as no-go results.  With this understanding, let us proceed to briefly survey developments in other string theories.

\subsection{Type IIA flux compactifications}

In type IIB string theory, non-perturbative effects have proved necessary to stabilize all the moduli,\footnote{As explained in \S\ref{ss:lvs}, cf.~\eqref{eq:thisispertk}, purely perturbative stabilization is a logical possibility in type IIB flux compactifications, but has not yet been achieved in a controlled example.} but in type IIA string theory,  
all moduli can be stabilized with fluxes alone. 
The best-understood construction is due to DeWolfe, Giryavets, Kachru, and Taylor (DGKT) \cite{DeWolfe:2005uu}. Let us briefly review the main points of this scenario.

The starting point is the bosonic ten-dimensional action for \emph{massive} type IIA supergravity with massless Neveu-Schwarz-Neveu-Schwarz bosonic fields including  the metric 
 $g_{MN}$, the dilaton $\phi$ and the antisymmetric tensor $ B_{MN}$,  with field strength $H_3= \d B_2$, and Ramond-Ramond fields $C_1, C_3$ with field strengths $F_2, F_4$, as well as the Romans mass term $F_0$. 

Compactifying massive type IIA supergravity on the orientifold $T^6/\mathbb Z_3^2$ with orientifold action $\O=\Omega_{\text{ws}} (-1)^{F_L}\sigma$, with $\Omega_{\text{ws}}$ worldsheet parity, $(-1)^{F_L}$ left-moving fermion number, and $\sigma$ an involution acting on the three complex coordinates $x_i$ as $\sigma: x_i\rightarrow - \bar x_i$, one finds 
a four-dimensional model with $\mathcal{N}=1$ supersymmetry and an O6-plane filling the four-dimensional spacetime and wrapping a three-cycle of $T^6$. 
The geometric moduli of this toroidal orientifold
are the three untwisted moduli and nine blow-up modes.

One can include fluxes $F_0, F_2, F_4$ satisfying the quantization condition:
\be
\int F_q=(2\pi)^{q-1}\alpha'^{(q-1)/2} f_q, \qquad f_q \in \mathbb Z\,\quad q=0,2,4\,,
\ee
with a similar expression for $H_3$ in terms of integers $h_3$. 
To simplify matters we set $f_2=0$, since $f_2$ is not essential for moduli stabilization, and we define the parameters:
\be
m_0=\frac{f_0}{2\pi \sqrt{2\alpha'}}\,, \qquad p= (2\pi)^2 \alpha' h_3\,, \qquad e^i=\frac{\kappa^{1/3}}{\sqrt{2}}\left(2\pi \sqrt{\alpha'} \right)^3 f_4^i\,,
\ee
with the index $i=1,2,3$  labelling the different four-cycles of $T^6$. Here $\kappa$ stands for the normalization of the triple intersection number  $\kappa=\int \omega_1\wedge \omega_2 \wedge \omega_3$, with $\omega_1,\omega_2,\omega_3$ a basis of the $T^6$ two-forms. 

The presence  of the O6-plane gives rise to a tadpole that can be cancelled by $H_3$ and $F_0$ fluxes satisfying
\be
m_0p = -2\pi \sqrt{2\alpha'}\,,
\ee
with no conditions on the $F_4$ fluxes $f_4^i$.

The four-dimensional fields include
the dilaton $\phi$ and the untwisted moduli $T_i$, which we write as
\be
T_i=b_i+ i t_i, \qquad \tau= \xi-i e^{-D}\,,
\ee
with  the volume and four-dimensional dilaton given by
\be
\V=\kappa_{ijk}t_it_jt_k=:\kappa t_1 t_2 t_3, \qquad e^D=\frac{e^\phi}{\sqrt{\V}}\,,
\ee
and with axionic fields $C_3=\sqrt{2}\xi\,\text{Re}\,\Omega$ and $B_2= b_i\omega_i$.

Incorporating 
the fluxes 
in the ten-dimensional action, one finds a nontrivial scalar potential.
The axions are minimized at 
\begin{equation}
     b_i=0\,, \qquad \xi=e_0/p\,, \qquad \text{with}~~e_0=\int F_6\,.
\end{equation}
The scalar potential for the dilaton and the untwisted moduli is then
\be
V=\frac{p^2e^{2\phi}}{4\V^2}+\frac12 \left(\sum_{i=1}^3 e_i^2 t_i^2\right) \frac{e^{4\phi}}{2\V}+\frac{m_0^2 e^{4\phi}}{2\V}-\sqrt{2} \left | m_0 p\right| \frac{e^{3\phi}}{\V^{3/2}}\,.
\ee
The four terms come from $\left | H_3 \right |^2$, $\left | F_4 \right |^2$, $\left | F_0 \right |^2$, and the orientifold, respectively. The potential is simple enough to allow an analytic solution for the minimum:
\be\label{solutions}
t_i=\frac{1}{\left |e_i \right |} \sqrt{\frac{5}{3}\left |\frac{e_1e_2e_3}{\kappa m_0}\right |}\,, \qquad e^D= |p|\sqrt{\frac{27}{160}\left |\frac{\kappa m_0}{e_1e_2e_3}\right |}\,.
\ee
The minimum can also be found using the $\mathcal{N}=1$ effective field theory, with the K\"ahler potential
\be\label{eq:dgktk}
K=-\log \V- 4\log \Bigl(i(\tau-\bar \tau)\Bigr)\,.
\ee
The flux superpotential is: 
\begin{eqnarray}\label{eq:dgktw}
W&=& -p \tau + e_0 + \int J_c\wedge F_4-\frac12 \int J_c\wedge J_c\wedge F_2-\frac{F_0}{6}\int J_c\wedge J_c \wedge J_c\,, \nonumber \\ 
&=& -p\tau + e_0+e_i T^i+\frac12 \kappa_{ijk} m^i T^j T^k-\frac{m_0}{6}\kappa_{ijk} T^iT^j T^k\,,
\end{eqnarray}
where $J_c$ is the complexified K\"ahler form $J_c=B_2+i J= \sum_{i=1}^{h^{1,1}} T_i \omega_i$. The parameters $m^i$ are proportional to the $F_2$ fluxes, which we have set to zero. The blow-up modes can be incorporated by writing the volume as
\begin{equation}\label{eq:dgch}
\V=\kappa t_1t_2t_3+\beta\sum_{A=4}^{12} t_A^3\,. 
\end{equation} 
Using \eqref{eq:dgktk} and \eqref{eq:dgktw} in \eqref{eq:vkw}, one recovers (\ref{solutions}). The minimum for the blow-up modes can be computed giving:
\be
t_A=-\sqrt{\frac{-10f_A}{3\beta m_0}}
\ee
where $f_A$ is the $F_4$ flux on the corresponding blown-up four-cycle. Provided that
$f_A\ll e_i$, the blow-up modes are small enough to justify the toroidal orientifold 
approximation, in the sense that expansion around the singularity is justified, with the blow-up modes hierarchically smaller than the string scale. 
The negative signs of the $t_A$ solutions are determined by the K\"ahler cone conditions. Comparing \eqref{eq:dgch} and \eqref{eq:scf}, we see that the geometry has a Swiss cheese form.

Let us summarize key features of the DGKT model:
\begin{itemize}
    \item The value of the potential at the minimum is negative,
   
    \be
V=- 2 \sqrt{\left| \frac{ m_0 e_1 e_2 e_3}{15 \kappa } \right|}~ e^{4D}
    \ee
    so the solution is an AdS vacuum.
    \item Contrary to the IIB case, the fluxes $e_i$ are not bounded by tadpole conditions, and so the number of solutions is infinite.
    \item The solutions can be supersymmetric or non-supersymmetric \cite{DeWolfe:2005uu,Marchesano:2019hfb}. 
    \item Scaling the fluxes $e_i\rightarrow \lambda e_i$ yields $\V\rightarrow \lambda^{3/2}\V$, $e^\phi \rightarrow \lambda^{-3/4} e^\phi$ and $V\rightarrow \lambda^{-9/2}V$, which shows that there is a hierarchy of scales: the larger the fluxes, the larger the volume and the smaller the string coupling.
\end{itemize}
In the DGKT solution the effects of the orientifold planes are accounted for in the smeared approximation.
The corresponding localized solutions were studied in  \cite{Junghans:2020acz,Marchesano:2020qvg}, to first order in an expansion around the large flux limit.

The potential CFT dual of the DGKT model has an interesting property: the operators dual to some of the stabilized moduli have integer conformal dimensions \cite{Apers:2022tfm,Apers:2022zjx,Apers:2022vfp,Quirant:2022fpn} 
(see also the recent extension \cite{Andriot:2023fss}, as well as \cite{Plauschinn:2022ztd} for similar considerations in the type IIB mirror).

Moving beyond the AdS$_4$ vacua of DGKT, 
a proposal for classical de Sitter vacua of 
massive\footnote{Massless type IIA supergravity has proved to be a comparatively difficult setting for constructing stabilized vacua.  However, see \cite{Cribiori:2021djm}, which presented scale-separated vacua of massless type IIA, with controllably small backreaction of the orientifold planes.} type IIA supergravity, in compactifications on a circle times a negatively curved space, was presented in \cite{Cordova:2018dbb} (see also \cite{Cordova:2019cvf}).  These configurations contain localized and backreacted O8-planes, and corrections from string theory are large near the singular sources.\footnote{It was argued in
\cite{Cribiori:2019clo} that the solutions of \cite{Cordova:2018dbb} are incompatible with the
integrated supergravity equations of motion.}

\subsection{Heterotic strings}
The first attempts at moduli stabilization were within the context of the heterotic string. 
Starting with  the  $E_8\times E_8$ string with the Standard Model group  inside one of the $E_8$ factors, the second $E_8$ provides a hidden sector that could lead to a gaugino condensate  superpotential.
In the  heterotic string the tree-level  gauge kinetic function is  $f=S$,
where $S$ is the heterotic dilaton field, related to the string coupling  by $g_s^{-2}=\langle \hbox{Re}S \rangle$.  If the condensing gauge group in the hidden sector is $G \subseteq E_8$, the gaugino condensate superpotential is
\begin{equation}\label{eq:hetcond}
    W_{\lambda\lambda} =\mathcal{A}\,e^{-\frac{8\pi^2}{c(G)} S}\,.
\end{equation}
In the presence of $H_3$ flux, one then has \cite{Derendinger:1985kk, Dine:1985rz}:
\be \label{eq:hetwprob}
W=\int H \wedge \Omega + \mathcal{A}\,e^{-\frac{8\pi^2}{c(G)} S}\,.
\ee
This can be combined with the standard K\"ahler potential\footnote{Fluxes and  non-perturbative  effects were combined with $\alpha'$ corrections to the K\"ahler potential to find a version of LVS in the heterotic string in \cite{Cicoli:2013rwa}.} for $S$ and the overall K\"ahler modulus field $T$,
\be
K=-3\log\left(T+\overline T \right) - \log \left( S+\overline S \right)\,,
\ee
to determine the scalar potential for $S$ and $T$. However, quantization of $H_3$ fluxes appears to only allow a minimum at strong coupling, beyond the domain of validity of the EFT \cite{Dine:1985rz}. Furthermore, the modulus $T$ remains unfixed.

Three proposals were made to address these problems. First, assuming that the hidden gauge group is a product of at least two gauge groups that allow gaugino condensation, the superpotential for $S$ could be  of the racetrack form \cite{Krasnikov:1987jj, Dixon:1990ds, Casas:1990qi} 
\begin{equation}
W(S)=\sum_i \mathcal{A}_i\,e^{-c_iS}\,,
\end{equation}
where the exponentials compete with each other to give rise to a weak coupling minimum with $\langle S \rangle \gg 1$.

The second idea was to invoke $T$-duality of the EFT, which for a single complex field $T$ is the $SL(2,\mathbb Z)$ action on $T$ \cite{Ferrara:1989qb, Font:1990nt,Ferrara:1990ei}, with $S$ being invariant. Imposing that the K\"ahler-invariant combination $K+\log|W|^2$ is invariant then implies that  the   non-perturbative superpotential should transform as a modular form of weight $3$ \cite{Font:1990nt,Ferrara:1990ei}:
\be
W(S,T)=\frac{F(j(T), S)}{\eta(T)^6}\,, \qquad W\rightarrow (icT+d)^{-3}W\,, \qquad T\rightarrow \frac{aT-ib}{icT+d}\,, \quad ad-bc=1\,,
\ee
where $\eta(T)$ is the Dedekind $\eta$ function and $j(T)$ is the absolute modular invariant function. Concrete calculations of the one-loop threshold  corrections to the
holomorphic gauge kinetic function give  $f=S+\alpha\log\eta(T)$ \cite{Dixon:1990pc}. This leads to moduli stabilization of the $T$ field at $\langle T\rangle=1.2$ \cite{Font:1990nt} in string units (see also \cite{Ferrara:1990ei}), with a negative cosmological constant, also of string scale, and with broken supersymmetry.

Generalizations to functions $F\bigl(j(T), S\bigr)=\Omega(S) G_4^m G_6^n P(j)/\eta^{8m+12n}$, where $G_4, G_6$ are the holomorphic Eisenstein functions of  weight $4$ and $6$ respectively, and $P(j(T)$ is a polynomial of $j(T)$, were studied in \cite{Cvetic:1991qm}, and found to generate potentials with  perturbatively stable AdS vacua. Most of these configurations have minima at the fixed points $T=1$ and 
$T=e^{i\pi/6}$, which are always extrema and automatically preserve supersymmetry, but there exist other configurations in which supersymmetry is broken \cite{Cvetic:1991qm}. Recent generalizations to include concrete toroidal orbifold models \cite{Parameswaran:2010ec} and potential de Sitter minima have been studied in \cite{Leedom:2022zdm,Knapp-Perez:2023nty}.

In Calabi-Yau  
compactifications of the heterotic string, $H_3$ flux
can at most fix the complex structure moduli.
This limitation has motivated studying the heterotic string compactified on more general spaces, including
$SU(3)$ structure and half-flat manifolds \cite{deCarlos:2005kh, Anderson:2011cza,Klaput:2012vv}. Furthermore, in addition to the geometric moduli, 
heterotic  compactifications have
vector bundle moduli (see e.g.~\cite{Buchbinder:2013dna}) 
which are singlets of the corresponding gauge group for $(0,2)$ or non-standard-embedding compactifications. Stabilizing them is also a challenge, similar to that of open string moduli in type II strings. 
A difference with type II models is the fact that in heterotic models the gauge sector is in the bulk, and so there is no route to a modular approach that separates the task of moduli stabilization from that of constructing the Standard Model.  In the heterotic string the two challenges need to be addressed simultaneously.

A third approach \cite{Gukov:2003cy} began by revisiting the apparent obstacle to using \eqref{eq:hetwprob} to stabilize at weak coupling, which is that for an integrally-quantized three-form $dB \in H^3(X,\mathbb{Z})$, the superpotential term $\int \d B \wedge \Omega$ cannot be small.
However, in the heterotic string the three-form flux is given by 
\begin{equation}
H = \d B + \frac{\alpha'}{4}\Omega_3(A)-\frac{\alpha'}{4}\Omega_3(\omega)\,, 
\end{equation}
where for a gauge connection $A$ we define
the Chern-Simons three-form $\Omega_3(A)$ by
\begin{equation}
\Omega_3(A):= A \wedge \d A + \frac{2}{3} A \wedge A \wedge A\,.   
\end{equation}
Similarly, $\Omega_3(\omega)$  is the Chern-Simons three-form built from the spin connection $\omega$.
Integrating over a three-cycle $Q \in H_3(X,\mathbb{Z})$, we define the Chern-Simons invariant
\begin{equation}
\text{CS}(A,Q):=\int_Q     \Omega_3(A)\,.
\end{equation}
In general, $\text{CS}(A,Q)$ is a rational number, not necessarily an integer.
The proposal of 
\cite{Gukov:2003cy}
was to consider $X_6$ with $Q \subset X_6$ supporting fractional Chern-Simons invariants, take $\d B=0$, and embed the spin connection in the gauge connection, in such a way that the total flux superpotential is small.
Writing $\check{\Omega}$ for the three-cycle Poincar\'{e} dual to $\Omega$, we have, for $\d B=\Omega_3(\omega)=0$,
\begin{equation}\label{eq:wtotcs}
\int H \wedge \Omega = -\frac{\alpha'}{4} \int_{\check{\Omega}} \Omega_3(A) = -\frac{\alpha'}{4}\,\text{CS}(A,\check{\Omega})\,,
\end{equation} and the object on the right is not restricted by integer quantization.

Examples were given in \cite{Gukov:2003cy} in which $A$ is a connection on a flat bundle with nontrivial holonomies, i.e.~where $A$ is characterized by its Wilson lines, and $\int H \wedge \Omega$ is smaller than integral quantization would have required.
Stabilization of all moduli was studied in 
\cite{Cicoli:2013rwa}, while $\text{CS}(A,Q)$ was very carefully computed in a number of explicit examples, some with flat connections (Wilson lines) and others with curvature, in \cite{Apruzzi:2014dza,Anderson:2020ebu}.
These constructions illustrate the inevitable connection between the visible sector and moduli stabilization in the heterotic string: some of the choices of visible sector Wilson lines proposed in the literature to break the visible $E_8$ to the Standard Model actually introduce nonvanishing fractional Chern-Simons invariants, and so break supersymmetry via \eqref{eq:wtotcs} \cite{Gukov:2003cy}.

\subsection{Type I}
Even though type I theories can be thought of as orientifolded versions of type II, there are mechanisms that are most readily developed in the type I context. Fluxes of three-forms combined with magnetic fluxes give a rich range of possibilities for moduli stabilization. Concrete models in terms of magnetized 
D9-branes with toroidal orbifold compactifications have been studied in \cite{Antoniadis:2004pp,Antoniadis:2005nu}; see also \cite{Bianchi:2007fx}.

\subsection{M-theory}\label{ss:mth}

Four-dimensional $\mathcal{N}=1$ vacua result from
compactifications of eleven-dimensional supergravity, which is the low-energy limit of M-theory, on seven-manifolds of $G_2$ holonomy \cite{Duff:2002rw}.

Suppose that $X_7$ is a compact manifold of $G_2$ holonomy, with $N=b_3(X_7)$, and let $\{\Sigma_i\}$, $i=1,\ldots,N$ be a basis of independent three-cycles.
Defining $\ell_M^9 = 4\pi\kappa_{11}^2$ and introducing the $G_2$-invariant three-form $\Phi$, we can write the geometric moduli as \cite{Acharya:2005ez}
\begin{equation}
    z_i \equiv t_i + i s_i := \int_{\Sigma_i} C_3 + i \int_{\Sigma_i} \Phi\,.
\end{equation} 
Then the volume $\V_{7} = {\text{Vol}}(X_7)/\ell_M^7$ is a homogeneous function of the $s_i$, of degree $7/3$ \cite{Beasley:2002db}, and the K\"ahler potential is
\be
K=-3\ln\Bigl(4\pi^{1/3}\,\V_7(s_1,\ldots s_N)\Bigr)\,.
\ee
The $G_4$ flux of M-theory gives rise to a classical superpotential \cite{Gukov:1999gr,Beasley:2002db} --- for statistics of the resulting flux vacua, see \cite{Acharya:2005ez}.
Additional nonperturbative contributions to the superpotential arise from gaugino condensation in gauge sectors supported on singularities in $X_7$, as well as from Euclidean M2-branes wrapping three-cycles calibrated by $\Phi$ (see e.g.~\cite{Braun:2018fdp}).
 
The phenomenology of $G_2$ compactifications was explored in \cite{Acharya:2007rc,Acharya:2012tw}, in a class of scenarios in which the gaugino condensate superpotential alone, with vanishing fluxes, was argued to stabilize all moduli.

Overall, despite much study of the geometry and physics of manifolds of $G_2$ holonomy (see e.g.~\cite{Joycebook,Atiyah:2001qf,Acharya:2001gy,Halverson:2014tya,Halverson:2015vta,Braun:2016igl} and references therein), the details of moduli stabilization  are not yet as fully developed, especially in the more realistic setups with singularities, as in the string theory case.

\subsection{General mechanisms}\label{sec:genmech}

We have now seen many proposals for moduli stabilization from different corners of the string/M-theory moduli space, some of them with the potential to give rise to de Sitter solutions.
Even though the implementations differ, the core structures are similar,
and the strategy is always to incorporate a sufficiently rich collection of sources that are present in string theory.

We have extensively discussed the role of fluxes, D-branes, anti-D-branes, orientifolds, and quantum effects.
Let us now consider a more complete list of possible sources:
following the discussion in \cite{Silverstein:2004id}, we classify the various sources of non-derivative terms in the four-dimensional action from the higher-dimensional theory, without assuming that the starting point is supersymmetric or that the compactification is Ricci-flat.
\begin{itemize}
\item {\bf Non-critical strings}. Strings in dimensions above the critical dimension provide a positive contribution to the vacuum energy \cite{Silverstein:2004id},
\be
V_{\text{nc}}\propto \frac{\left ( D-D_{\text{crit}}\right ) e^{2\phi}}{\V}\,,
\ee
and can give rise to anti-de Sitter and de Sitter vacua in four dimensions \cite{Silverstein:2001xn,Maloney:2002rr}; see \cite{Junghans:2023lpo} for a recent critical assessment of this mechanism.

\item{\bf Curvature of compact space}. In the preceding sections we largely restricted to compactifications admitting Ricci-flat metrics, for which the contribution of the internal space to the Einstein-Hilbert action vanishes.  In a more general compactification on a compact six-manifold $X_6$ with curvature $\mathcal{R}_6$, one has \cite{Silverstein:2004id}
\be
V_{\text{curv}}\propto -\frac{e^{2\phi}}{\V^2} \int_{X_6} \mathcal{R}_6 \,.
\ee
Thus, the curvature contribution to the four-dimensional potential is \emph{negative} if $X_6$ is a positively-curved space, such as a sphere, and is positive if $X_6$ is a negatively-curved space, such 
a product of Riemann surfaces of genus $g>1$. Hyperbolic compactifications have been explored as a means of obtaining de Sitter solutions from eleven dimensions in \cite{DeLuca:2021pej} (for previous work in this direction see \cite{Silverstein:2007ac}).  An important lesson from the two-dimensional case is that negatively-curved manifolds are much more numerous than positively-curved manifolds.

\item{\bf Fluxes}. As we have seen, fluxes of antisymmetric tensor fields make positive contributions to the scalar potential.
The positivity of the flux contribution is a direct consequence of the positivity of the corresponding kinetic terms in the higher-dimensional theory.

\item{\bf Branes}. In principle branes contribute  positively to the scalar potential, but due to their BPS nature this contribution often cancels against other sources in a full solution.  Antibranes, as well as any non-BPS branes, contribute positively to the scalar potential.

\item{\bf Orientifolds}. Orientifolds contribute a similar amount as D-branes, but with a negative sign if they have negative tension.

\item{\bf Perturbative corrections}. We have seen that such corrections contribute with different powers of $e^\phi$ and $1/\V$, and the sign can be positive or negative.

\item{\bf Non-perturbative effects}. Non-perturbative effects contribute terms of order $e^{-c\rho}$ where $\rho$ is the modulus (K\"ahler or dilaton) measuring the gauge coupling. Again either sign is possible.
\end{itemize}

It is important to emphasize that the effects enumerated above exist, and it is not justified to ignore them. 
Combining some or all of these effects can naturally give rise to vacua for the moduli fields at the level of the EFT analysis.

However, the devil is in the details: finding minima in the
regime of validity of the EFT is highly nontrivial.
There is an enormous space of possible models, very little of which has been populated to date with full-fledged and explicit constructions.
The cause of this state of affairs is not that no such constructions can exist, but just that quantum gravity is relatively difficult, and writing down a mathematical model of a realistic universe takes a certain amount of work.  Moreover, although the concrete realizations produced to date may look baroque, the general principles are clear and clean.

\section{Cosmology}

Arguably the most important application of moduli stabilization is to the study of cosmology in string theory.
The dark universe involves three main unknowns: dark matter and dark energy today, and the inflationary energy at very early times.  Moduli are natural candidates for all three unknowns.
Moreover, the dynamics of moduli fields encodes the evolution of the extra dimensions of string theory, and can impact early stages of cosmic history.

Inflation, a transient period of accelerated expansion, is the leading scenario to describe early-universe cosmology.  Inflationary theory has impressive achievements, including solving the horizon and flatness problems of the Friedmann-Lema\^{\i}tre-Robertson-Walker (FLRW) cosmology and, more importantly, explaining the observed pattern of density perturbations imprinted in the cosmic microwave background (CMB). However, inflation itself does not have an explanation: it is a framework in search of a theory. String theory may be that theory. 

Having a period of accelerated expansion tends to dilute whatever physics existed at high energies, so one could imagine that 
string-theoretical signatures might be hidden from potential observations. Fortunately, the moduli tend to be light enough to survive after inflation, and may play a relevant role after inflation: not only in the early universe, between inflation and nucleosynthesis, but also at late times in the form of dark energy. This motivates the study of cosmology within moduli stabilization.\footnote{String theory may have cosmological signatures beyond inflation (see for instance the recent review \cite{Cicoli:2023opf} and references therein). However, the implementation of alternatives to inflation is challenging: EFT descriptions are often insufficient, and the correspondence with observations is less developed than in the case of inflation.
In this sense inflation appears to be preferred both by experimental evidence and by theoretical considerations.  Even so, it is prudent to explore and develop alternatives.}

\subsection{Inflation}
  
Inflation was historically understood as a mechanism, and a phenomenon, in quantum field theory coupled to general relativity.
If one temporarily and counterfactually disregards all effects of the quantization of gravity, inflation is a striking success: its predictive and explanatory power far exceeds the new questions that it raises.  In particular, inflation predicted that the universe should be spatially flat to good approximation, and it predicted that the CMB should have fluctuations that are nearly, but not exactly, scale-invariant; nearly Gaussian; and correlated on superhorizon scales.  All these predictions were decisively confirmed at the start of the era of precision cosmology. 

Inflation led to new questions: what is the inflaton field? What dynamics led to the initial conditions for inflation?  
String theory has the potential to address these questions: in particular, it is replete with candidates for the inflaton field, in the form of the many axions and moduli found in typical compactifications.
At the same time, string theory sharpens one of the central problems of inflationary cosmology: why is the inflaton potential flat enough to support prolonged accelerated expansion?

To see this, consider 
a single real scalar
field $\phi$, a candidate inflaton, coupled to general relativity, whose Lagrangian density at tree level is
\begin{equation}\label{eq:ltreeint}
\mathcal{L} = \frac{M_{\text{pl}}^2}{2}\,\mathcal{R} -\frac{1}{2}\left(\partial\phi\right)^2 - V_0(\phi)\,,
\end{equation}
where $V_0(\phi)$ is the tree-level (i.e., classical) scalar potential.  Including terms with at most two derivatives, the quantum-corrected Lagrangian density --- incorporating the effects of loops of the light fields, which are the inflaton and the graviton --- takes the form
\begin{equation}\label{eq:leffint}
\mathcal{L} = \frac{M_{\text{pl}}^2}{2}\,\mathcal{R} -\frac{1}{2} f(\phi) \left(\partial\phi\right)^2 - V(\phi)\,,
\end{equation}
where $f(\phi)$ and $V(\phi)$ are two functions of $\phi$, and we have used the freedom to rescale to Einstein frame to absorb a possible additional function of $\phi$ from the Einstein-Hilbert term.  The effect of $f(\phi)$ is generally small (see \cite{Baumann:2014nda} for a detailed discussion), and so we will set $f(\phi) \to 1$ henceforth.  

A sufficient condition for the theory \eqref{eq:leffint} to support a prolonged period of inflation is that the slope and curvature of the potential are small compared to the Plack mass: given suitable initial conditions, the potential energy can then dominate over the kinetic energy, with $p \approx -\rho$, and can do so for many $e$-folds of expansion.  
Specifically, \emph{slow-roll inflation} is possible if
\begin{equation}\label{eq:srcond}
\varepsilon := \frac{M_{\text{pl}}^2}{2}\biggl(\frac{V'}{V}\biggr)^2 \ll 1 \qquad \text{and} \qquad \eta :=  M_{\text{pl}}^2 \frac{V''}{V} \ll 1\,,
\end{equation} with primes denoting derivatives $\frac{d}{d\phi}$.

In general there is no systematic relation between $V_0(\phi)$ and $V(\phi)$, so having the classical potential $V_0(\phi)$ fulfill the slow-roll conditions \eqref{eq:srcond} does not imply that the full potential $V(\phi)$ will do so.  In other words, quantum corrections to the potential can affect whether inflation is possible.

In the important case that $V_0(\phi)$ preserves an approximate shift symmetry, so will $V(\phi)$.  In particular, if 
\begin{equation}
    V_0(\phi) = \frac{1}{2} m_0^2 \phi^2\,,
\end{equation} with $m_0 \ll M_{\text{pl}}$, then one finds
\begin{equation} \label{eq:vintcorr}
    V(\phi) = V_0(\phi) +\mathcal{O}\bigl(V_0/M_{\text{pl}}^4\bigr) \approx V_0(\phi)\,.
\end{equation}  
Thus, a small inflaton mass is radiatively stable: its smallness is not spoiled by loops of the light fields.  

An argument that was historically influential, but leads to erroneous conclusions, states that the correction term $\mathcal{O}\bigl(V_0/M_{\text{pl}}^4\bigr)$ in \eqref{eq:vintcorr} captures the leading quantum gravity corrections to the effective theory \eqref{eq:ltreeint}, and so inflation with the potential $\frac{1}{2}m_0^2\phi^2$ is not impacted by the quantization of gravity.
The logical flaw is the following: \emph{if} \eqref{eq:ltreeint} were the classical theory for $\phi$ and the massless graviton, obtained after integrating out all the massive degrees of freedom of quantum gravity, then indeed the remaining corrections from loops of the light fields would be negligible.  But deep principles of quantum gravity appear to forbid exact shift symmetries,\footnote{Some of the sharpest arguments against exact global symmetries in quantum gravity were developed in the context of the Weak Gravity Conjecture \cite{Arkani-Hamed:2006emk}; see the reviews \cite{Palti:2019pca,vanBeest:2021lhn,Agmon:2022thq,Harlow:2022ich}.} such as would occur with $m_0 \to 0$, and so the question of how small $m_0$ can be is  already a quantum gravity question: it can only be resolved with knowledge of the ultraviolet completion of gravity.

The terminology is one cause of the confusion.  The Wilsonian EFT obtained by integrating out heavy fields, and high-momentum modes of light fields, is reasonably termed the \emph{classical}  theory of the light fields, in which one can still compute quantum corrections from loops of the light fields.  However, the nature and couplings of the heavy fields, which in turn impact the form of the classical EFT, are dictated by the ultraviolet completion, i.e.~by the \emph{quantum}  theory of gravity.

To further illustrate this point, consider two candidate theories of quantum gravity, $\mathscr{T}_1$ and $\mathscr{T}_2$, each with some spectrum of massive states with $M>M_{\text{pl}}$.  Suppose that at low energies $\mathscr{T}_1$ contains a light field $\phi$ with an accidentally flat potential $V_{\mathscr{T}_1}$, such that
$\varepsilon\bigl(V_{\mathscr{T}_1}\bigr), \eta\bigl(V_{\mathscr{T}_1}\bigr) \ll 1$: then slow-roll inflation is possible in the theory $\mathscr{T}_1$.
If the spectra of heavy states in $\mathscr{T}_1$ and $\mathscr{T}_2$ differ only by factors of order unity, one would be tempted to call  $\mathscr{T}_1$ and $\mathscr{T}_2$ `similar' ultraviolet completions of gravity, and to claim that their low-energy phenomenology should be similar as well.   However, the slow-roll parameter $\eta$ is highly sensitive to the Planck-scale spectrum, and one generally finds $\eta\bigl(V_{\mathscr{T}_2}\bigr) \approx 1$, so that inflation does not occur in the theory $\mathscr{T}_2$.
Thus, whether inflation occurs or not can be changed by very subtle changes to the spectrum of states at masses $M > M_{\text{pl}}$.

In summary, quantum gravity effects do not decouple from inflation, and the $\eta$ parameter is sensitive to the ultraviolet completion. This fact is termed the \emph{eta problem}.  In some older papers this issue is called the \emph{supergravity eta problem}, but it is in no way special to supergravity.  Indeed, the eta problem is nothing more than the statement that an inflaton mass that is small compared to the cutoff scale is unnatural in effective field theory. 

Another statement of the problem is as follows: if the low-energy potential is
\begin{equation} \label{eq:etacorr}
    V(\phi) = V_0(\phi)\Biggl(1 + \sum_{\delta} c_{\delta}\Bigl(\frac{\phi}{M_{\text{pl}}}\Bigr)^{\delta}\Biggr)
\end{equation} with Wilson coefficients $c_{\delta}$, 
then in the absence of any additional structure, one needs to know the $c_{\delta}$ with $\delta \lesssim 2$ to at least $\mathcal{O}(1\%)$ accuracy in order to exhibit slow-roll inflation. 
That is, slow-roll inflation is affected by Planck-suppressed operators up to dimension $\Delta = \delta +4 \approx 6$.

This state of affairs is an extraordinary opportunity for string theory.  Knowledge of the structure of quantum gravity is essential
in order to interpret the results of precision CMB observations made over the past two decades.  

\subsection{Inflation in string theory}\label{ss:inflationst}

The arguments just presented are general facts about a low-energy effective field theory of quantum gravity, and did not rely on details of string theory.  However, having a concrete ultraviolet completion in hand dramatically sharpens the picture.\footnote{Comprehensive treatments of inflation in string theory appear in \cite{Baumann:2014nda, Cicoli:2023opf}; here we will limit ourselves to highlighting developments closely linked to moduli stabilization.}

Early works on inflation in string theory, with closed string moduli serving as the inflaton, include \cite{Binetruy:1986ss,Banks:1995dp}.
The discovery of D-branes led to the idea of brane inflation \cite{Dvali:1998pa}, developed in detail in \cite{Burgess:2001fx} (see also \cite{Dvali:2001fw,Alexander:2001ks}), in which the inflaton is the separation of a brane-antibrane pair.
The end of inflation is triggered by condensation of the open string tachyon, potentially leaving cosmic superstrings as relics \cite{Sarangi:2002yt}.
However, these studies appeared before the emergence of concrete scenarios for the stabilization of all moduli, and so the treatment of moduli couplings was necessarily incomplete.

The first analysis of inflation 
in a compactification with stabilized moduli appeared in \cite{Kachru:2003sx}, which considered a D3-brane moving towards an anti-D3-brane in a Klebanov-Strassler throat region, in the context of a KKLT vacuum.  The first key finding of \cite{Kachru:2003sx} was that warping caused the Coulomb interaction of the D3-brane and anti-D3-brane to be extremely flat.\footnote{Warping also allowed for cosmic superstrings with tension far below existing limits \cite{Kachru:2003sx,Copeland:2003bj}.}  However, a more significant finding, with implications for a much wider range of models, was that moduli stabilization reintroduced the eta problem \cite{Kachru:2003sx}.  Specifically, stabilization of the K\"ahler moduli by nonperturbative effects introduced new inflaton mass terms, in such a way that the total inflaton potential could be flat enough for prolonged slow-roll inflation only if competing terms could be balanced against each other in a modest (percent-level) fine-tuning.

The principal model-dependent corrections to the inflaton mass in the scenario of \cite{Kachru:2003sx} come from the inflaton-dependence of Pfaffian factors in the non-perturbative superpotential \eqref{eq:wrigidnotpure}.  The effects on the inflaton potential of non-perturbative superpotential terms supported on D7-branes in the throat region were computed in \cite{Baumann:2006cd}
(building on \cite{Berg:2004ek}) and \cite{Baumann:2007ah}, while the general form of contributions from the bulk of the compactification was determined in \cite{Baumann:2008kq} and \cite{Baumann:2010sx}.
The resulting phenomenology was analyzed in \cite{Agarwal:2011wm,McAllister:2012am}; for details see \cite{Baumann:2014nda}.
For a recent alternative implementation of brane inflation see \cite{Burgess:2022nbx}.

Closed string moduli are also promising inflaton candidates.
In the context of LVS, K\"ahler moduli can drive inflation \cite{Conlon:2005jm,Cicoli:2008gp}. In particular, for fibered Calabi-Yau compactifications with fiber modulus $\tau_f$, the volume can be written as
$\V\propto \tau_f F(\tau_i)+G(\tau_i)$ with  $\tau_i\neq \tau_f$, $i=1,\ldots, h_{11}-1$ and $F,G$ homogeneous functions of degree $1/2$ and $3/2$, respectively. The fiber modulus $\tau_f$ can be an inflaton candidate: the scalar potential for the canonically-normalized scalar field $\phi$, defined as $\tau_f\propto e^{\alpha\phi}$, $\alpha=1/\sqrt{3}$, is
\be
V(\phi)=V_0\Bigl(1-ae^{-\alpha \phi}+\ldots\Bigr)\,,
\ee
where $V_0$ and $a$ are computable constants. The ellipses include terms that are subdominant terms at large $\phi$.
Fibre{\texttrademark}  
inflation models lead to a potentially-detectable primordial tensor signal, with a robust prediction for the spectral index $n_s=0.969$, a tensor-to-scalar ratio $r=0.007$, and a field displacement $\Delta\phi\simeq 5$ in Planck units. These models will be confronted by experiment in the next decade.

The approximate flatness of the potential for $\phi$ is a consequence of classical scaling symmetries, as discussed in \S\ref{sec:pert} and \S\ref{sec:cyc}, 
for which $\phi$ can be seen as a pseudo-Goldstone boson \cite{Burgess:2014tja, Burgess:2016owb}.  Even so, the K\"ahler moduli inflation models of 
\cite{Conlon:2005jm,Cicoli:2008gp} 
are sensitive to the form of loop corrections to the K\"ahler potential. For a recent estimate of such corrections see \cite{Cicoli:2023njy}.

If the inflaton is an axion, the axion shift symmetry can protect the potential against corrections \cite{Freese:1990rb}.  
Embedding the original natural inflation idea of \cite{Freese:1990rb} in string theory proved difficult because the requisite super-Planckian decay constants --- of individual axions --- have not been found in controlled parameter regimes
\cite{Banks:2003sx,Arkani-Hamed:2006emk}.\footnote{Strictly speaking, single axion periodicities somewhat greater than $M_{\text{pl}}$ have been exhibited in the Lagrangians of consistent compactifications \cite{Bachlechner:2014gfa,Conlon:2016aea}, but the moduli potentials in these models do not allow the dynamics of large-field inflation: the axion energy in such a case would backreact disastrously.}
Ideas for evading this problem include
alignment of two or more axions \cite{Kim:2004rp} and
collective displacement of many axions \cite{Dimopoulos:2005ac,Easther:2005zr}.
Axion inflation models in string theory and F-theory include \cite{Grimm:2007hs,Grimm:2014vva}.

Axion monodromy \cite{Silverstein:2008sg} achieves large-field inflation by winding through many periods of an axion with sub-Planckian periodicity.
The axion monodromy model of \cite{McAllister:2008hb} involves a type IIB flux compactification on a Calabi-Yau orientifold with $h^{1,1}_{-}>0$: the inflaton is the two-form $C_2$ (see Table \ref{tab:multiplets}) on a cycle wrapped by an NS5-brane/anti-NS5-brane pair, and the inflaton potential is \cite{McAllister:2008hb,Flauger:2009ab}
\begin{equation}\label{eq:linearv}
    V(\phi) = \mu^3 \phi + \Lambda^4\,\text{cos}\biggl(\frac{\phi}{f}\biggr)\,,
\end{equation} where $\mu$, $f$, and $\Lambda$ are parameters with the dimensions of mass.  The periodic modulation in \eqref{eq:linearv} can give rise to oscillatory signatures in the CMB \cite{Flauger:2009ab}.
The mechanism of \cite{McAllister:2008hb} is compatible with the KKLT scenario for moduli stabilization: the shift symmetry of $C_2$ protects the inflaton from some (but not all) potentially dangerous terms.

The order parameter measuring the progress of axion monodromy inflation is the reduction of \emph{monodromy charge} as the inflaton configuration unwinds.  For example, in \cite{McAllister:2008hb} the monodromy charge is D3-brane charge accumulated on the NS5-brane pair, while in \cite{Ibanez:2014swa} it is D3-brane charge induced on moving D7-branes.  A very general problem in such models is that the stress-energy of monodromy charge backreacts on the internal space
\cite{McAllister:2008hb,Flauger:2009ab,McAllister:2016vzi,Kim:2018vgz}.
As a result, one finds that some form of fine-tuning is required in all fully-realized\footnote{Oversimplified four-dimensional EFT models of axion monodromy often omit the effects of backreaction of monodromy charge, but this is akin to assuming a shift symmetry, as in \eqref{eq:vintcorr}.} axion monodromy scenarios, though the extent of the difficulty is model-dependent.  To date, there is no fully explicit realizaion of axion monodromy inflation in a compact model, but there is also no known obstacle other than the complexity of the configuration: see e.g.~\cite{Marchesano:2014mla,Hebecker:2014kva,Blumenhagen:2014gta,Hebecker:2014eua,Arends:2014qca,Blumenhagen:2014nba,McAllister:2014mpa,Hebecker:2015rya} for progress in this direction.

\subsection{After inflation}

One of the most important aspects of string moduli is that they naturally change the early history of the universe even after a period of inflation. Inflation tends to wash out physics at energies higher than the inflation scale, but moduli can survive in some circumstances. 

In supersymmetry-breaking compactifications one can estimate the typical moduli masses to be $m_\phi \simeq  m_{3/2}$, since the moduli couple gravitationally and are massless in the absence of supersymmetry breaking (in Minkowski space).
In concrete cases this estimate is modified as follows:
\be
\hbox{KKLT}:\quad m_\phi\simeq \ln\left(\frac{M_{\text{pl}}}{m_{3/2}}\right)m_{3/2}, \qquad\hbox{LVS}:\quad m_\phi\simeq \left(\frac{m_{3/2}}{M_{\text{pl}}}\right)^{1/2} m_{3/2}\,,
\ee
with $\phi$ the lightest modulus (e.g. the volume modulus in LVS) and $m_{3/2}=\frac{W_0}{\V}M_{\text{pl}}$.

Because the moduli have gravitational-strength couplings, their decay rates are suppressed by the Planck mass, with
\be
\Gamma\simeq \frac{m_\phi^3}{16\pi M_{\text{pl}}^2}\,,
\ee
and the corresponding reheating temperature from moduli decay is of order
\be
T_{\text{rh}}\simeq \left(\frac{m_\phi}{M_{\text{pl}}}\right)^{1/2} m_\phi\,.
\ee
These properties of the moduli have a number of important post-inflationary implications:

\begin{itemize}
    \item {\bf Moduli domination}. Because the moduli are light but not relativistic, and because their decay rate is suppressed by powers of the Planck mass, they tend to be very stable, and can dominate the energy density of the universe as in matter domination, with $\rho\sim 1/a(t)^3$. This means that 
    in theories with moduli, inflation is very often followed by a period of moduli domination, unlike the standard Big Bang cosmology.

    \item {\bf Cosmological moduli problem (CMP)} \cite{Coughlan:1983ci, Banks:1993en, deCarlos:1993wie}. In order to avoid spoiling the success of Big Bang nucleosynthesis (BBN), the reheating temperature from moduli decays has to satisfy  $T_{\text{rh}}\gtrsim 1$ MeV, which implies $m_\phi\gtrsim 30$ TeV. In the naive estimate that $m_\phi \simeq M_{\text{soft}}\simeq m_{3/2}$ it would thus be difficult to have $m_\phi \gtrsim 30 $ TeV while imposing $M_{\text{soft}}\simeq 1$ TeV in order to address the hierarchy problem.
    
    In concrete KKLT and LVS scenarios the soft masses depend on the gravitino mass in different ways, depending on the location of the Standard Model (on D3-branes or D7-branes) and the degree of sequestering of the supersymmetry-breaking sector. For instance as we have seen in the previous section (see Table \ref{tab:soft}): 
    \begin{eqnarray}
    \hbox{KKLT}: && \quad m_\phi\simeq \ln^2\left(\frac{M_{\text{pl}}}{m_{3/2}}
    \right)M_{\text{soft}},\nonumber \\
    \hbox{LVS}: && \quad m_\phi \simeq M_{\text{soft}}\quad \hbox{or}\quad  m_\phi \simeq\left(\frac{M_{\text{pl}}}{m_{3/2}}\right)^{1/2} M_{\text{soft}}
    \end{eqnarray}
    Therefore the cosmological moduli problem may be ameliorated in some scenarios. Furthermore, the lack of evidence for supersymmetry at the TeV scale has turned this problem into a feature, namely moduli domination.

    \item{\bf Overshoot problem and kination}. 
    After inflation, which is thought to occur at relatively high energies,
    moduli fields will evolve towards the 
    minimum of their scalar potential. Since the vacuum has to describe physics at low energies, there is in general a hierarchy between the initial and final energies. Furthermore, since universal fields including the overall volume and dilaton have runaway minima, it is then possible that the evolution of some modulus field $\phi$ is energetic enough to pass over the barrier between the physical four-dimensional vacuum and the vacuum at infinity
    \cite{Brustein:1992nk}.  This difficulty is known as the overshoot problem.

    Fortunately, there are ways to address this problem \cite{Brustein:2004jp,Conlon:2008cj, Conlon:2022pnx,Apers:2022cyl}. 
    If the kinetic energy of the evolving modulus dominates over other sources, the resulting period is known as kination. The FLRW equations show that the kinetic energy density during kination dilutes as $\rho\propto 1/a(t)^6$, which is much faster than other sources of energy ($1/a^4$ for radiation or $1/a^3$ for matter).
    Thus, these other sources will end up dominating, and the corresponding friction term 
    in the scalar field equation,
    \be
    \ddot\phi+3H\dot\phi+V'(\phi)=0\,,
    \ee
    is enough to efficiently avoid overshooting. This can be explicitly computed by considering the set of equations as a dynamical system in which for simple enough potentials, like $V\sim e^{-\lambda \phi}$ in the rolling region (such as in LVS), there is an attractor solution that avoids overshooting.
    
    \item{\bf Dark radiation}.
 
Another consequence of moduli for post-inflationary physics is that  the last modulus to decay, rather than the inflaton field itself,  is responsible for (the final stage of) reheating. This differs from the simplest inflationary cosmology.  The decay products tend to include, in addition to the Standard Model fields, other light 
bosons resulting from the compactification,
such as ultralight axions or dark photons, which may contribute to dark radiation \cite{Cicoli:2012aq,Higaki:2012ar}.\footnote{For a recent review with further references see \cite{Cicoli:2023opf}.  Dark radiation constraints from ultralight axions were considered in \cite{Gendler:2023kjt}, and found not to become more severe as the number of axions increases.} 
The bounds on dark radiation are very tight. These are parameterized by the effective number of neutrinos $N_{\text{eff}}$. It is known experimentally that any contribution to this number from fields beyond the Standard Model is bounded by   
\cite{Planck:2018vyg}
\be 
N_{\text{eff}}^{\text{exp}}=2.99 \pm 0.17, \qquad N_{\text{eff}}^{\text{SM}}=3.04, \qquad  \Delta N_{\text{eff}}<0.2\,,
\ee
leaving almost no room for extra contributions. This is a strong experimental constraint on string theory models, and is also 
an opportunity. For instance, the axion partner of the volume modulus in LVS may be a candidate for dark radiation, since it is always relativistic, as its mass is of order $m\simeq e^{-c\V^{2/3}}$: see \cite{Cicoli:2022fzy} for a recent analysis of dark radiation in LVS.

\item{\bf Oscillons and moduli stars}.
    As the scalar fields evolve toward their vacuum state, in addition to the coherent oscillations that lead to moduli domination, they may give rise to inhomogeneities known as oscillons or oscillatons (or boson stars): these are non-topological solitonic objects that are unstable but long-lived. Independent of gravity, the structure of the scalar potential away from the minimum may be sensitive to the non-linear self-interactions of the scalar field. In an expansion around the minimum the potential can be written as:
    \be
    V=V_0+m^2\phi^2+\frac{g\phi^n}{\Lambda^{n-4}}+\ldots\,,
    \ee
    with $n$ the first non-vanishing power and $\Lambda$ a cutoff. If $g<0$, then the interaction is attractive and tends to give rise to oscillons of mass $M\simeq \Lambda^2/m$ and size $R\sim 1/m$. The mechanism for their formation may be either tachyonic oscillations or parametric resonance. Once gravity is relevant ($\Lambda\simeq M_{\text{pl}}$), it contributes to the attraction and gives rise to oscillatons or boson stars (see \cite{Visinelli:2021uve} for recent review and references). 
    
    These objects could be formed by the KKLT modulus, or by the small modulus in LVS, but not by the volume or fiber moduli \cite{Antusch:2017flz, Krippendorf:2018tei}. Numerical simulations have been performed illustrating the structure of these inhomogeneities and in particular the spectrum of gravitational waves that they produce. Even though the spectrum varies, the corresponding frequency of the gravitational waves is of order $\omega \simeq 1$ GHz, which is well beyond the reach of LIGO/VIRGO and LISA. This has motivated recent efforts to search 
    for ultra-high-frequency gravitational waves.

    High-energy theories, including string theories, furnish other potential sources of gravitational waves that could test energies as high as the GUT scale, $M\simeq 10^{-2} M_{\text{pl}}$. 
    The sensitivity required to probe these frequencies is beyond current technology, but the fact that there are no known astrophysical sources for such high-frequency gravitational waves 
    makes these waves very appealing future targets: a detection would shed light on physics at energies far beyond the reach of colliders and at the same time probe very early cosmology. For a comprehensive review of ultra-high-frequency gravitational waves see \cite{Aggarwal:2020olq}. 
    
\end{itemize}

\subsection{Dark energy}

Observations of supernovae and of the cosmic microwave background show 
that the expansion of the Universe is accelerating.
The current dark energy density is 
\begin{equation}\label{eq:rhode}
    \rho \approx 10^{-123} M_{\text{pl}}^4\,,
\end{equation}
with an equation of state relating $\rho$ and pressure $p$ \cite{Planck:2018vyg}:
\be
w=\frac{p}{\rho}=-1.03 \pm 0.03\,.
\ee
The most straightforward explanation is that the dark energy density comes from a cosmological constant,  $p=-\rho$, that is positive but extremely small.
This is the origin of the now-standard $\Lambda$CDM model for cosmology, where $\Lambda$ stands for the cosmological constant and CDM for cold dark matter.

The cosmological constant problem is the task of explaining why \eqref{eq:rhode} is so small.
At present there is no generally-accepted dynamical solution to the cosmological constant problem, in quantum field theory or in string theory.
However, string theory has provided a framework:
anthropic selection in the landscape of flux compactifications.\footnote{Those who prefer to avoid anthropic arguments are encouraged to propose an alternative in the form of a dynamical explanation of the smallness of the dark energy. For recent attempts see for instance \cite{Kaloper:2023xfl,Burgess:2021obw}.}

The idea that the cosmological constant might be determined by anthropic selection was
advanced by Weinberg, who argued that requiring the formation of galaxies imposes the constraint \cite{Weinberg:1987dv}
\begin{equation}\label{eq:weinbound}
-10^{-123}\lesssim \rho/M_{\text{pl}}^4 \lesssim 3\times 10^{-121}\,.
\end{equation}
However, even if one accepts anthropic\footnote{A conceptually similar approach that does not explicitly invoke observers appears in \cite{Bousso:2007kq}.} selection as an approach, it is also necessary to have a family of theories or vacua, among which are some that obey \eqref{eq:weinbound} and can be selected. 

The next key idea
was the Bousso-Polchinski
\cite{Bousso:2000xa} observation that 
compactifications admitting many choices of quantized flux furnish a densely-spaced `discretuum' of vacua, potentially including some with $\rho$ near the observed value (cf.~\cite{Feng:2000if}).
The toy model proposed by Bousso and Polchinski takes the form (in units $M_{\text{pl}}=1$)
\begin{equation}\label{eq:bptoy}
    \Lambda =  \Lambda_{\text{bare}} + \vec{q}\,^{\mathsf{T}}\cdot\vec{q}\,,
\end{equation} where $\vec{q} \in \mathbb{Z}^N$ represents a vector of flux integers. 
If the bare cosmological constant $\Lambda_{\text{bare}}$ is negative (and, we suppose, of order unity), then finding a configuration with small $|\Lambda|$ from \eqref{eq:bptoy} amounts to finding a lattice vector $\vec{q}$ whose length is extremely close to $\sqrt{|\Lambda_{\text{bare}}|}$.
For $N \gg 1$ this is typically possible in principle, but actually finding such a vector is exponentially costly \cite{Denef:2006ad}.
 
Bousso and Polchinski did not propose a complete microscopic realization of their scenario
in \cite{Bousso:2000xa}, but a framework came soon after, in the form of
the landscape of flux vacua in type IIB compactifications, with moduli stabilized as in 
the KKLT scenario.  The fluxes in question are $F_3$ and $H_3$, and the large dimension $N$ of the flux lattice is $N=2h^{2,1}$, which is of order hundreds in typical Calabi-Yau threefolds.

The AdS$_4$ vacua found in \cite{Demirtas:2021nlu} and reviewed in \S\ref{sec:explicitkklt} are incarnations of the KKLT scenario, and can have exponentially small cosmological constants.
However, the vacua of 
\cite{Demirtas:2021nlu} are \emph{not} realizations of the Bousso-Polchinski mechanism as written in \eqref{eq:bptoy}: the bare cosmological constant $\Lambda_{\text{bare}}$ vanishes in the vacua of \cite{Demirtas:2021nlu}, and $N$ need not be large: examples exist with $N=4$ and $5$ in which $|\rho| \ll  10^{-123} M_{\text{pl}}^4$.  Moreover, the moduli mass scale in \cite{Demirtas:2021nlu} is itself extremely small, whereas if one finds small $\Lambda$ via \eqref{eq:bptoy} it remains possible that all mass scales other than $\Lambda$ itself are large.  Thus, the constructions of 
\cite{Demirtas:2021nlu} constitute progress in achieving scale separation in stabilized vacua, but not in solving the cosmological constant problem.
 
Although the outlines are in place, a proper implementation of the anthropic approach to the cosmological constant problem will require a better understanding of how to populate the landscape through vacuum decays, and how to define a measure on the string landscape (see e.g.~\cite{Bousso:2008hz}). Vacuum decays in gravitational theories have been studied for more than forty years \cite{Coleman:1980aw,Brown:1988kg,Fischler:1990pk}, and even though these studies were carried out using semiclassical techniques, there is a coherent picture that fits well with string theory.
The decay proceeds by means of the nucleation of a bubble of one vacuum in the background of the original vacuum. In the string landscape there would be a huge number of vacua being constantly nucleated, providing an example of eternal inflation (for a review of eternal inflation see \cite{Guth:2007ng}). In type IIB flux compactifications, the bubble wall  would correspond to fivebranes wrapping the relevant three-cycles.\footnote{Ensuring that moduli remain stabilized across a bubble nucleation transition is an obstacle to populating a landscape dynamically in type IIB compactifications.}  The Euclidean formalism of \cite{Coleman:1980aw} then seems to imply that the bubble universe would be an open universe.  If true, this would be a concrete prediction of the landscape (see for instance \cite{Freivogel:2005vv}). Populating this landscape could allow one of the bubbles to correspond to a universe with a cosmological constant consistent with current observations.

Overall, the measure problem and the dynamics of populating the landscape are complex and subtle questions that are not well-understood, but that could have significant impact on our understanding of cosmology.  For recent discussions see for instance \cite{Hebecker:2018ofv,Hebecker:2020aqr,McNamara:2020uza,DeAlwis:2019rxg,Cespedes:2020xpn,Cespedes:2023jdk,Friedrich:2023tid,Carifio:2017nyb,Khoury:2019yoo,Giudice:2021viw,Kartvelishvili:2020thd,Khoury:2022ish}.
 
A cosmological constant appears to be the simplest explanation for the dark energy, but the reality may be more complicated.   
Current observations are also consistent with the idea of quintessence, in which an ultralight scalar field is slowly  rolling today, driving accelerated expansion.
Note that even though runaway potentials are generic in string theory, quintessence is not, 
because
the potential needs to be very flat for the field to roll  slowly today. 
There have been several attempts to implement quintessence in flux compactifications:
see for instance \cite{Hellerman:2001yi,Fischler:2001yj,Svrcek:2006hf,Panda:2010uq} for early works, and \cite{Denef:2018etk,Cicoli:2018kdo,Hebecker:2019csg,Cicoli:2021fsd, Cicoli:2021skd,Cicoli:2023opf} for recent discussions of quintessence in string theory.

\section{Axions}\label{sec:axions}
 
In an effective field theory, the lightest --- and hence, often the most important --- fields are those whose mass terms are controlled by symmetries.  The four-dimensional effective theories resulting from Calabi-Yau orientifold compactifications of superstring theories always include the massless graviton, protected by general covariance, but also a number of axions,\footnote{Some authors reserve the term `axion' for a particular pseudoscalar coupled to QCD, and refer to other shift-symmetric pseudoscalars as `axion-like particles', but we will not follow this usage: we will speak of many axions, one of which is the QCD axion.} i.e.~pseudoscalars $\theta$ with approximate shift symmetries $\theta \to \theta + const$.
As we have reviewed, moduli masses receive quantum corrections that typically put moduli out of reach of terrestrial and astrophysical measurements: the hard task in moduli stabilization is accurately computing the moduli masses, not establishing that they are generically large compared to Standard Model scales. Finally, gauge bosons and chiral fermions are found in many, but not all, compactifications.
In this sense, the generic prediction of Calabi-Yau flux compactifications, in the long-wavelength limit, is \emph{four-dimensional Einstein gravity coupled to axion fields}.

Axions are the subject of an immense array of experimental searches: through their interactions with light and matter on Earth, in the sun, in supernovae, in galaxies, and in the CMB, as well as through their gravitational effects on structure formation.  Because there are so many paths to improved upper limits (or to a detection), in different environments and relying on different physics and technology, it is clear that axion science will advance rapidly in the coming decades.\footnote{See \cite{Marsh:2015xka} for a review of the physics of axions.}  Combining this state of affairs with the near-inevitability of axions in string theory, we conclude that axions furnish one of the most promising paths to testing string theory.

\subsection{Axions in string theory}\label{sec:axionstring}

Axions in string theory\footnote{See e.g.~\cite{Choi:1997an,Banks:2003sx,Conlon:2006tq,Svrcek:2006yi} for early works on axion couplings in string theory.} result from dimensional reduction of $p$-form potentials on $p$-cycles.  Consider a term in the ten-dimensional action of the form
\begin{equation}\label{eq:scp}
S_{C_p} = \int \text{d} C_p \wedge \star \text{d}C_p\,.
\end{equation}
The action \eqref{eq:scp} is minimized if $\d C_p=0$, and the equation of motion reads\footnote{To reach the form \eqref{eq:delcp}, one imposes the gauge-fixing condition $\text{d}^{\dagger}C_p=0$, where $\text{d}^{\dagger}:=- \star \text{d} \star$ is the adjoint exterior derivative.}
\begin{equation}\label{eq:delcp}
\Delta C_p = 0\,,
\end{equation}
where $\Delta$ is the Laplacian acting on $p$-forms.
The solutions to \eqref{eq:delcp} correspond, by Hodge's theorem, to cohomology classes in $H^p(X,\mathbb{R})$.

Topologically trivial changes in the field configuration of the potential $C_p$ --- i.e., changes $C_p \to C_p + \text{d} A_{p-1}$, with $A_{p-1}$ a $(p-1)$-form, are gauge redundancies, and in particular the action \eqref{eq:scp} is invariant.
However, a change in the cohomology class of $C_p$, $C_p \to C_p + c\, \omega_p$, with $c \in \mathbb{R}$ and with $\omega_p \in H^p(X,\mathbb{Z})$,
can be detected by a suitable object charged under $C_p$.  The Wilson line in Maxwell theory is an example for $p=1$.

In type IIB string theory, we will focus on the case $p=4$, with $C_4$ the Ramond-Ramond four-form.  
We take $\omega_a$ to be a basis of $H^4(X,\mathbb{Z})$, with  $\omega^a$ the dual homology basis. 
Euclidean D3-branes wrapping four-cycles $D$ couple to $C_4$ as in \eqref{eq:sed3},
and so writing an integral divisor class $D = d_a \omega^a$ in terms of integers $d_a$, we have
\begin{equation}\label{eq:axionsed3}
\frac{1}{2\pi}\,S_{\text{ED3}} = \text{Vol}(D) + i d_a\int_{\omega^a} C_4\,.
\end{equation}
Under $C_p \to C_p + c^a \omega_a$, the Euclidean D3-brane action \eqref{eq:axionsed3} shifts as
\begin{equation}
S_{\text{ED3}} \to S_{\text{ED3}} + 2\pi i c^a d_a\,.
\end{equation}
Integral shifts remain a symmetry even in the presence of Euclidean D3-branes, but the continuous shift symmetry is broken.

To examine the corresponding four-dimensional axions, we make the ansatz 
\begin{equation}
C_4 = \theta^a(x) \omega_a\,,
\end{equation} with $\omega_a$ a basis of $H^4(X,\mathbb{Z})$,
and find
\begin{equation}
S_{C_p} = - M_{\text{pl}}^2 \int \d^{4} x \sqrt{-g}\, \partial_{\mu} \theta^a \partial_{\nu}\theta^b \int_X \omega_a \wedge \star \omega_b\,.
\end{equation} 
The pairing 
\begin{equation}\label{eq:omk}
\int_X \omega_a \wedge \star \omega_b = \frac{1}{2} K^{\text{tree}}_{ab} = \frac{1}{4} \frac{\partial}{\partial \tau_a}\frac{\partial}{\partial \tau_b} K_{\text{tree}} = -\frac{1}{2} \frac{\partial}{\partial \tau_a}\frac{\partial}{\partial \tau_b}\,\text{log}(\mathcal{V})
\end{equation}
thus defines the matrix of kinetic couplings of the axions $\theta^a$.
In \eqref{eq:omk} we have compared to the tree-level kinetic terms derived in \S\ref{sss:ktree}, using \eqref{eq:ktree} and  \eqref{eq:calvolis}, and the derivatives are to be evaluated using \eqref{eq:tauis}.

The axions $\theta^a$ have no non-derivative interactions to any order in the $g_s$ and $\alpha'$ expansions: their potential arises exclusively from non-perturbative effects.  This structure is a consequence of the ten-dimensional gauge symmetry, and does not rely on supersymmetry.  Moduli, in contrast, do generally receive contributions to their potential at (almost) all orders in both perturbative expansions.  

At first glance one might guess that perturbative quantum effects should be easier to compute than non-perturbative quantum effects --- if so, the moduli potential would be more accessible to theoretical understanding than the axion potential.
It is true that some of the leading perturbative corrections to the effective action are well-understood, and it is also true that a genuinely exact and systematic computation of the non-perturbative potential for axions has not yet been performed.
Even so, the axion shift symmetries are so powerful that at the time of writing, much more is known about the structure of the axion potential, and the resulting physics at low energies, than about the corresponding questions for moduli.

\subsection{Kreuzer-Skarke axiverse}

Let us try to understand this state of affairs in the well-studied case of $C_4$ axions, which result from the expansion $C_4 = \theta^a \omega_a$ with $\omega_a$ a basis for the orientifold-even subspace $H^{4}_{+}(X_6,\mathbb{Z})$.
The actions $S$ of the Euclidean D3-branes that give the dominant potential terms are determined by the volumes of the wrapped four-cycles $D$, up to logarithmic corrections from Pfaffians involving the other moduli:
\begin{equation}
\text{Re}(S) = 2\pi \text{Vol}(D) - \text{log}\bigl(\mathcal{A}(z_i,\tau)\bigr)\,.
\end{equation}
If one makes the further assumption that supersymmetry is weakly broken by the expectation value $W_0$ of the flux superpotential, then the dominant Euclidean D3-branes are those that contribute to the superpotential.\footnote{We set aside the exotic possibility of parametrically large recombination of non-holomorphic cycles \cite{Demirtas:2019lfi}.}  These Euclidean D3-branes wrap holomorphic four-cycles, which are calibrated by the K\"ahler form $J$ and so have volumes that are given algebraically in terms of the data of the K\"ahler parameters and intersection numbers, 
\begin{equation}
\text{Vol}(D_a) = \frac{1}{2} \int_{D_a} J \wedge J = \frac{1}{2}\kappa_{abc}t^b t^c\,.
\end{equation}
The geometric data of four-cycle volumes is then readily accessible. In contrast, computing the volume of a non-holomorphic four-cycle would require knowing, and integrating, the volume form determined by the Calabi-Yau metric: this could be attempted by building on the metrics obtained in  \cite{Ashmore:2019wzb,Cui:2019uhy,Anderson:2020hux,Douglas:2020hpv,Larfors:2022nep,Berglund:2022gvm,Gerdes:2022nzr,Ashmore:2023ajy}, but has not been carried out.

In view of the results of \S\ref{sss:wnp}, computing $W_{\text{ED3}}$ at a given point in $T^{\star}$ in K\"ahler moduli space involves enumerating the smallest effective divisors at $T^{\star}$ that support two fermion zero modes, as in \eqref{eq:fullrigid}.
Thus, the leading exponentials are determined by knowing topological and geometric data.  From the leading exponentials alone one cannot compute the exact axion potential, but nonetheless one can obtain enough information to make meaningful comparisons to experiments.

In summary, the four-dimensional effective action for axions can be computed with knowledge of the topological data of a Calabi-Yau orientifold, together with the specification of a point $T^{\star}$ in K\"ahler moduli space.
In a complete top-down derivation one would compute the potential for K\"ahler moduli, find its minimum $T_{\text{min}}$, and study axion couplings there.  At present K\"ahler moduli stabilization can be performed only in certain lamppost regions of moduli space, where for example the KKLT or LVS approaches are applicable.  One approach is to fix a scenario and study the axion couplings, as done in \cite{Cicoli:2012sz} for LVS.\footnote{A recent treatment appears in \cite{Broeckel:2021dpz}.}  

Alternatively, one can take advantage of the fact that computing the data of axion effective theories is feasible in a much broader swath of moduli space than the lampposts where moduli stabilization is understood.  
One can thus aim to characterize axion couplings in string theory by searching for results that hold everywhere in K\"ahler moduli space, or that are at least approximately independent of location.\footnote{Alternatively, one can pursue constraints that result from sampling the moduli space according to a natural measure.}

More precisely, one seeks results that hold, or hold to some level of approximation, everywhere in the subregion of moduli space where computations are controlled.  Near the walls of the extended K\"ahler cone, $\alpha'$ corrections to the effective action are large, and the axion kinetic term is not necessarily well-approximated by the couplings $\int \omega \wedge \star \omega$ that result from the leading-order term $\int \d C_4 \wedge \star \d C_4$ in the ten-dimensional action.  Moreover, near loci where divisors shrink, the associated instanton series are poorly controlled, and the axion potential is not a sum of exponentially small periodic terms.

Following \cite{Demirtas:2018akl}, we define the \emph{stretched K\"ahler cone} as the subregion of the K\"ahler cone in which all effective curves have volume $\ge 1$:
\begin{equation}\label{eq:skc}
\mathcal{K} := \biggl\{ J \in H^{1,1}(X_6,\mathbb{R}) \Bigr\vert ~\int_{\mathcal{C}} J > 1 ~~\forall~\mathcal{C}\biggr\}\,,
\end{equation}
where the condition is imposed for all effective curves $C \in H_2(X_6,\mathbb{Z})$, and we work in units of $\ell_s^2 = (2\pi)^2\alpha^\prime$.

Within the stretched K\"ahler cone, and in the approximation that the scale of supersymmetry breaking is small compared to the cutoff of the effective theory, the axion kinetic term is determined by classical geometric data, and the dominant potential terms 
breaking the PQ symmetries arise from the Euclidean D3-brane superpotential $W_{\text{ED3}}$.  Thus, a systematic computation of the leading terms in the effective theory of $C_4$ axions\footnote{The masses and couplings of the saxions $\tau_a = \text{Vol}(D_a)$ vary across the K\"ahler moduli space.  A reasonable expectation is that near the walls of the K\"ahler cone, competition among perturbative 
corrections to the K\"ahler potential, as in \eqref{eq:thisispertk}, could lead to stabilization of the saxions, though computing the saxion potential in detail is presently out of reach.  If the saxions are stabilized by perturbative effects, they will be much heavier than the axions, and will be comparatively decoupled from low-energy phenomena.} reduces to an algorithmic procedure in computational geometry.  
In the case that the Calabi-Yau threefold $X$ is a hypersurface in a toric variety, this procedure is actually feasible, and can even be automated: see the discussion in \S\ref{sec:computation}.

The \emph{Kreuzer-Skarke axiverse} \cite{Demirtas:2018akl} is the resulting ensemble of axion effective theories from
Calabi-Yau threefold hypersurface compactifications of type IIB string theory, with $1 \le h^{1,1} \le 491$.
A striking property of this ensemble is that the cycle volumes manifest hierarchies that are polynomial in $h^{1,1}$.  An intuitive explanation is that complicated topology does not easily fit into a small (and Ricci-flat) space, while a quantitative explanation in terms of systems of linear inequalities is given in \cite{Demirtas:2018akl}. 
One empirical result is that --- at a reference location in moduli space, the tip of the stretched K\"ahler cone 
--- certain 
divisor volumes\footnote{More precisely, in every basis of $H_4(X,\mathbb{Z}$, at least one (and often, more than one) basis element has size at least as large as \eqref{eq:ksa}: see \cite{Demirtas:2018akl} for further details.}  scale as
\begin{equation}\label{eq:ksa}
\text{Vol}(D_A) \propto \bigl(h^{1,1}\bigr)^4\,,
\end{equation} though it should be understood that the scatter around \eqref{eq:ksa} is significant \cite{Demirtas:2018akl}.
 
As a result of \eqref{eq:ksa}, the Kreuzer-Skarke axiverse has important qualitative differences from that envisioned in the first discussions of the string axiverse \cite{Arvanitaki:2009fg}.
In both cases there are hundreds of axions $\phi_a$ with masses $m_a$ spanning a huge range of scales,\footnote{Light axions are both theoretically natural, because their mass originates from a non-perturbative effect, and experimentally allowed: 
light parity-even spinless fields (i.e., scalars) are strongly constrained by experimental limits on fifth forces, but light pseudoscalars are not.} but the distributions of decay constants $f_a$ are very different.
The original scenario of \cite{Arvanitaki:2009fg} proposed decay constants clustered around a high scale, 
\begin{equation}
  \text{{\emph{String Axiverse}}~\cite{Arvanitaki:2009fg}:~~}  f_a \sim M_{\text{GUT}} \approx 2 \times 10^{16}~\text{GeV}\,.
\end{equation}
In simple Calabi-Yau compactifications that have only a single axion, $f_a$ is indeed typically large,
and so an axiverse modeled as $N \gg 1$ copies of such single axions would have a narrow distribution of decay constants $f_a$.  However, it turns out that topologically complex Calabi-Yau threefolds manifest geometric hierarchies, as in \eqref{eq:ksa}.
As a consequence, the decay constants $f_a$ in the Kreuzer-Skarke axiverse spread over a wide range of scales, with 
\begin{equation}\label{eq:frange}
 \text{{\emph{Kreuzer-Skarke Axiverse}}~\cite{Demirtas:2018akl}:~~}   \mathcal{O}(10^9)~\text{GeV} \lesssim f_a \lesssim \mathcal{O}(10^{16})~\text{GeV}\,.
\end{equation} This finding changes the axion phenomenology as well as the prospects for detection, as we now explain.

\subsection{Strong CP problem}
One important issue that can be understood in this approach is the strong CP problem.  The neutron electric dipole moment is so small that it has never been measured.  In quantum field theory without gravity, its smallness can be explained, as suggested by Peccei and Quinn (PQ), by introducing an axion that dynamically relaxes the CP-breaking $\theta$-angle of QCD.  However, this solution is very sensitive to quantum gravity corrections: even rather high-dimension Planck-suppressed operators spoil the PQ mechanism.  This weakness of the PQ mechanism is called the \emph{quality problem}.

Because the success or failure of the PQ mechanism for solving the strong CP problem hinges on quantum gravity effects, one should compute these effects directly in an ultraviolet completion of gravity.  This was achieved in \cite{Demirtas:2021gsq}, in the setting of type IIB compactifications on orientifolds of Calabi-Yau hypersurfaces, in the geometric regime.  If QCD is realized on a stack of D7-branes wrapping a four-cycle $D_{QCD}$, the QCD axion is  
\begin{equation}
\theta_{\text{QCD}} = \int_{D_{\text{QCD}}} C_4\,.
\end{equation}
The terms in the effective action that spoil PQ quality arise from instantons --- specifically, Euclidean D3-branes --- that carry PQ charge.  Applying the methods reviewed in \S\ref{sec:computation}, one can enumerate the leading Euclidean D3-brane superpotential terms, which for a four-cycle $D_A$ take the form
\begin{equation}\label{eq:wed3pq}
W_{\text{ED3}} = \mathcal{A}_A e^{-2\pi Q^{a}_{~A} T_a}\,.
\end{equation}
The importance of \eqref{eq:wed3pq} at a given point in K\"ahler moduli space falls exponentially with $\text{Vol}(D_A) = \text{Re}(T_A)$.
Inside the stretched K\"ahler cone --- i.e., in the region of moduli space where the $\alpha'$ expansion is under control 
--- the four-cycle volumes $\text{Vol}(D_A)$ are hierarchical: a few are $\approx \mathcal{O}(1) \ell_s^4$ but a majority scale as in \eqref{eq:ksa}.  The resulting mass of the QCD axion, and hence the shift of the QCD $\theta$-angle $\bar{\theta}$ away from the CP-conserving value $\bar{\theta}=0$, was computed\footnote{The phenomenological scaling of \eqref{eq:ksa} helps explain the result of \cite{Demirtas:2021gsq}, but was not used as an input in \cite{Demirtas:2021gsq}: the four-cycle volumes were directly computed in each case.} in a large ensemble of explicit geometries in \cite{Demirtas:2021gsq}, with the result
\begin{equation}\label{eq:pqscale}
    \bar{\theta} \propto \text{exp}\Bigl( - c N^4 \Bigr)\,,
\end{equation} with $N = h^{1,1}$ the number of axions, and equivalently the number of K\"ahler moduli.  
We reiterate that \eqref{eq:pqscale} contains explicit upper bounds on the leading quantum gravity contributions to $\bar{\theta}$.\footnote{The technical assumptions entering \eqref{eq:pqscale} are laid out in \cite{Demirtas:2021gsq}, and are fairly modest: supersymmetry breaking should be well below the compactification scale, and there should be no light states charged under QCD except those from the Standard Model and their superpartners.}
Thus, in models with $N \gtrsim 15$ axions, everywhere in moduli space where the $\alpha'$ expansion is under control, the strong CP problem is solved by the Peccei-Quinn mechanism.

It does not follow that (type IIB) string theory predicts that the strong CP problem is automatically solved in general, because we have no knowledge of $\bar{\theta}$ in the regime where cycles are small and instanton corrections are large.  Rather, one can say that Calabi-Yau orientifold compactifications of type IIB string theory, in the geometric regime, furnish an ultraviolet completion of gravity that exhibits an automatic solution of the strong CP problem.  Although this solution applies only if $h^{1,1} \gtrsim 15$, the overwhelming majority of known Calabi-Yau threefolds do have $h^{1,1} > 15$.\footnote{Strictly speaking, this statement applies to counts of Calabi-Yau threefolds
that are potentially inequivalent, but have not yet been proven to be distinct: cf.~\cite{Gendler:2023ujl}.}

\subsection{Axion-photon couplings}

An impressive collection of experiments are searching for axions through their couplings to the photon, for example through X-ray spectrum oscillations (see e.g.~\cite{Marsh:2017yvc,Reynolds:2019uqt,Reynes:2021bpe}), CMB birefringence, or production in the Sun or the Milky Way.
For $N$ canonically-normalized axions $\varphi_a$ with decay constants $f_a$, the couplings can be written
\begin{equation}
    \mathcal{L} \supset -\frac{1}{4}g_{a\gamma\gamma} \varphi^a\,F_{\mu\nu}\tilde{F}^{\mu\nu}\,,
\end{equation}
with
\begin{equation}
    g_{a\gamma\gamma} = \frac{\alpha_{\text{EM}}}{2\pi f_a}\Theta_a~~\text{(no sum)}\,.
\end{equation} where $\Theta_a$ is a dimensionless coupling that results from a mixing angle.

Let us temporarily suppose that the mixing angles $\Theta_a$ are $\mathcal{O}(1)$, or at least are not hierarchically small.  Then the finding that 
in Calabi-Yau compactifications with many axions \cite{Demirtas:2018akl,Mehta:2021pwf},
the decay constants spread down to small scales, $f \sim \mathcal{O}(10^9)~\text{GeV}$, as in \eqref{eq:frange}, suggests that such theories could easily be ruled out by near-future limits on axion-photon couplings.

It turns out, however, that the mixing angles $\Theta_a$ are \emph{not} of order unity \cite{Halverson:2019cmy,Gendler:2023kjt}.
Two mechanisms are responsible: first, the kinetic matrix is very sparse, as a result of the structure of divisor intersections \cite{Halverson:2019cmy}.  Second, Euclidean D3-branes wrapping the four-cycle that supports QED generate a small mass scale, and any axions lighter than this \emph{light threshold} have hierarchically suppressed couplings to the photon \cite{Gendler:2023kjt}.

The picture that emerges is of an axiverse that is mostly dark, but in which a handful of axions have couplings to photons that put them in reach of ongoing observations \cite{Marsh:2015xka,Irastorza:2018dyq,Adams:2022pbo,Rogers:2023ezo}.

\subsection{Black hole superradiance}

Black hole superradiance is another phenomenon that connects the string axiverse to observations \cite{Arvanitaki:2010sy,Brito:2015oca}.
An axion field whose Compton wavelength is comparable to the Schwarzschild radius of a Kerr black hole will develop a superradiant instability and extract angular momentum from the black hole, unless nonlinear self-interactions of the axion disrupt the development of a condensate.  Thus, upper bounds on the spins of astrophysical black holes set limits on the existence of axions in a range of mass $m$ and interaction strength (i.e., decay constant) $f$.  These limits do not involve any cosmological assumptions: the axions need not be produced in cosmic history, or persist as a relic population. The axions are created from the vacuum by the black hole, so what is tested is whether such fields are present in the Lagrangian.
For stellar-mass black holes, for which the data is better at present, the exclusions are in the vicinity of $m \approx 10^{-10}\,\text{eV}$, and apply for $f \gtrsim 10^{14}~\text{GeV}$.

Black hole superradiance in string theory was studied in \cite{Mehta:2021pwf}: the spectrum of $C_4$ masses and decay constants was computed in an ensemble of type IIB compactifications on Calabi-Yau threefold hypersurfaces, and were compared to measured black hole spins.  For this purpose the K\"ahler moduli were taken at certain reference locations, such as the tip of the stretched K\"ahler cone.  A considerable fraction of geometries at these points in moduli space are excluded by observations.

\section{Computational advances}\label{sec:computation}

The study of flux compactifications rests on computation. 
To construct a Calabi-Yau threefold $X_6$ and determine
the leading-order data of the $\mathcal{N}=2$ effective theory that results from a type IIB compactification on $X_6$, one can start by identifying a suitable ambient space, such as a projective space or a more general toric variety; defining the Calabi-Yau as a subvariety; and computing the Hodge numbers and intersection numbers.  One then determines the region of moduli space where this description is valid, for example by computing the K\"ahler cone associated to the given geometric phase.
The exact prepotential for the vector multiplet moduli space is obtained by computing the periods of the $(3,0)$ form on $X_6$, while computing the periods of the $(3,0)$ form on the mirror $\widetilde{X}_6$ and making use of mirror symmetry furnishes a subset of non-perturbative corrections to the hypermultiplet moduli space.  
Proceeding to an $\mathcal{N}=1$ compactification requires a further series of computations.  One identifies an orientifold involution, makes a choice of quantized fluxes, evaluates the flux superpotential in terms of the periods, 
and enumerates terms in the non-perturbative superpotential by computing the topology of effective divisors.  

The above computation yields
that part of the data of the effective theory 
that 
is either classical, holomorphic, or both
--- in particular, the K\"aher potential at leading order in $g_s$ and $\alpha'$, and the classical plus non-perturbative superpotential.  
Approximating the effective theory using only classical plus holomorphic data is a useful 
starting point at weak coupling and large volume, where perturbative corrections are small.
But it is also important to recognize the disparity between the task of computing classical plus holomorphic data, and that of computing perturbative corrections.
The steps in the former undertaking, as enumerated above, are all well-specified computations in algebraic geometry and topology.  In some cases these might be intractable, or become expensive when the dimension of moduli space is large, but they can at least be stated precisely to a person (or computer) unfamiliar with string theory.  Moreover, the questions are generally algebraic or topological in character, rather than analytic.

In contrast, string perturbation theory is incompletely formulated, especially in general flux backgrounds.
Sigma model perturbation theory is hardly better, and evaluating curvature corrections eventually requires computing the curvature of the Ricci-flat metric.  
Significant progress in characterizing perturbative corrections may require fundamental advances in string theory, and at a minimum will require the application of expertise in highly technical areas of string theory.  In summary, the task of computing perturbative corrections remains a \emph{physics} problem, whereas that of computing the classical plus holomorphic data has been sharpened --- by efforts beginning in the early days of supergravity 
and string theory --- into a mathematics problem.

The past decade has seen extraordinary progress in solving this mathematics problem, and doing so automatically, at scale, and in the general case. 
Approaches to the problem 
can be usefully classified into three generations.  In the first, pencil and paper were the dominant tool, so the 
the number of relevant moduli was necessarily very small: examples include classic works on mirror symmetry for two-parameter models \cite{Candelas:1993dm,Candelas:1994hw}.
The second generation is characterized by increasingly sophisticated computer assistance, but with algorithms whose expense grows exponentially in the number of moduli, so that $\mathcal{O}(10)$ moduli is the upper limit (e.g., \cite{Altman:2014bfa}).
At this stage some codes were purpose-built for research in string theory (e.g.,~\texttt{cohomCalg} \cite{Blumenhagen:2010pv}), while others used general-purpose computational geometry and computational algebra software such as \texttt{Sage} and PALP \cite{Kreuzer:2002uu}.

The third, and current, generation is characterized by the emergence of new algorithms whose cost is only polynomial, allowing for the first time the study of examples with hundreds or thousands of moduli, as well as the construction of large ensembles of compactifications, for example across the entire Kreuzer-Skarke list.
Third-generation approaches to computing intersection numbers, Mori and K\"ahler cones, and other classical data of Calabi-Yau threefold hypersurfaces are built into version 1.0 of the software \texttt{CYTools} \cite{Demirtas:2022hqf}.  Third-generation algorithms for computing period integrals and Gopakumar-Vafa invariants through the mirror map appeared in \cite{Demirtas:2023als}, while orientifolding at this level was demonstrated in \cite{Moritz:2023jdb}; both capabilities will be integrated into a future release of \texttt{CYTools}.
Another characteristic of the third generation of tools is the application of methods of machine learning, which has led to significant progress in e.g.~the computation of Ricci-flat metrics on compact Calabi-Yau threefolds  \cite{Ashmore:2019wzb,Cui:2019uhy,Anderson:2020hux,Douglas:2020hpv,Larfors:2022nep,Berglund:2022gvm,Gerdes:2022nzr,Ashmore:2023ajy}. Machine learning has also been applied to explore other aspects of Calabi-Yau geometry in e.g.~\cite{Bull:2018uow,Berman:2021mcw,Jejjala:2022lxh,Berglund:2023ztk,Erbin:2021hmx,Klaewer:2018sfl,Cole:2019enn,Cole:2021nnt,Krippendorf:2021uxu}; for a review, see \cite{Ruehle:2020jrk}.

Toric geometry is at the heart of much recent progress in computing topological data of string compactifications.  A fine, regular, star triangulation of any of the 473{,}800{,}776 reflexive polytopes in the Kreuzer-Skarke list \cite{Kreuzer:2000xy} defines a toric variety in which the generic anticanonical hypersurface is a smooth Calabi-Yau threefold.
Much of the topology of the hypersurface is inherited from the ambient toric variety, and can be read off from the triangulation.  Toric geometry thus provides a combinatorial approach to Calabi-Yau compactifications.  Advances in triangulation algorithms \cite{Demirtas:2022hqf,MacFadden:2023cyf}, 
combined with applications of general results in toric geometry
\cite{Demirtas:2023als,Moritz:2023jdb},
have led to polynomial-time approaches to problems that were inaccessible with general computational algebraic geometry methods.

All the classical data, and nearly all the holomorphic data, of a type IIB compactification\footnote{Related capabilities for F-theory models are provided by \texttt{FTheoryTools} \cite{FTheoryTools}.} on a Calabi-Yau threefold hypersurface is accessible by the above methods, but the Pfaffian prefactors of Euclidean D3-brane or gaugino condensate superpotential terms require separate treatment.  In toroidal models the full moduli-dependence of these factors can be computed on the worldsheet \cite{Berg:2004ek}.
The dependence on D3-brane positions is well-understood in local models \cite{Baumann:2006th}, and was recently computed in an orientifold of a compact elliptic Calabi-Yau threefold \cite{Kim:2022uni}.  A systematic method for computing the moduli dependence of the Pfaffian of any rigid Euclidean Dp-brane was developed in \cite{Kim:2023cbh}.  The 
normalization of the contribution of a rigid Euclidean Dp-brane was 
fixed in \cite{Alexandrov:2022mmy}, building on a series of works treating the normalization of Euclidean Dp-brane amplitudes \cite{Sen:2020cef,Sen:2021tpp,Alexandrov:2021dyl}.

Even once the data of an effective theory is in hand, 
exploring the resulting landscape of vacua is computationally challenging: one often faces exponentially many possible choices of flux, and a scalar potential depending on hundreds of moduli and axions.\footnote{It was argued in \cite{Acharya:2006zw, Grimm:2020cda, Bakker:2021uqw} that although the flux landscape is very large, it is finite, but the arguments are not yet conclusive.  The construction and enumeration of flux vacua was pursued in e.g.~\cite{Braun:2020jrx,Bena:2021wyr,Grimm:2021ckh,Tsagkaris:2022apo,Plauschinn:2023hjw}, and techniques for computation in axion landscapes were developed in \cite{Gendler:2023hwg}.}  The \texttt{JAXVacua} framework \cite{Dubey:2023dvu},
which deploys modern computational methods including automatic differentiation and just-in-time compilation, marks a significant advance in the computational frontier of moduli stabilization.  For recent applications, see~\cite{Krippendorf:2023idy,Ebelt:2023clh}.

\section{Conclusions}

The past two decades have seen striking advances in 
moduli stabilization, which
is a 
prerequisite for
any attempt 
to connect string theory to
low-energy experiments. 
In this work we have surveyed the leading approaches towards the still-distant goal 
of stabilization in 
cosmologically realistic vacua.
 
The moduli stabilization mechanisms that we have reviewed
invoke a common set of sources that contribute to lifting the moduli spaces found in Calabi-Yau vacuum configurations.
These sources include quantized fluxes, D-brane and orientifold configurations, and perturbative and non-perturbative quantum effects.
Crucially, these sources are not ad-hoc modifications of the theory: they are instead necessary components of 
general compactifications. 
Setting the fluxes to zero, rather than to one of the vast number of nonzero values allowed by consistency, is a fine-tuning. 
Similarly, as in any physical system, quantum effects are not optional. 
Compactifications in which all such sources are omitted generally preserve more supersymmetry, are easier to analyze, and hence were the first to be understood.  This state of affairs 
is akin to the historical development of cosmology, in which maximally symmetric vacuum spacetimes were understood long before realistic inhomogeneous universes.  
Likewise, the rich landscape of modern string compactifications is built on the older foundation of moduli spaces of highly-supersymmetric solutions.

The art of moduli stabilization lies not in finding contributions to the scalar potential for the moduli --- indeed, such terms are unavoidable in generic compactifications --- but in identifying resulting minima that occur in a region of computational control.
The Dine-Seiberg problem indicates that arbitrarily precise control is only guaranteed in the region of moduli space where the scalar potential is a runaway, and so non-trivial vacua most naturally  appear in regions of strong coupling.\footnote{Nature has been kind to us in furnishing weak couplings at high energies: recall that with the spectrum of the Standard Model, all three gauge couplings are weak  at energies close to the Planck scale (the QED Landau pole is at exponentially larger energies).  This suggests that perhaps 
the ultraviolet-complete theory will allow weakly-coupled vacua at high energies.}

Fortunately, the landscape of topological choices --- of quantized fluxes, D-brane configurations, and the underlying compactification topology --- is vast enough to allow for solutions that \emph{are} controllable.  The specific moduli stabilization scenarios that we reviewed, such as KKLT and LVS, involve strategic choices that cause minima to appear in computable parameter regimes.  The necessary choices are discrete fine-tunings, but they are achievable.
For example, the selection of flux quanta giving a small classical flux superpotential in the KKLT scenario leads to a vacuum in which the  Calabi-Yau volume is large, and this can be accomplished in actual examples \cite{Demirtas:2021nlu}.

The task of ensuring that all approximations are under control remains the key technical frontier of moduli stabilization.  With each advance in understanding corrections to the four-dimensional EFTs of string compactifications, one can reduce the theory errors in existing constructions, or pursue new constructions that furnish a broader and less fine-tuned class of stabilized vacua.  
Despite the achievements of recent years, the number of actual solutions that have been constructed is extremely small compared to the astronomical number of compactifications known to exist. 
Moreover, much of what is now understood applies under particular lampposts, such as the geometric regime in Ricci-flat compactifications.
Whole realms of vacua lie undiscovered.

Finally, arguably the most important problem in moduli stabilization is obtaining de Sitter solutions that account for the observed acceleration of the universe.  
We have seen no meaningful evidence of any  
fundamental obstacle to the existence of de Sitter vacua.
However, much work remains to establish a landscape of de Sitter vacua in 
totally explicit and precisely-controlled compactifications of string theory.

The new computational techniques reviewed in \S\ref{sec:computation} may prove crucial for addressing these questions, for systematically constructing more general classes of flux compactifications, and eventually for establishing the structure of cosmological solutions of string theory.

\section*{Acknowledgments}
We would like to thank Carlo Angelantonj and Ignatios Antoniadis, and all the editors of the Handbook on Quantum Gravity, for inviting us to write this review, and for their infinite patience through many delays. We also thank Shamit Kachru for early collaboration on this project.
We are grateful to Michele Cicoli, Joe Conlon, Naomi Gendler, Jim Halverson, Manki Kim, Sven Krippendorf,  David J.E.~Marsh, Raffaele Savelli, Elijah Sheridan, Pramod Shukla, Roberto Valandro, and Timm Wrase for comments on a preliminary draft, and we particularly thank Jakob Moritz and Andreas Schachner for many corrections.  The work of L.M.~was supported in part by NSF grant PHY–2014071.

\bibliographystyle{utphys}
\bibliography{refs}

\end{document}